\documentclass{llncs}

\usepackage{common}
\include{ms.bib}
\usepackage{chngcntr}
\usepackage{url}

\newcounter{ex}
\newenvironment{exmp}{
\refstepcounter{ex}%
\par\vspace{.8em}\hfill\framebox{\small Example~\theex}\nopagebreak
\\[-3.23em]\nopagebreak}
{\relax}

\begin{document}

\setlength{\pdfpageheight}{\paperheight}
\setlength{\pdfpagewidth}{\paperwidth}

%

\newcommand{\clj}[1]{\lstinline{#1}}
\newcommand{\java}[1]{\lstinline{#1}}
\newcommand{\rkt}[1]{\lstinline{#1}}





\title{Practical Optional Types for Clojure}

\author{\ma{\text{Ambrose Bonnaire-Sergeant}^\dagger\text{, Rowan Davies*}, \text{Sam Tobin-Hochstadt}^\dagger}}
\institute{Indiana University$\dagger$; Omnia Team, Commonwealth Bank of Australia*
  \{abonnair,samth\}@indiana.edu, Rowan.Davies@cba.com.au}

\maketitle

\begin{abstract}
  Typed Clojure is an optional type system for Clojure, a dynamic
language in the Lisp family that targets the JVM. Typed Clojure
enables Clojure programmers to gain greater confidence in the
correctness of their code via static type checking while remaining in
the Clojure world, and has acquired significant adoption in the
Clojure community. Typed Clojure repurposes Typed Racket's
\emph{occurrence typing}, an approach to statically reasoning about
predicate tests, and also includes several new type system features to
handle existing Clojure idioms.

In this paper, we describe Typed Clojure and present these type system
extensions, focusing on three features widely used in Clojure. 
 First, multimethods provide extensible
operations, and their Clojure semantics turns out to have a surprising
synergy with the underlying occurrence typing framework.
Second, Java
interoperability is central to Clojure's mission but introduces
challenges such as ubiquitous \texttt{null}; Typed Clojure handles
Java interoperability while ensuring the absence of null-pointer
exceptions in typed programs. 
Third, Clojure programmers
idiomatically use immutable dictionaries for data structures; Typed
Clojure handles this with multiple forms of
heterogeneous dictionary types.

We provide a formal model of the Typed Clojure type system
incorporating these and other features, with a proof of
soundness. Additionally, Typed Clojure is now in use by numerous
corporations and developers working with Clojure, and we present
a quantitative analysis on the use of type system features
in two substantial code bases.


\end{abstract}



\section{Clojure with static typing}


The popularity of dynamically-typed languages in software
development, combined with a recognition that types often improve
programmer productivity, software reliability, and performance, has
led to the recent development of a wide variety of optional and
gradual type systems aimed at checking existing programs written in
existing languages.  These include  TypeScript~\cite{typescript} and Flow~\cite{flow} for
JavaScript, Hack~\cite{hack} for PHP, and mypy~\cite{mypy}
for Python among the optional systems, and Typed Racket~\cite{TF08}, Reticulated
Python~\cite{Vitousek14}, and Gradualtalk~\cite{gradualtalk} among gradually-typed systems.\footnote{We
  use ``gradual typing'' for systems like Typed Racket with sound
  interoperation between typed and untyped code; Typed Clojure or
 TypeScript which don't
  enforce type invariants we describe as ``optionally typed''.}

One key lesson of these systems, indeed a lesson known to early
developers of optional type systems such as Strongtalk, is that type
systems for existing languages must be designed to work with the
features and idioms of the target language. Often this takes the form
of a core language, be it of functions or classes and objects,
together with extensions to handle distinctive language features.

We synthesize these lessons to present \emph{Typed Clojure}, an
optional type system for Clojure. 
Clojure is a dynamically
typed language in the Lisp family---built on the Java Virtual
Machine (JVM)---which has recently gained popularity as an alternative
JVM language.  It offers the flexibility of a Lisp dialect, including
macros, emphasizes a functional style via
immutable data structures, and provides
interoperability with existing Java code, allowing programmers to use
existing Java libraries without leaving Clojure.
Since its initial release in 2007, Clojure has been widely adopted for
``backend'' development in places where its support for parallelism,
functional programming, and Lisp-influenced abstraction is desired on
the JVM. As a result, there is an extensive base of existing untyped
programs whose developers can benefit from Typed Clojure,
an experience we discuss in this paper.

Since Clojure is a language in the
Lisp family, we apply the lessons of Typed Racket, an existing gradual type
system for Racket, to the core of Typed Clojure, consisting of an extended
$\lambda$-calculus over a variety of base types shared between all Lisp systems.
Furthermore, Typed Racket's \emph{occurrence typing} has proved
necessary for type checking realistic Clojure programs.

\begin{figure*}[t!]
  \normalsize
\begin{lstlisting}
(*typed ann pname [(U File String) -> (U nil String)] typed*)
(defmulti pname class)  ; multimethod dispatching on class of argument
(defmethod pname String [s] (*invoke pname (*interop new File s interop*) invoke*)) ; String case 
(defmethod pname File [f] (*interop .getName f interop*)) ; File case, static null check
(*invoke pname "STAINS/JELLY" invoke*) ;=> "JELLY" :- (U nil Str)
\end{lstlisting}
\caption{A simple Typed Clojure program (delimiters: {\color{interop}Java interoperation (green)}, 
  {\color{types}type annotation (blue)},
  {\color{invoke}function invocation (black)}, {\color{red}collection literal (red)}, {\color{mygray}other (gray)})}
\label{fig:ex1}
\end{figure*}

However, Clojure goes beyond Racket in many ways, requiring several
new type system features which we detail in this paper.
Most significantly, Clojure supports, and Clojure developers use,
\textbf{multimethods} to structure their code in extensible
fashion. Furthermore, since Clojure is an untyped language, dispatch
within multimethods is determined by application of dynamic predicates
to argument values. 
Fortunately, the dynamic dispatch used by multimethods has surprising
symmetry with the conditional dispatch handled by occurrence
typing. Typed Clojure is therefore able to effectively handle complex
and highly dynamic dispatch as present in existing Clojure programs. 

But multimethods are not the only Clojure feature crucial to type
checking existing programs. As a language built on the Java Virtual
Machine, Clojure provides flexible and transparent access to existing
Java libraries, and \textbf{Clojure/Java interoperation} is found in almost
every significant Clojure code base. Typed Clojure therefore builds in
an understanding of the Java type system and handles interoperation
appropriately. Notably, \texttt{null} is a distinct type in Typed Clojure,
designed to automatically rule out null-pointer exceptions.

An example of these features is given in
\figref{fig:ex1}. Here, the \clj{pname} multimethod dispatches
on the \clj{class} of the argument---for \clj{String}s,
the first method implementation is called, for \clj{File}s, the
second. The \clj{String} method calls
a \clj{File} constructor, returning a non-nil \clj{File} instance---the 
\clj{getName} method 
on \clj{File} requires a non-nil target, returning a nilable
type.  

Finally, flexible, high-performance immutable dictionaries
are the most common Clojure data structure.
Simply treating them as uniformly-typed
key-value mappings would be insufficient for existing
programs and programming styles. Instead, Typed Clojure provides a
flexible \textbf{heterogenous map} type, in which specific entries can be specified. 

While these features may seem disparate, they are unified in important
ways. First, they leverage the type system mechanisms
inherited from Typed Racket---multimethods when using 
dispatch via predicates, Java interoperation for handling
\texttt{null} tests, and heterogenous maps using union types and
reasoning about subcomponents of data. Second,
they are crucial features for handling Clojure code in
practice. Typed Clojure's use in real Clojure deployments would not be
possible without effective handling of these three Clojure features. 


Our main contributions are as follows:

\begin{enumerate}
  \item We motivate and describe  Typed Clojure, an optional
    type system for Clojure that understands existing Clojure idioms.
  \item We present a sound formal model for three crucial type
    system features: multi-methods, Java
    interoperability, and heterogenous maps.
  \item We evaluate the use of Typed Clojure features on existing
    Typed Clojure code, including both open source and in-house systems.
\end{enumerate}




\noindent
 The remainder of this paper begins with an example-driven
 presentation of the main type system features in
 \secref{sec:overview}. We then incrementally present a core calculus
 for Typed Clojure covering all of these features together in
 \secref{sec:formal} and prove type soundness
 (\secref{sec:metatheory}). We then 
 present an empirical analysis of significant code bases written
 in \coretyped{}---the full implementation of Typed Clojure---in \secref{sec:experience}. 
 Finally, we discuss related work and conclude.

\section{Overview of Typed Clojure}

\label{sec:overview}

We now begin a tour of the central features of Typed Clojure,
beginning with Clojure itself. Our presentation
uses the full Typed Clojure system to illustrate key type system
ideas,\footnote{Full examples: \url{https://github.com/typedclojure/esop16}} before studying the core features in detail in
\secref{sec:formal}.

\subsection{Clojure}

Clojure~\cite{Hic08} is a Lisp that runs on the
Java Virtual Machine with support for concurrent programming
and immutable data structures in a mostly-functional
style.
%
Clojure provides easy interoperation with existing Java libraries, with Java values being like any other Clojure value. 
However, this smooth interoperability comes at the cost of pervasive \java{null}, which leads to the possibility of null pointer exceptions---a drawback we address in Typed Clojure.

%
%
%

\subsection{Typed Clojure}

A simple one-argument function \clj{greet} is annotated with \clj{ann} to take and return strings.

\begin{lstlisting}
(*typed ann  greet [Str -> Str] typed*)
(defn greet [n] (*invoke str "Hello, " n "!" invoke*))
(*invoke greet "Grace" invoke*) ;=> "Hello, Grace!" :- Str
\end{lstlisting}
Providing \clj{nil} (exactly Java's \java{null})
is a static type error---\clj{nil} is not a string.
\begin{lstlisting}
(*invoke greet nil invoke*) ; Type Error: Expected Str, given nil
\end{lstlisting}

\paragraph{Unions} To allow \clj{nil}, we use \emph{ad-hoc unions} (\clj{nil} and \clj{false}
are logically false).
\begin{lstlisting}
(*typed ann  greet-nil [(U nil Str) -> Str] typed*)
(defn greet-nil [n] (*invoke str "Hello" (when n (*invoke str ", " n invoke*)) "!" invoke*))
(*invoke greet-nil "Donald" invoke*) ;=> "Hello, Donald!" :- Str 
(*invoke greet-nil nil invoke*)      ;=> "Hello!"         :- Str
\end{lstlisting}
Typed Clojure prevents well-typed code from dereferencing \clj{nil}.

\paragraph{Flow analysis} Occurrence typing~\cite{TF10}
models type-based control flow.
In \clj{greetings}, a branch ensures \clj{repeat}
is never passed \clj{nil}.
\begin{lstlisting}
(*typed ann  greetings [Str (U nil Int) -> Str] typed*)
(defn greetings [n i]
  (*invoke str "Hello, " (when i (*invoke apply str (*invoke repeat i "hello, " invoke*) invoke*)) n "!" invoke*))
(*invoke greetings "Donald" 2 invoke*)  ;=> "Hello, hello, hello, Donald!" :- Str
(*invoke greetings "Grace" nil invoke*) ;=> "Hello, Grace!"                :- Str
\end{lstlisting}
Removing the branch is a static type error---\clj{repeat} cannot be passed \clj{nil}.
\begin{lstlisting}
(*typed ann  greetings-bad [Str (U nil Int) -> Str] typed*)
(defn greetings-bad [n i]           ; Expected Int, given (U nil Int)
  (*invoke str "Hello, " (*invoke apply str (*invoke repeat i "hello, " invoke*) invoke*) n "!" invoke*))
\end{lstlisting}

\subsection{Java interoperability}
\label{sec:overviewjavainterop}

Clojure can interact with Java constructors, methods, and fields.
This program calls the \clj{getParent} on a constructed
\clj{File}
instance, returning a nullable string.

\begin{exmp}
\begin{lstlisting}
(*interop .getParent (*interop new File "a/b" interop*) interop*)  ;=> "a" :- (U nil Str)
\end{lstlisting}
\label{example:getparent-direct-constructor}
\end{exmp}
Typed Clojure can integrate with the Clojure compiler to avoid expensive reflective 
calls like \clj{getParent}, however if a specific overload cannot be found based on the
surrounding static context, a type error is thrown.
\begin{lstlisting}
(fn [f] (*interop .getParent f interop*)) ; Type Error: Unresolved interop: getParent
\end{lstlisting}
Function arguments default to \clj{Any}, which is similar to a union of all types. Ascribing
a parameter type allows Typed Clojure to find a specific method.


\begin{exmp}
\begin{lstlisting}
(*typed ann parent [(U nil File) -> (U nil Str)] typed*)
(defn parent [f] (if f (*interop .getParent f interop*) nil))
\end{lstlisting}
\label{example:parent-if}
\end{exmp}


The conditional guards from dereferencing \clj{nil}, and---as before---removing 
it is a static type error, as typed code could possibly dereference \clj{nil}.
\begin{lstlisting}
(defn parent-bad-in [f :- (U nil File)]
  (*interop .getParent f interop*)) ; Type Error: Cannot call instance method on nil.
\end{lstlisting}

Typed Clojure rejects programs that assume methods cannot return \clj{nil}.
\begin{lstlisting}
(defn parent-bad-out [f :- File] :- Str
  (*interop .getParent f interop*)) ; Type Error: Expected Str, given (U nil Str).
\end{lstlisting}
Method targets can never be \clj{nil}.
Typed Clojure also prevents passing \clj{nil} as Java method or
constructor arguments by default---this restriction can be
adjusted per method.

%
%
%
In contrast, JVM invariants guarantee constructors return non-null.\footnote{\url{http://docs.oracle.com/javase/specs/jls/se7/html/jls-15.html#jls-15.9.4}}
\begin{exmp}
\begin{lstlisting}
(*invoke parent (*interop new File s interop*) invoke*)
\end{lstlisting}
\end{exmp}

\subsection{Multimethods}

\label{sec:multioverview}

\emph{Multimethods} are a kind of extensible function---combining a \emph{dispatch function} with 
one or more \emph{methods}---widely used to define Clojure operations.

\paragraph{Value-based dispatch}
This simple multimethod takes a keyword (\clj{Kw}) and says hello in different languages.

\begin{exmp}
\begin{lstlisting}
(*typed ann hi [Kw -> Str] typed*) ; multimethod type
(defmulti hi identity) ; dispatch function `identity`
(defmethod hi :en [_] "hello") ; method for `:en`
(defmethod hi :fr [_] "bonjour") ; method for `:fr`
(defmethod hi :default [_] "um...") ; default method
\end{lstlisting}
\label{example:hi-multimethod}
\end{exmp}

When invoked, the arguments are first supplied to the dispatch function---\clj{identity}---yielding
a \emph{dispatch value}. A method is then chosen
based on the dispatch value, to which the arguments are then passed to return a value.
\begin{lstlisting}
(*invoke map hi [*vec :en :fr :bocce vec*] invoke*) ;=> (*list "hello" "bonjour" "um..." list*)
\end{lstlisting}
For example, 
\clj{(*invoke hi :en invoke*)} evaluates to \clj{"hello"}---it executes
the \clj{:en} method
because \clj{(*invoke = (*invoke identity :en invoke*) :en invoke*)} is true
and \clj{(*invoke = (*invoke identity :en invoke*) :fr invoke*)} is false.

Dispatching based on literal values enables certain forms of method
definition, but this is only part of the story for multimethod dispatch.

\paragraph{Class-based dispatch}
For class values, multimethods can choose methods based on subclassing
relationships.
Recall the multimethod from \figref{fig:ex1}. 
%
The dispatch function \clj{class}
dictates 
whether the \clj{String} or \clj{File} method is chosen.
%
The multimethod dispatch rules use
\clj{isa?}, a hybrid predicate which is both a subclassing check for classes and
an equality check for other values.

\begin{lstlisting}
(*invoke isa? :en :en invoke*)       ;=> true
(*invoke isa? String Object invoke*) ;=> true
\end{lstlisting}
The current dispatch value and---in turn---each method's associated dispatch value
is supplied to \clj{isa?}. If exactly one method returns true, it is chosen.
For example,
the call
  \clj{(*invoke pname "STAINS/JELLY" invoke*)}
picks the \clj{String} method because \clj{(*invoke isa? String String invoke*)}
is true, and
\clj{(*invoke isa? String File invoke*)}
is not.

\subsection{Heterogeneous hash-maps}

The most common way to represent compound data in Clojure 
are immutable hash-maps, typicially with keyword keys.
Keywords double as functions that
look themselves up in a map, or return \clj{nil} if absent.
\begin{exmp}
\begin{lstlisting}
(def breakfast {*map :en "waffles" :fr "croissants" map*})
(*invoke :en breakfast invoke*)    ;=> "waffles" :- Str
(*invoke :bocce breakfast invoke*) ;=> nil       :- nil
\end{lstlisting}
\label{example:breakfastcomplete}
\end{exmp}

\emph{HMap types} describe the most common usages of
keyword-keyed maps.
\begin{lstlisting}
breakfast ; :- (HMap :mandatory {:en Str, :fr Str}, :complete? true)
\end{lstlisting}
This says
\clj{:en} and \clj{:fr} are known entries mapped to strings,
and the map is fully specified---that is, no other entries exist---by \clj{:complete?} being \clj{true}.

HMap types default to partial specification, with
\clj{'\{:en Str :fr Str\}} abbreviating \clj{(HMap :mandatory \{:en Str, :fr Str\})}.
\begin{exmp}
\begin{lstlisting}
(*typed ann lunch '{:en Str :fr Str} typed*)
(def lunch {*map :en "muffin" :fr "baguette" map*})
(*invoke :bocce lunch invoke*) ;=> nil :- Any ; less accurate type
\end{lstlisting}
\label{example:lunchpartial}
\end{exmp}

\paragraph{HMaps in practice} The next example is extracted from a production system at CircleCI,
a company with a large production Typed Clojure system
(\secref{sec:casestudy} presents a case study and empirical
result from this code base).

\newpage

\begin{exmp}
\begin{lstlisting}
(*typed defalias RawKeyPair ; extra keys disallowed
  (HMap :mandatory {:pub RawKey, :priv RawKey}, 
        :complete? true) typed*)
(*typed defalias EncKeyPair ; extra keys disallowed
  (HMap :mandatory {:pub RawKey, :enc-priv EncKey}, :complete? true) typed*)

(*typed ann enc-keypair [RawKeyPair -> EncKeyPair] typed*)
(defn enc-keypair [kp]
  (*invoke assoc (*invoke dissoc kp :priv invoke*) :enc-priv (*invoke encrypt (*invoke :priv kp invoke*) invoke*) invoke*))
\end{lstlisting}
\label{example:circleci}
\end{exmp}
%
%
As \clj{EncKeyPair} is fully specified, we remove extra keys like \clj{:priv}
via \clj{dissoc}, which returns a new map that is the first argument without the
entry named by the second argument. Notice removing \clj{dissoc} causes a type error.
\begin{lstlisting}
(defn enc-keypair-bad [kp] ; Type error: :priv disallowed
  (*invoke assoc kp :enc-priv (*invoke encrypt (*invoke :priv kp invoke*) invoke*) invoke*))
\end{lstlisting}

\subsection{HMaps and multimethods, joined at the hip}

HMaps and multimethods are the primary ways for representing
and dispatching on data respectively, and so are intrinsically linked.
As type system designers, we must
search for a compositional approach that can anticipate
any combination of these features.

Thankfully, occurrence typing, originally designed for reasoning about
\clj{if} tests, provides the compositional approach we need.
By extending the system with
a handful of rules based on HMaps and other functions, 
we can automatically cover both easy cases and those
that compose rules in arbitrary ways.

Futhermore, this approach extends to multimethod dispatch by reusing
occurrence typing's approach to conditionals
and
encoding a small number of rules to handle
the \clj{isa?}-based dispatch.
In practice, conditional-based control flow typing
extends to multimethod dispatch, and vice-versa.

We first demonstrate a very common, simple dispatch style,
then move on to deeper structural dispatching where occurrence typing's
compositionality shines.

\paragraph{HMaps and unions} Partially specified HMap's with a common dispatch key
combine naturally with ad-hoc unions.
An \clj{Order} is one of three kinds of HMaps.


\begin{lstlisting}
(*typed defalias Order "A meal order, tracking dessert quantities."
  (U '{:Meal ':lunch, :desserts Int} '{:Meal ':dinner :desserts Int}
     '{:Meal ':combo :meal1 Order :meal2 Order}) typed*)
\end{lstlisting}

The \clj{:Meal} entry is common to each HMap, always mapped to a known keyword singleton
type.
It's natural to dispatch on the \clj{class} of an instance---it's similarly
natural to dispatch on a known entry like \clj{:Meal}.

\newpage

\begin{exmp}
\begin{lstlisting}
(*typed ann desserts [Order -> Int] typed*)
(defmulti desserts :Meal)  ; dispatch on :Meal entry
(defmethod desserts :lunch [o] (*invoke :desserts o invoke*))
(defmethod desserts :dinner [o] (*invoke :desserts o invoke*))
(defmethod desserts :combo [o] 
  (*invoke + (*invoke desserts (*invoke :meal1 o invoke*) invoke*) (*invoke desserts (*invoke :meal2 o invoke*) invoke*) invoke*))

(*invoke desserts {*map :Meal :combo, :meal1 {*map :Meal :lunch :desserts 1 map*}, 
           :meal2 {*map :Meal :dinner :desserts 2 map*} map*} invoke*) ;=> 3
\end{lstlisting}
\label{example:desserts-on-meal}
\end{exmp}
The \clj{:combo} method is verified to only structurally recur
on \clj{Order}s. This is achieved because we learn the argument \clj{o} must
be of type
\clj{'\{:Meal ':combo\}}
since
\clj{(isa? (:Meal o) :combo)}
is true. Combining this
with the fact that \clj{o} is an \clj{Order}
eliminates possibility of \clj{:lunch} and \clj{:dinner}
orders, simplifying \clj{o} to
\clj{'\{:Meal ':combo :meal1 Order :meal2 Order\}}
which contains appropriate arguments for both recursive calls.

\paragraph{Nested dispatch}
A more exotic dispatch mechanism for \clj{desserts}
might be on the \clj{class} of the \clj{:desserts} key.
If the result is a number, then we know the \clj{:desserts}
key is a number, otherwise the input is a \clj{:combo} meal.
We have already seen dispatch on \clj{class} and on keywords
in isolation---occurrence typing automatically understands
control flow that combines its simple building blocks.

The first method has dispatch value \clj{Long}, a subtype
of \clj{Int}, and the second method has \clj{nil}, the sentinel value for a failed map lookup.
In practice, \clj{:lunch} and \clj{:dinner} meals will dispatch to the \clj{Long}
method, but Typed Clojure infers a slightly more general type due to the definition
of \clj{:combo} meals.

\begin{exmp}
\begin{lstlisting}
(*typed ann desserts' [Order -> Int] typed*)
(defmulti desserts' 
  (fn [o :- Order] (*invoke class (*invoke :desserts o invoke*) invoke*)))
(defmethod desserts' Long [o] 
;o :- (U '{:Meal (U ':dinner ':lunch), :desserts Int}
;       '{:Meal ':combo, :desserts Int, :meal1 Order, :meal2 Order})
  (*invoke :desserts o invoke*))
(defmethod desserts' nil [o]
  ; o :- '{:Meal ':combo, :meal1 Order, :meal2 Order}
  (*invoke + (*invoke desserts' (*invoke :meal1 o invoke*) invoke*) (*invoke desserts' (*invoke :meal2 o invoke*) invoke*) invoke*))
\end{lstlisting}
\label{example:desserts-on-class}
\end{exmp}
%

In the \clj{Long} method, Typed Clojure learns that
its argument is at least of type \clj{'\{:desserts Long\}}---since
\clj{(*invoke isa? (*invoke class (*invoke :desserts o invoke*) invoke*) Long invoke*)}
must be true.
%
Here the
\clj{:desserts} entry
\emph{must} be present and mapped to a \clj{Long}---even in a \clj{:combo} meal,
which does not specify \clj{:desserts}
as present or absent.

In the \clj{nil} method,
\clj{(*invoke isa? (*invoke class (*invoke :desserts o invoke*) invoke*) nil invoke*)}
must be true---which implies \clj{(*invoke class (*invoke :desserts o invoke*) invoke*)} is \clj{nil}.
Since lookups on missing keys return \clj{nil}, either
\begin{itemize}
  \item \clj{o} maps the \clj{:desserts} entry to \clj{nil}, like the value \clj{\{:desserts nil\}}, or
  \item \clj{o} is missing a \clj{:desserts} entry.
\end{itemize}
We can express this type with the \clj{:absent-keys} HMap option
\begin{lstlisting}
(U '{:desserts nil} (HMap :absent-keys #{:desserts}))
\end{lstlisting}
This eliminates non-\clj{:combo} meals
since their \clj{'\{:desserts Int\}} type does not agree
with this new information (because \clj{:desserts}
is neither \clj{nil} or absent).


\paragraph{From multiple to arbitrary dispatch}
Clojure multimethod dispatch, and Typed Clojure's handling of it, goes
even further, supporting dispatch on multiple arguments via vectors.
Dispatch on multiple arguments is beyond the scope of this paper,
but the same intuition applies---adding support for multiple dispatch
admits arbitrary combinations and nestings
of it and previous dispatch rules.

\section{A Formal Model of \lambdatc{}}

\label{sec:formal}

After demonstrating the core features of Typed Clojure, 
we link them together in a formal model called
\lambdatc{}.
Building on occurrence typing,
we incrementally add each
novel feature of Typed Clojure to the formalism,
interleaving presentation of syntax, typing rules, operational semantics,
and subtyping.

\subsection{Core type system}
\label{sec:coretypesystem}

We start with a review of
occurrence typing~\cite{TF10}, the foundation of \lambdatc{}.

\paragraph{Expressions} Syntax is given in \figref{main:figure:termsyntax}. Expressions \e{} 
include variables \x{}, values \v{},
applications, abstractions, conditionals, and let expressions.
All binding forms introduce fresh variables---a subtle but important point since our type environments
are not simply dictionaries.
Values include booleans \bool{}, \nil{}, class literals {\class{}}, keywords \k{},
integers {\nat{}},
constants {\const{}}, and strings \str{}. Lexical closures {\closure {\openv{}} {\abs {\x{}} {\t{}} {\e{}}}}
close value environments \openv{}---which map bindings to values---over functions.

\paragraph{Types} Types \s{} or \t{} 
include the top type \Top,
\emph{untagged} unions {\Unionsplice {\overrightarrow{\t{}}}}, 
singletons ${\Value \singletonmeta{}}$,
and class instances \class{}.
We abbreviate the classes
\Booleanlong{} to \Boolean{}, 
\Keywordlong{} to \Keyword{},
\NumberFull{}  to \Number{},
\StringFull{}  to \String{}, and 
\FileFull{}  to \File{}.
We also abbreviate the types
\EmptyUnion{}     to \Bot{}, 
{\ValueNil}       to \Nil{}, 
{\ValueTrue}      to \True, and
{\ValueFalse} to {\False}.
The difference between the types
\Value{\class{}} and \class{} is subtle.
The former is inhabited by class literals like \Keyword{} and the result of 
\appexp{\classconst{}}{\makekw{a}}---the latter by \emph{instances} of classes,
like a keyword literal \makekw{a}, an instance of the type \Keyword{}.
Function types 
$
{\ArrowOne {\x{}} {\s{}}
             {\t{}}
             {\filterset {\prop{}} {\prop{}}}
             {\object{}}}
$
contain \emph{latent} (terminology from~\cite{Lucassen88polymorphiceffect}) propositions \prop{}, object \object{}, and return type
\t{},
which may refer to the function argument \x{}.
They are instantiated with the
actual object of the argument in applications. 

\paragraph{Objects}
Each expression is associated with 
a symbolic representation
called an \emph{object}.
For example,
  variable \makelocal{m} has object \makelocal{m};
  $\appexpone{\ccclass{\appexp{\makekw{lunch}}{\makelocal{m}}}}$ has object ${\path{\classpe{}}{\path{\keype{\makekw{lunch}}}{\makelocal{m}}}}$; and $42$ has the \emph{empty} object \emptyobject{} since it is unimportant in our system.
\figref{main:figure:termsyntax} gives the syntax for objects \object{}---non-empty objects 
\path{\pathelem{}}{\x{}} combine of a root variable \x{} and a \emph{path} \pathelem{},
which consists of
a possibly-empty sequence of \emph{path elements} (\pesyntax{}) applied right-to-left from the root variable.
We use two path elements---{\classpe{}} and {\keype{k}}---representing the results
of calling \classconst{} and looking up a keyword $k$, respectively.

\paragraph{Propositions with a logical system}
In standard type systems, association lists often
track the types of variables, like in LC-Let and LC-Local.
\begin{mathpar}
\infer [LC-Let]
{ \judgementtwo {{\propenv{}}}
                {\e{1}} {\s{}}
  \\
  \judgementtwo {{\propenv{}},\x{} \mapsto {\s{}}}
                {\e{2}} {\t{}}
}
{ 
  \judgementtwo {\propenv{}} 
            {\letexp{\x{}}{\e{1}}{\e{2}}} {\t{}}
           }

\infer [LC-Local]
{ {\propenv{}}(\x{}) = {\t{}}
}
{ \judgementtwo {\propenv{}} 
            {\x{}} {\t{}}
           }
\end{mathpar}

Occurrence typing instead pairs \emph{logical formulas},
that can reason about arbitrary non-empty objects,
with a \emph{proof system}.
The logical statement {\isprop{\s{}}{\x{}}} says
variable $x$ is of type \s{}. 
\begin{mathpar}
\infer [T0-Let]
{ \judgementtwo {{\propenv{}}}
                {\e{1}} {\s{}}
  \\
  \judgementtwo {{\propenv{}},\isprop{\s{}}{\x{}}}
                {\e{2}} {\t{}}
}
{ 
  \judgementtwo {\propenv{}} 
            {\letexp{\x{}}{\e{1}}{\e{2}}}
            {\t{}}
           }

\infer [T0-Local]
{ \inpropenv {\propenv{}} {\isprop {\t{}} {\x{}}}}
{ \judgementtwo {\propenv{}} 
            {\x{}} {\t{}}
           }
\end{mathpar}
In T0-Local, 
$
{ \inpropenv {\propenv{}} {\isprop {\t{}}{\x{}}}}
$
appeals to the proof system to solve for \t{}.

\begin{figure}[t!]
  \footnotesize
$$
\begin{array}{lrll}
  \e{} &::=& \x{}
                      \alt \v{} 
                      \alt {\comb {\e{}} {\e{}}} 
                      \alt {\abs {\x{}} {\t{}} {\e{}}} 
                      \alt {\ifexp {\e{}} {\e{}} {\e{}}}
                      \alt {\letexp {\x{}} {\e{}} {\e{}}}
                      &\mbox{Expressions} \\
  \v{} &::=&          \singletonmeta{}
                      \alt {\nat{}}
                      \alt {\const{}}
                      \alt {\str{}}
                      \alt {\closure {\openv{}} {\abs {\x{}} {\t{}} {\e{}}}}
                &\mbox{Values} \\
                \constantssyntax{}\\
  \s{}, \t{}    &::=& \Top 
                      \alt {\Unionsplice {\overrightarrow{\t{}}}}
                      \alt
                      {\ArrowOne {\x{}} {\t{}}
                                   {\t{}}
                                   {\filterset {\prop{}} {\prop{}}}
                                   {\object{}}}
                      \alt {\Value \singletonmeta{}} 
                      \alt \trdiff{\class{}}
                &\mbox{Types} \\
  \singletonallsyntax{}
                \\ \\
  \occurrencetypingsyntax{}\\
  \propenvsyntax{}\\
  \openvsyntax{}
\end{array}
$$
\caption{Syntax of Terms, Types, Propositions and Objects}
\label{main:figure:termsyntax}
\end{figure}

We further extend logical statements to \emph{propositional logic}.
\figref{main:figure:termsyntax} describes the syntax
for propositions \prop{},
consisting of positive and negative \emph{type propositions} 
about non-empty objects---{\isprop {\t{}} {\path {\pathelem{}} {\x{}}}}
and {\notprop {\t{}} {\path {\pathelem{}} {\x{}}}}
respectively---the latter pronounced ``the object {\path {\pathelem{}} {\x{}}} is \emph{not} of type \t{}''.
The other propositions are standard logical connectives: implications, conjunctions,
disjunctions, and the trivial (\topprop{}) and impossible (\botprop{}) propositions.
The full proof system judgement
$
{ \inpropenv {\propenv{}} {\prop{}} }
$
says \emph{proposition environment} {\propenv{}} proves proposition \prop{}.

Each expression is associated with two propositions---when expression
\e{1} is in test position like
\ifexp{\e{1}}{\e{2}}{\e{3}},
the type system extracts \e{1}'s `then' and `else' proposition to check
\e{2} and \e{3} respectively.
For example, in \ifexp{\makelocal{o}}{\e{2}}{\e{3}}
we learn variable {\makelocal{o}} is true in \e{2} via {\makelocal{o}}'s `then' proposition $\notprop{\falsy{}}{\makelocal{o}}$, and 
that {\makelocal{o}} is false in \e{3} via {\makelocal{o}}'s `else' proposition $\isprop{\falsy{}}{\makelocal{o}}$.

To illustrate, recall \egref{example:desserts-on-meal}.
The parameter \makelocal{o} is of type $\Order$,
written
{\isprop{\Order}{\makelocal{o}}}
as a proposition.
In the ${\makekw{combo}}$ method, we know
${\appexp{\makekw{Meal}}{\makelocal{o}}}$ is ${\makekw{combo}}$,
based on multimethod dispatch rules. This is written
  {\isprop{\Value{\makekw{combo}}}{\path{\keype{\makekw{Meal}}}{\makelocal{o}}}},
pronounced ``the ${\makekw{Meal}}$ path of variable \makelocal{o} is of type
{\Value{\makekw{combo}}}''.

To attain the type of \makelocal{o}, 
we must solve for \t{} in
$
{ \inpropenv 
  {\propenv{}}
  {\isprop {\t{}} {\makelocal{o}}}}
$,
under proposition environment
$
\propenv{} = {{\isprop{\Order}{\makelocal{o}}},
    {\isprop{\Value{\makekw{combo}}}{\path{\keype{\makekw{Meal}}}{\makelocal{o}}}}}
$
which deduces \t{} to be a {\makekw{combo}} meal.
The logical
system \emph{combines} pieces of type information to deduce more accurate types for lexical
bindings---this is explained in \secref{formalmodel:proofsystem}.

%

\begin{figure*}[t]
\footnotesize
  \begin{mathpar}
        {\TLocal}
        {\TAbs}
        {\TIf}
    \\
    \begin{array}{c}
    {\TKw}\\
      {\TNum}\\
    \end{array}
    \begin{array}{c}
      {\TNil}\\
      {\TFalse}\\
    {\TConst}
    \end{array}
    \begin{array}{c}
    {\TStr}\\
    {\TClass}\\
    {\TTrue}
  \end{array}
        \\

    {\TLet}
    \\

    {\TApp}\ \ 
    {\TSubsume}
    \\
  \end{mathpar}
  \caption{Core typing rules}
  \label{main:figure:othertypingrules}
\end{figure*}

\begin{figure}
  \footnotesize
  \begin{mathpar}

\SUnionSuper{}\ \ \ 
\SUnionSub{}\ \ \ 
\SFunMono{}\ \ \ 
\begin{array}{l}
\SObject{}\\
\SClass{}\\
\SSBool{}
\end{array}

\SFun{}
\begin{array}{l}
    \SRefl{}\ \ \ 
    \STop{}\\
\SSKw{}
\end{array}

  \end{mathpar}
  \caption{Core subtyping rules}
  \label{main:figure:subtyping}
\end{figure}

\begin{figure}
\begin{mathpar}
    \BIfTrue{}

    \BIfFalse{}
\end{mathpar}
  \caption{Select core semantics}
\label{main:figure:coresemantics}
\end{figure}

\paragraph{Typing judgment}

We formalize our system following Tobin-Hochstadt and Felleisen \cite{TF10}.
The typing judgment 
$
{\judgementrewrite   {\propenv}
              {\e{}} {\t{}}
  {\filterset {\thenprop {\prop{}}}
              {\elseprop {\prop{}}}}
  {\object{}}
  {\ep{}}}
$
says expression \e{} rewrites to \ep{}, which
is of type \t{} in the 
proposition environment $\propenv{}$, with 
`then' proposition {\thenprop {\prop{}}}, `else' proposition {\elseprop {\prop{}}}
and object \object{}. 

We write 
{\judgementtworewrite{\propenv}{\e{}} {\t{}}{\ep{}} 
to mean 
{\judgementrewrite   {\propenv}
              {\e{}} {\t{}}
  {\filterset {\thenprop {\propp{}}}
              {\elseprop {\propp{}}}}
  {\objectp{}}
  {\ep{}}}
for some {\thenprop {\propp{}}}, {\elseprop {\propp{}}}
and {\objectp{}},
and
  abbreviate self rewriting judgements
{\judgementrewrite   {\propenv}
              {\e{}} {\t{}}
  {\filterset {\thenprop {\prop{}}}
              {\elseprop {\prop{}}}}
  {\object{}}
  {\e{}}}
to
{\judgementselfrewrite   {\propenv}
              {\e{}} {\t{}}
  {\filterset {\thenprop {\prop{}}}
              {\elseprop {\prop{}}}}
  {\object{}}}.

\paragraph{Typing rules}

The core typing rules are
given as \figref{main:figure:othertypingrules}. We introduce
the interesting rules with the complement number predicate
as a running example.
\begin{equation}
\abs{\makelocal{d}}{\Top}{\ifexp{\appexp{\numberhuh{}}{\makelocal{d}}}{\false{}}{\true{}}}
\end{equation}

The lambda rule T-Abs introduces \isprop{\s{}}{\x{}}} = \isprop{\Top}{\makelocal{d}}
to check the body.
With \propenv{} = \isprop{\Top}{\makelocal{d}},
T-If first checks the test \e{1} = {\appexp{\numberhuh{}}{\makelocal{d}}}
via the T-App rule, with three steps.

First, in T-App the operator \e{} = \numberhuh{} is checked with T-Const, which
uses 
\constanttypeliteral{} (\figref{main:figure:constanttyping}, dynamic semantics in the supplemental material)
to type constants.
\numberhuh{} is a predicate over numbers, and
\classconst{} returns its argument's class.

Resuming {\appexp{\numberhuh{}}{\makelocal{d}}},
in T-App the operand \ep{} = \makelocal{d} is checked with
T-Local as
\begin{equation}
\judgementselfrewrite{\propenv{}}
                     {\makelocal{d}}
                     {\Top}
                     {\filterset{\notprop{\falsy}{\makelocal{d}}}
                                {\isprop{\falsy}{\makelocal{d}}}}
                     {\makelocal{d}}
\end{equation}
which encodes the type, proposition, and object information
about variables. The proposition {\notprop{\falsy}{\makelocal{d}}}
says ``it is not the case that variable {\makelocal{d}} is of type {\falsy}'';
{\isprop{\falsy}{\makelocal{d}}} says ``{\makelocal{d}} is of type {\falsy}''.

Finally, the T-App rule substitutes the operand's object \objectp{}
for the parameter \x{} in the latent type, propositions, and object. The proposition
{\isprop{\Number{}}{\makelocal{d}}} says ``{\makelocal{d}} is of type {\Number{}}'';
{\notprop{\Number{}}{\makelocal{d}}} says ``it is not the case that {\makelocal{d}}
is of type {\Number{}}''. The object {\makelocal{d}} is the symbolic representation
of what the expression {\makelocal{d}} evaluates to.
\begin{equation}
\judgementselfrewrite{\propenv{}}
  {\appexp{\numberhuh{}}{\makelocal{d}}}
  {\Boolean{}}
  {\filterset{\isprop{\Number{}}{\makelocal{d}}}
             {\notprop{\Number{}}{\makelocal{d}}}}
  {\emptyobject{}}
\end{equation}
To demonstrate, the `then' proposition---in T-App {\replacefor {\thenprop{\prop{}}} {\objectp{}} {\x{}}}---substitutes
the latent `then' proposition of \constanttype{\numberhuh{}} with 
\makelocal{d}, giving
{\replacefor {\isprop{\Number{}}{\x{}}} {\makelocal{d}} {\x{}}} =
{\isprop{\Number{}}{\makelocal{d}}}.

To check the branches of {\ifexp{\appexp{\numberhuh{}}{\makelocal{d}}}{\false{}}{\true{}}},
T-If
introduces \thenprop{\prop{1}} = \isprop{\Number{}}{\makelocal{d}}
to check \e{2} = {\false{}},
and \elseprop{\prop{1}} = \notprop{\Number{}}{\makelocal{d}}
to check 
\e{3} = \true{}.
The branches are first checked with T-False and T-True respectively,
the T-Subsume premises
\inpropenv {\propenv{}, {\thenprop {\prop{}}}} {\thenprop {\propp{}}}
and
\inpropenv {\propenv{}, {\elseprop {\prop{}}}} {\elseprop {\propp{}}}
allow us to pick compatible propositions for both branches.
$$
\begin{array}{c}
\judgementselfrewrite{\propenv{},{\isprop{\Number{}}{\makelocal{d}}}}
  {\false{}}
  {\Boolean{}}
  {\filterset{\notprop{\Number{}}{\makelocal{d}}}
             {\isprop{\Number{}}{\makelocal{d}}}}
  {\emptyobject{}}
  \\
\judgementselfrewrite{\propenv{},{\notprop{\Number{}}{\makelocal{d}}}}
  {\true{}}
  {\Boolean{}}
  {\filterset{\notprop{\Number{}}{\makelocal{d}}}
             {\isprop{\Number{}}{\makelocal{d}}}}
  {\emptyobject{}}
\end{array}
$$
%

Finally T-Abs assigns a type to the overall function:
$$
{\judgementselfrewrite{}{\abs{\makelocal{d}}{\Top}{\ifexp{\appexp{\numberhuh{}}{\makelocal{d}}}{\false{}}{\true{}}}}
                    {\ArrowOne {\makelocal{d}} {\Top{}}
                                      {\Boolean{}}
                                      {\filterset {\notprop{\Number{}}{\makelocal{d}}}
                                                  {\isprop{\Number{}}{\makelocal{d}}}}
                                      {\emptyobject{}}}
                    {\filterset {\topprop{}}
                                {\botprop{}}}
                    {\emptyobject{}}}
$$

\paragraph{Subtyping}
\figref{main:figure:subtyping} presents subtyping
as a reflexive and transitive relation with top type \Top. 
Singleton types are instances of their respective classes---boolean singleton types
are of type \Boolean{}, class literals are instances of \Class{} and keywords are
instances of \Keyword{}.
Instances of classes \class{} are subtypes of \Object{}. Function types 
are subtypes of \IFn{}. All types except for \Nil{} are subtypes of \Object{},
so \Top{} is similar to {\Union{\Nil}{\Object}}.
Function subtyping is contravariant left of the arrow---latent propositions, object
and result type are covariant.
Subtyping for untagged unions is standard.

\paragraph{Operational semantics} We define the dynamic semantics for \lambdatc{}
in a big-step style using an environment, following~\cite{TF10}.
We include both errors and a \wrong{} value, which is provably ruled out by the
type system.
The main judgment is \opsem{\openv{}}{\e{}}{\definedreduction{}}
which states that \e{} evaluates to answer \definedreduction{} in environment
\openv{}. We chose to omit the core rules (included in supplemental material)
however a notable difference is \nil{} is a false value, which affects the
semantics of \ifliteral{} (\figref{main:figure:coresemantics}).

\subsection{Java Interoperability}

\begin{figure}
  \footnotesize
  $$
  \begin{altgrammar}
    \e{} &::=& \ldots   {\fieldexp {\fld{}} {\e{}}} \alt {\methodexp {\mth{}} {\e{}} {\overrightarrow{\e{}}}}
                      \alt {\newexp {\class{}} {\overrightarrow{\e{}}}}
                      &\mbox{Expressions}\\
     &\alt& \nonreflectiveexpsyntax{} &\mbox{Non-reflective Expressions}\\

    \v{} &::=& \ldots \alt {\classvalue{\classhint{}} {\overrightarrow {\classfieldpair{\fld{}} {\v{}}}}}
    &\mbox{Values} \\

    \classtableallsyntax{}
  \end{altgrammar}
  $$
  \begin{mathpar}
    {\TNew}

    {\TMethod}

    {\TField}

  \end{mathpar}
 \begin{mathpar}
  \begin{altgrammar}
    \convertjavatypenil{}
  \end{altgrammar}
  \begin{altgrammar}
    \convertjavatypenonnil{}
  \end{altgrammar}
  \begin{altgrammar}
    \converttctype{}
  \end{altgrammar}
\end{mathpar}
  \begin{mathpar}
    \BField{}\ \ \ 
    \BNew{}

    \BMethod{}
  \end{mathpar}
  \caption{Java Interoperability Syntax, Typing and Operational Semantics}
  \label{main:figure:javatyping}
\end{figure}

\begin{figure}
  $$
\constanttypefigure{}
  $$
  \caption{Constant typing}
  \label{main:figure:constanttyping}
\end{figure}

We present Java interoperability in a restricted setting without class inheritance,
overloading or Java Generics.
We extend the syntax in \figref{main:figure:javatyping} with Java field lookups and calls to
methods and constructors. 
To prevent ambiguity between zero-argument methods and fields,
we use Clojure's primitive ``dot'' syntax:
field accesses are written \fieldexp{\fld{}}{\e{}}
and method calls $\methodexp{\mth{}}{\e{}}{\overrightarrow{e}}$.

In \egref{example:getparent-direct-constructor}, \clj{(*interop .getParent (*interop new File "a/b" interop*) interop*)}
translates to
\begin{equation}  \label{eq:unresolved}
  \qquad {\methodexp {\getparent{}} {\newexp {\File{}} {\makestr{a/b}}} {}}
\end{equation}

But both the constructor and method are unresolved.
We introduce \emph{non-reflective} expressions for specifying exact Java overloads.
\begin{equation} \label{eq:resolved}
\qquad {\methodstaticexp {\File} {} {\String} {\getparent{}} {\newstaticexp {\String} {\File{}} {\File{}} {\makestr{a/b}}} {}}
\end{equation}
From the left, the one-argument constructor for \File takes a \String, and the 
\getparent{} method of
\File{} takes zero arguments
and
returns a \String.

We now walk through this conversion.

\paragraph{Constructors} First we check and convert {\newexp {\File{}} {\makestr{a/b}}} to {\newstaticexp {\String} {\File{}} {\File{}} {\makestr{a/b}}}.
The T-New typing rule checks and rewrites constructors.
To check
{\newexp {\File{}} {\makestr{a/b}}}
we first resolve the constructor overload in the class table---there is at most one
to simplify presentation.
With \classhint{1} = \String,
we convert to a nilable type the argument with \t{1} = \Union{\Nil}{\String}
and type check {\makestr{a/b}} against \t{1}.
Typed Clojure defaults to allowing non-nilable arguments, but this
can be overridden, so we model the more general case.
The return Java type \File is converted to a non-nil
Typed Clojure type \t{} = \File for the return type,
and the propositions say constructors can never be false---constructors
can never produce the internal boolean value that Clojure uses for \false{}, or \nil{}.
Finally, the constructor rewrites to {\newstaticexp {\String} {\File{}} {\File{}} {\makestr{a/b}}}.

\paragraph{Methods} Next we convert {\methodexp {\getparent{}} {\newstaticexp {\String} {\File{}} {\File{}} {\makestr{a/b}}} {}}
to the non-reflective expression
{\methodstaticexp {\File} {} {\String} {\getparent{}} {\newstaticexp {\String} {\File{}} {\File{}} {\makestr{a/b}}} {}}.
The T-Method rule for unresolved methods
checks {\methodexp {\getparent{}} {\newstaticexp {\String} {\File{}} {\File{}} {\makestr{a/b}}} {}}.
We verify the target type \s{} = \File is non-nil by T-New.
The overload is chosen from the class table based on \classhint{1} = \File---there is at most one.
The nilable return type \t{} = \Union{\Nil}{\String} is given, and the 
entire expression rewrites to expression \ref{eq:resolved}.
%

The T-Field rule (\figref{main:figure:javatyping}) is like T-Method, but without arguments.

The evaluation rules B-Field, B-New and B-Method (\figref{main:figure:javatyping}) simply evaluate their
arguments and call the relevant JVM operation, which we do not model---\secref{sec:metatheory}
states our exact assumptions.
There are no evaluation rules for reflective Java interoperability, since there are no typing
rules that rewrite to reflective calls.

\subsection{Multimethod preliminaries: \isaliteral}

\label{sec:isaformal}

We now consider the \isaliteral{} operation, a core part of the multimethod dispatch mechanism. 
Recalling the examples in \secref{sec:multioverview},
\isaliteral{} is
a subclassing test for classes, but otherwise is an equality test.
%
The T-IsA rule uses \isacompareliteral{}
(\figref{main:figure:mmsyntax}), a metafunction which produces the propositions for
\isaliteral{} expressions.

To demonstrate the first \isacompareliteral{} case,
the expression
\isaapp{\appexp{\classconst{}}{\x{}}}{\Keyword}
is true if \x{} is a keyword, otherwise false.
When checked with T-IsA,
the object of the left subexpression \object{} = {\path{\classpe{}}{\x{}}}
(which starts with the {\classpe{}} path element)
and the type of the right subexpression \t{} = {\Value{\Keyword}} (a singleton class type)
together trigger the first \isacompareliteral{} case
\isacompare{\s{}}{\path{\classpe{}}{\x{}}}{\Value{\Keyword}}{\filterset{\isprop{\Keyword}{\x{}}}{\notprop{\Keyword}{\x{}}}},
giving propositions that correspond to our informal description {\filterset{\thenprop{\prop{}}}{\elseprop{\prop{}}}} = {\filterset{\isprop{\Keyword}{\x{}}}{\notprop{\Keyword}{\x{}}}}.

The second \isacompareliteral{} case captures the simple equality mode for non-class singleton types.
For example,
the expression
\isaapp{\x{}}{\makekw{en}} produces true 
when \x{} evaluates to {\makekw{en}}, otherwise it produces false.
Using T-IsA,
it has the propositions {\filterset{\thenprop{\prop{}}}{\elseprop{\prop{}}}} = 
\isacompare{}{\x{}}{\Value{\makekw{en}}}{\filterset {\isprop {\Value{\makekw{en}}}{\x{}}}{\notprop{\Value{\makekw{en}}}{\x{}}}}
since \object{} = {\x{}} and \t{} = {\Value{\makekw{en}}}.
The side condition on the second \isacompareliteral{} case ensures we are in equality mode---if \x{} can possibly be a class in 
\isaapp{\x{}}{\Object{}}, \isacompareliteral{} uses its conservative default case,
since if \x{} is a class literal, subclassing mode could be triggered.
Capture-avoiding substitution of objects {\replacefor {} {\object{}} {\x{}}} used in this case erases propositions
that would otherwise have \emptyobject{} substituted in for their objects---it
is defined in the appendix.

The operational behavior of \isaliteral{} is given by B-IsA (\figref{main:figure:mmsyntax}). \isaopsemliteral{} explicitly handles classes in the second case.



\subsection{Multimethods}

\begin{figure}
  \footnotesize
$$
\begin{altgrammar}
  \e{} &::=& \ldots \alt {\createmultiexp {\t{}} {\e{}}} \alt
             {\extendmultiexp {\e{}} {\e{}} {\e{}}}
             \alt {\isaapp {\e{}} {\e{}}} &\mbox{Expressions} \\
  \v{} &::=& \ldots \alt {\multi {\v{}} {\disptable{}}}
                &\mbox{Values} \\
 \disptablesyntax{} \\
  \s{}, \t{} &::=& \ldots \alt {\MultiFntype{\t{}}{\t{}}}
                &\mbox{Types}
\end{altgrammar}
$$
  \begin{mathpar}
    \TDefMulti{}

    \TDefMethod{}

    \TIsA{}
  \end{mathpar}
  \begin{mathpar}
    \isapropsfigure{}
  \end{mathpar}
  \begin{mathpar}
    \Multisubtyping{}

    \BDefMulti{}
  \end{mathpar}
  \begin{mathpar}
    \BDefMethod{}
  \end{mathpar}
  \getmethodfigure{}
\begin{mathpar}
  {\BIsA{}}
  {\isaopsemfigure{}}
  \\
\BBetaMulti{}
\end{mathpar}
\caption{Multimethod Syntax, Typing and Operational Semantics}
\label{main:figure:mmsyntax}
\end{figure}

\figref{main:figure:mmsyntax} presents \emph{immutable} multimethods without default methods to ease presentation.
\figref{main:figure:mmexample} translates the mutable \egref{example:hi-multimethod} to \lambdatc{}.

\begin{figure}
$\letexp{hi_0} {\createmultiexp {\ArrowOne{\x{}}{\Keyword}{\String}{\filterset{\topprop{}}{\topprop{}}}{\emptyobject{}}} {\abs{\x{}}{\Keyword}{\x{}}}}
  {\\\text{\quad}
    \letexp{hi_1} {\extendmultiexp {hi_0} {\makekw{en}} {\abs {\x{}} {\Keyword} {\makestr{hello}}}}
      {\\\text{\quad\quad}
        \letexp{hi_2} {\extendmultiexp {hi_1} {\makekw{fr}} {\abs {\x{}} {\Keyword} {\makestr{bonjour}}}}
        {\\\text{\quad\quad\quad
          \appexp{hi_2}{\makekw{en}}}}}}
$
\caption{Multimethod example}
\label{main:figure:mmexample}
\end{figure}
%
%
%
%
To check 
{\createmultiexp {\ArrowTwo{\x{}}{\Keyword}{\String}} {\abs{\x{}}{\Keyword}{\x{}}}},
we note
{\createmultiexp {\s{}} {\e{}}} creates a multimethod with \emph{interface type} \s{}, and dispatch function \e{}
of type \sp{},
producing a value of type
{\MultiFntype {\s{}} {\sp{}}}. 
The T-DefMulti typing rule checks the dispatch function, and
verifies both the interface and dispatch type's domain agree.
Our example checks with \t{} = \Keyword, interface type \s{} = {\ArrowTwo{\x{}}{\Keyword}{\String}},
dispatch function type \sp{} = {\ArrowOne{\x{}}{\Keyword}{\Keyword}{\filterset{\topprop{}}{\topprop{}}}{\x{}}}, and overall type
$
{\MultiFntype {\ArrowTwo{\x{}}{\Keyword}{\String}}
              {\ArrowOne{\x{}}{\Keyword}{\Keyword}{\filterset{\topprop{}}{\topprop{}}}{\x{}}}}
$.

Next, we show how to check
$
{\extendmultiexp {hi_0} {\makekw{en}} {\abs {\x{}} {\Keyword} {\makestr{hello}}}}
$.
The expression 
{\extendmultiexp {\e{m}} {\e{v}} {\e{f}}} creates a new multimethod
that extends multimethod \e{m}'s dispatch table, mapping dispatch value
\e{v} to method \e{f}. The T-DefMulti typing rule
checks \e{m} is a multimethod with dispatch function type \t{d},
then calculates the extra information we know based on the current
dispatch value {\thenprop{\proppp{}}}, which is assumed when checking the method
body.
Our example checks with \e{m} being of type
$
{\MultiFntype {\ArrowTwo{\x{}}{\Keyword}{\String}}
              {\ArrowOne{\x{}}{\Keyword}{\Keyword}{\filterset{\topprop{}}{\topprop{}}}{\x{}}}}
$
with \objectp{} = {\x{}} (from below the arrow on the right argument of the previous type) and \t{v} = \Value{\makekw{en}}. 
Then {\thenprop{\proppp{}}} = 
{\isprop {\Value{\makekw{en}}}{\x{}}}
from
$
\isacompare{}{\x{}}{\Value{\makekw{en}}}
{\filterset {\isprop {\Value{\makekw{en}}}{\x{}}}{\notprop{\Value{\makekw{en}}}{\x{}}}}
$
(see \secref{sec:isaformal}).
Since \t{} = \Keyword{}, we check the method body with
$
\judgement{{\isprop{\Keyword}{\x{}}},{\isprop {\Value{\makekw{en}}}{\x{}}}}
  {\makestr{hello}}
  {\String}{\filterset{\topprop{}}{\topprop{}}}{\emptyobject{}}
$.
Finally from the interface type \t{m}, we know \thenprop{\prop{}} = \elseprop{\prop{}} = \topprop{},
and \object{} = \emptyobject{}, which also agrees with the method body, above.
Notice the overall type of a \defmethodliteral{} is the same as its first subexpression \e{m}.

It is worth noting the lack of special typing rules for overlapping methods---each
method is checked independently based on local type information.

\paragraph{Subtyping}
Multimethods are functions, via S-PMultiFn,
which says a multimethod can be upcast to its interface type. 
Multimethod call sites are then handled by T-App via T-Subsume. Other rules are given
in \figref{main:figure:mmsyntax}. 

\paragraph{Semantics}
Multimethod definition semantics are also given 
in \figref{main:figure:mmsyntax}. 
B-DefMulti creates a multimethod with the given dispatch function and an empty dispatch table.
B-DefMethod produces a new multimethod with an extended dispatch table.

The overall dispatch mechanism is summarised by B-BetaMulti.
First the dispatch function \v{d} is applied to the argument \vp{} to obtain
the dispatch value \v{e}.
Based on \v{e},
the \getmethodliteral{} metafunction (\figref{main:figure:mmsyntax})
extracts a method \v{f} from the method table {\disptable{}}
and applies it to the original argument for the final result.

\subsection{Precise Types for Heterogeneous maps}
\label{sec:hmapformal}

\begin{figure}
  \footnotesize
  $$
  \begin{altgrammar}
    \e{} &::=& \ldots \alt \hmapexpressionsyntax{}
    &\mbox{Expressions} \\
    \v{} &::=& \ldots \alt {\emptymap{}}
    &\mbox{Values} \\
    \t{} &::=& \ldots \alt {\HMapgeneric {\mandatory{}} {\absent{}}}
    &\mbox{Types} \\
    \auxhmapsyntax{}\\
  \end{altgrammar}
  $$
  \begin{mathpar}
    {\TAssoc}

    {\TGetHMap}

    {\TGetAbsent}

    {\TGetHMapPartialDefault}
    \ \ \ 
  {\SHMapMono}
  \end{mathpar}
  \begin{mathpar}
  {\SHMapP}\ \ 
  {\SHMap}

  \end{mathpar}
  \begin{mathpar}
    {\BAssoc}\ \ 
    {\BGet}\ \ 
    {\BGetMissing}
  \end{mathpar}
  \caption{HMap Syntax, Typing and Operational Semantics}
  \label{main:figure:hmapsyntax}
\end{figure}

\begin{figure}
  \footnotesize
$$
\begin{array}{lr}
  \begin{array}{llll}
    \restrictfigure{}
  \end{array}
  \ \ 
  \begin{array}{llll}
    \removefigure{}
  \end{array}
\end{array}
$$
\caption{Restrict and remove}
\label{main:figure:restrictremove}
\end{figure}

\figref{main:figure:hmapsyntax}
presents
heterogeneous map types.
The type \HMapgeneric{\mandatory{}}{\absent{}}
contains {\mandatory{}}, a map of \emph{present} entries (mapping keywords to types),
\absent{}, a set of keyword keys that are known to be \emph{absent}
and
tag \completenessmeta{} which is either {\complete{}} (``complete'') if the map is fully specified by \mandatory{},
and {\partial{}} (``partial'') if there are \emph{unknown} entries.
The partially specified map of
\clj{lunch} in \egref{example:lunchpartial}
is written
\HMapp{\mandatoryset{\mandatoryentrynoarrow{\Valkw{en}}{\String}, {\mandatoryentrynoarrow{\Valkw{fr}}{\String}}}}{\emptyabsent{}}
(abbreviated \Lunch).
The type of the fully specified map
\clj{breakfast} in \egref{example:breakfastcomplete} elides the absent entries,
written
\HMapc{\mandatoryset{\mandatoryentrynoarrow{\Valkw{en}}{\String}, {\mandatoryentrynoarrow{\Valkw{fr}}{\String}}}}
(abbreviated \Breakfast).
To ease presentation, 
if an HMap has completeness tag \complete{} then \absent{} is elided and implicitly contains all keywords not in the domain of 
\mandatory{}---dissociating keys is not modelled, so the set of absent entries otherwise
never grows.
Keys cannot be both present and absent.

The metavariable \mapval{}
ranges over the runtime value of maps {\curlymapvaloverright{\k{}}{\v{}}},
usually written {\curlymapvaloverrightnoarrow{\k{}}{\v{}}}.
We 
only provide syntax for the empty map literal,
however when convenient we abbreviate non-empty map literals
to be a series of \assocliteral{} operations on the empty map.
We restrict lookup and extension to keyword keys. 

\paragraph{How to check}
A mandatory lookup is checked by T-GetHMap.
$$
\abs{\makelocal{b}}{\Breakfast}{\getexp{\makelocal{b}}{\makekw{en}}}
$$
The result type is \String, and the return object is \path{\keype{\makekw{en}}}{\makelocal{b}}.
The object {\replacefor {\path {\keype{k}} {\x{}}} {\object{}} {\x{}}}
is a symbolic representation for a keyword lookup of $k$ in \object{}.
The substitution for {\x{}} handles the case where \object{} is empty.
\begin{mathpar}
\begin{array}{rcl}
{\replacefor {\path {\keype{k}} {\x{}}} {\y{}} {\x{}}} &=& {\path {\keype{k}} {\y{}}} \\
\end{array}
\ \ \ \ \ \ \ 
\begin{array}{rcl}
{\replacefor {\path {\keype{k}} {\x{}}} {\emptyobject{}} {\x{}}} &=& \emptyobject{}
\end{array}
\end{mathpar}

An absent lookup is checked by T-GetHMapAbsent.
$$
\abs{\makelocal{b}}{\Breakfast}{\getexp{\makelocal{b}}{\makekw{bocce}}}
$$
The result type is \Nil---since \Breakfast is fully specified---with return object \path{\keype{\makekw{bocce}}}{\makelocal{b}}.

A lookup that is not present or absent is checked by
T-GetHMapPartialDefault.
$$
\abs{\makelocal{u}}{\Lunch}{\getexp{\makelocal{u}}{\makekw{bocce}}}
$$
The result type is \Top---since {\Lunch} has an unknown \makekw{bocce} entry---with return object \path{\keype{\makekw{bocce}}}{\makelocal{u}}.
Notice propositions are erased once they enter a HMap type.

For presentational reasons, lookups on unions of HMaps are only supported in T-GetHMap
and each element of the union must contain the relevant key.
$$
\abs{\makelocal{u}}{\Unionsplice{\Breakfast \Lunch}}{\getexp{\makelocal{u}}{\makekw{en}}}
$$
The result type is \String, and the return object is \path{\keype{\makekw{en}}}{\makelocal{u}}.
However, lookups of \makekw{bocce} on {\Unionsplice{\Breakfast \Lunch}} maps are unsupported.
This restriction still allows us to check many of the examples in \secref{sec:overview}---in
particular we can check 
\egref{example:desserts-on-meal}, as \makekw{Meal} is in common with both HMaps,
but cannot check \egref{example:desserts-on-class}
because a \makekw{combo} meal lacks a \makekw{desserts} entry.
Adding a rule to handle \egref{example:desserts-on-class} is otherwise straightforward.

Extending a map with T-AssocHMap preserves its completeness.
$$
\abs{\makelocal{b}}{\Breakfast}{\assocexp{\makelocal{b}}{\makekw{au}}{\makestr{beans}}}
$$
The result type is
$
\HMapc{\mandatoryset{\mandatoryentrynoarrow{\Valkw{en}}{\String}, {\mandatoryentrynoarrow{\Valkw{fr}}{\String}}
        ,{\mandatoryentrynoarrow{\Valkw{au}}{\String}}}}
$,
a complete map.
T-AssocHMap also enforces ${\k{}} \not\in {\absent{}}$ to prevent badly formed types.


\paragraph{Subtyping}
Subtyping for HMaps
designate \MapLiteral{} as a common supertype for all HMaps.
S-HMap says that HMaps are subtypes if they agree
on \completenessmeta{}, agree on mandatory entries with subtyping
and at least cover the absent keys of the supertype.
Complete maps are subtypes of partial maps
as long as they agree on the mandatory entries of the partial map via subtyping (S-HMapP).



The semantics for \getliteral{} and \assocliteral{} are straightforward.

\begin{figure}[t]
  $$
\begin{array}{llll}
\updatefigure{}
\end{array}
$$
\caption{Type update (the metavariable \propisnotmeta{} ranges over \t{} and \nottype{\t{}} (without variables), 
  \notsubtypein{}{\Nil{}}{\nottype{\t{}}} when \issubtypein{}{\Nil{}}{\t{}}, see
\figref{main:figure:restrictremove} for \restrictliteral{} and \removeliteral{}.
  )}
\label{main:figure:update}
\end{figure}


\subsection{Proof system}
\label{formalmodel:proofsystem}

The occurrence typing proof system uses standard propositional logic,
except for where nested information is combined. This is
handled by L-Update:
{  \footnotesize
  $$
\LUpdate{}
$$
}

It says
under \propenv{}, if object \path{\pathelemp{}}{\x{}} is of type \t{}, and 
an extension
\path{\pathelem{}}{\path{\pathelemp{}}{\x{}}}
is of possibly-negative type \propisnotmeta{}, then
{\update{\t{}}{\propisnotmeta{}}{\pathelem{}}}
is \path{\pathelemp{}}{\x{}}'s type under \propenv{}.

Recall \egref{example:desserts-on-meal}.
Solving
$
{ \inpropenv 
  {{\isprop{\Order}{\makelocal{o}}},
    {\isprop{\Value{\makekw{combo}}}{\path{\keype{\makekw{Meal}}}{\makelocal{o}}}}}
  {\isprop {\t{}} {\makelocal{o}}}}
$
uses L-Update, where \pathelem{} = {\emptypath{}} and \pathelemp{} = [{\keype{\makekw{Meal}}}].
$$
\inpropenv{\propenv{}}{\isprop{\update{\Order}{\Value{\makekw{combo}}}{[{\keype{\makekw{Meal}}}]}}{\makelocal{o}}}
$$
Since {\Order} is a union of HMaps, we structurally recur on the first case of \updateliteral{}
(\figref{main:figure:update}),
which preserves \pathelem{}.
Each initial recursion hits the first HMap case, since there is some \t{} such that
{\inmandatory{\k{}}{\t{}}{\mandatory{}}} and 
\completenessmeta{} accepts partial maps \partial{}.

To demonstrate,
\makekw{lunch} meals are handled by the first HMap case and
update to {\HMapp {\extendmandatoryset {\mandatory{}}{\Valkw{Meal}}{\sp{}}} {\emptyabsent{}}}
where \sp{} = {\update{\Valkw{lunch}}{\Valkw{combo}}{\emptypath{}}}
and \mandatory{} = \mandatoryset{\mandatoryentry{\Valkw{Meal}}{\Valkw{lunch}},{\mandatoryentry{\Valkw{desserts}}{\Number{}}}}.
\sp{} updates to \Bot via the penultimate \updateliteral{} case,
because \restrict{\Value{\makekw{lunch}}}{\Value{\makekw{combo}}} = \Bot
by the first \restrictliteral{} case.
The same happens to \makekw{dinner} meals,
leaving just the \makekw{combo} HMap. 

In \egref{example:desserts-on-class},
$
\inpropenv{\propenv{}}{\isprop{\update{\Order}{\Long}{[{\classpe{}}, {\keype{\makekw{desserts}}}]}}{\makelocal{o}}}
$
updates the argument in the {\Long} method.
This recurs twice for each meal to handle the {\classpe{}}
path element.

We describe the other \updateliteral{} cases.
The first \classpe{} case updates
to \class{} if \classconst{} returns \Value{\class{}}.
The second \keype{\k{}} case detects contradictions in absent
keys. 
The third \keype{\k{}} case updates unknown entries to be mapped to \t{} or absent.
The fourth \keype{\k{}} case updates unknown entries to be \emph{present}
when they do not overlap with \Nil{}.

%
%
%

\section{Metatheory}
\label{sec:metatheory}

We prove type soundness following Tobin-Hochstadt and Felleisen~\cite{TF10}.  Our model is extended
to include errors \errorvalv{} and a \wrong{} value, and we prove well-typed
programs do not go wrong; this is therefore a stronger theorem than
proved by Tobin-Hochstadt and Felleisen~\cite{TF10}. 
Errors behave like Java exceptions---they can be thrown and propagate ``upwards'' in the evaluation rules
(\errorvalv{} rules are deferred to the appendix).

Rather than modeling Java's dynamic semantics, a task of daunting
complexity, we instead make our assumptions about Java explicit. We
concede that method and constructor calls may diverge or error, but
assume they can never go wrong
(other assumptions given in the supplemental material).

{\javanewassumption{main}}



For the purposes of our soundness proof, we require that all values
are \emph{consistent}.  Consistency (defined in the supplemental
material) states that the types of closures are well-scoped---they do
not claim propositions about variables hidden in their closures.


We can now state our main lemma and soundness theorem.  The
metavariable \definedreduction{} ranges over \v{}, \errorvalv{} and
\wrong{}. Proofs are deferred to the supplemental material. 

\begin{lemma}\label{main:lemma:soundness}

  {\soundnesslemmahypothesis}
\end{lemma}

{\soundnesstheoremnoproof{main}}

%

\section{Experience}
\label{sec:experience}

Typed Clojure is implemented as \coretyped{}~\cite{coretyped},
which has seen wide usage.

\subsection{Implementation}

\coretyped{} provides preliminary integration with the Clojure compilation
pipeline, primarily to resolve Java interoperability.


The \coretyped{} implementation extends this paper in several key areas 
to handle checking real Clojure code, including an implementation
of Typed Racket's variable-arity polymorphism~\cite{stf-esop}, 
and support for other Clojure idioms like datatypes and protocols.
There is no integration with Java Generics, so only Java 1.4-style erased types are ``trusted''
by \coretyped{}.
Casts are needed to recover the discarded information, which---for collections---are 
then tracked via Clojure's universal sequence interface~\cite{CljSeqDoc}.

\subsection{Evaluation}
\label{sec:casestudy}

Throughout this paper, we have focused on three interrelated type
system features: heterogeneous maps, Java interoperability, and
multimethods. Our hypothesis is that these features are widely used in
existing Clojure programs in interconnecting ways, and that handling
them as we have done is required to type check realistic Clojure
programs.

To evaluate this hypothesis, we analyzed two existing \coretyped{}
code bases, one from the open-source community, and one from a company
that uses \coretyped{} in production. For our data gathering, we
instrumented the \coretyped{} type checker to record how often
various features were used (summarized in 
\figref{experience:featuretable}). 

\begin{figure*}[t]

\begin{tabular}{lll}
      \toprule

  & feeds2imap & CircleCI \\
  \midrule
  Total number of typed namespaces & 11 (825 LOC) & 87 (19,000 LOC) \\
  Total number of \clj{def} expressions & 93  & 1834 \\
  \tabitem
  checked & 52 (56\%) & 407 (22\%) \\
  \tabitem
  unchecked & 41 (44\%) & 1427 (78\%) \\
  Total number of Java interactions & 32 & 105 \\
  \tabitem
  static methods & 5 (16\%) & 26 (25\%) \\ 
  \tabitem
  instance methods & 20 (62\%) & 36 (34\%) \\
  \tabitem
  constructors & 6 (19\%) & 38 (36\%) \\
  \tabitem
  static fields & 1 (3\%) & 5 (5\%) \\
  Methods overriden to return non-nil & 0 & 35 \\
  Methods overriden to accept nil arguments & 0 & 1 \\
  Total HMap lookups & 27  & 328  \\
  \tabitem
  resolved to mandatory key & 20 (74\%) & 208 (64\%) \\
  \tabitem
  resolved to optional key & 6 (22\%) & 70 (21\%) \\
  \tabitem
  resolved of absent key & 0 (0\%) & 20 (6\%) \\
  \tabitem
  unresolved key & 1 (4\%) & 30 (9\%) \\
  Total number of \clj{defalias} expressions & 18  & 95 \\
  \tabitem
  contained HMap or union of HMap type & 7 (39\%)  & 62 (65\%) \\
  Total number of checked \clj{defmulti} expressions & 0  & 11 \\
  Total number of checked \clj{defmethod} expressions & 0  & 89 \\

\end{tabular}
\caption{Typed Clojure Features used in Practice}
\label{experience:featuretable}
\end{figure*}

\paragraph{feeds2imap}
feeds2imap\footnote{\url{https://github.com/frenchy64/feeds2imap.clj}}
is an open source library written in Typed Clojure. 
It provides an RSS reader using the \emph{javax.mail} framework.

Of 11 typed namespaces containing 825 lines of code, there are 32 Java interactions.
The majority are method calls, consisting of 20 (62\%) instance methods and 5 (16\%) static methods. 
The rest consists of 1 (3\%) static field access, and 6 (19\%) constructor calls---there are no instance field accesses.

There are 27 lookup operations on HMap types, of which 20 (74\%) resolve to mandatory entries, 6 (22\%) to optional entries, and 1 (4\%) is an unresolved lookup. 
No lookups involved fully specified maps.

From 93 \clj{def} expressions in typed code, 52 (56\%) are checked, with a rate of 1 Java interaction for 1.6 checked top-level definitions, and 1 HMap lookup to 1.9 checked top-level definitions.
That leaves 41 (44\%) unchecked vars, mainly due to partially complete porting to Typed Clojure, but in some cases due to unannotated third-party libraries.

No typed multimethods are defined or used. 
Of 18 total type aliases, 7 (39\%) contained one HMap type, and none contained unions of HMaps---on further inspection there was no HMap entry used to dictate control flow, often handled by multimethods.
This is unusual in our experience, and is perhaps explained by feeds2imap mainly wrapping existing \emph{javax.mail} functionality.

\paragraph{CircleCI}
CircleCI~\cite{CircleCI}
provides continuous integration services built with a mixture of open-
and closed-source software.
Typed Clojure was used at CircleCI in production systems for two years \cite{CircleCIUsesTC},
maintaining
87 namespaces and 19,000 lines of code,
an experience we summarise in \secref{sec:limitations}.

The CircleCI code base contains 11 checked multimethods.
 All 11 dispatch functions
are on a HMap key containing a keyword, in a similar style to
\egref{example:desserts-on-meal}.
Correspondingly, all 89 methods are associated with a keyword dispatch value.
The argument type was in all cases a single HMap type, however,
rather than a union type.
In our experience from porting other libraries, this is unusual.



Of 328 lookup operations on HMaps,
208 (64\%) resolve to mandatory keys,
70 (21\%) to optional keys,
20 (6\%) to absent keys, and
30 (9\%) lookups are unresolved.
%
Of 95 total type aliases defined with \clj{defalias},
62 (65\%) involved one or more HMap types.
%
%
%
%
%
%
%
%
Out of 105 Java interactions, 26 (25\%) are static methods, 36 (34\%)
are instance methods, 38 (36\%) are constructors, and 5 (5\%) are static
fields. 35 methods are overriden to return non-nil, and 1 method 
overridden to accept nil---suggesting that
\coretyped{} disallowing \clj{nil} as a method argument by default
is justified.

Of 464 checked top-level definitions (which consists of
57 \clj{defmethod} calls and 407 \clj{def} expressions),
1 HMap lookup occurs per 1.4 top-level definitions,
and 1 Java interaction occurs every 4.4 top-level definitions.

From 1834 \clj{def} expressions in typed code,
only 407 (22\%) were checked.
That leaves 1427 (78\%) which have unchecked definitions, either by an explicit \clj{:no-check} annotation
or \clj{tc-ignore} to suppress type checking,
or the \clj{warn-on-unannotated-vars} option, which skips \clj{def} expressions
that lack expected types via \clj{ann}.
From a brief investigation,
reasons include unannotated third-party libraries,
work-in-progress conversions to Typed Clojure,
unsupported Clojure idioms, 
and hard-to-check code.

\paragraph{Lessons}
Based on our empirical survey, HMaps and Java interoperability support
are vital features used on average more than once per typed
function. 
Multimethods are less common
in our case studies. The CircleCI code base contains only 26 multimethods total
in 55,000 lines of mixed untyped-typed Clojure code,
a low number in our experience.


%

\subsection{Further challenges}
\label{sec:limitations}

After a 2 year trial, the second case study decided to disabled type checking~\cite{CircleCIBlog}.
They were supportive of the fundamental ideas presented in this paper, but primarily
cited issues with the checker implementation in practice and would reconsider
type checking if they were resolved. This is also supported by \figref{experience:featuretable},
where 78\% of \clj{def} expressions are unchecked.

\smallsection{Performance}
Rechecking files with transitive dependencies is expensive since all dependencies must be rechecked.
We conjecture caching type state will significantly
improve re-checking performance,
though preserving static soundness in the context of arbitrary code reloading is a largely unexplored area.

\smallsection{Library annotations}
Annotations for external code are rarely available, so a large part of the
untyped-typed porting process is reverse engineering libraries.

\smallsection{Unsupported idioms}
While the current set of features is vital to checking Clojure code,
there is still much work to do.
For example, common Clojure functions are often too polymorphic for the current implementation
or theory to account for. The post-mortem~\cite{CircleCIBlog} contains more details.

\section{Related Work}

\paragraph{Multimethods} 
\cite{MS02} and collaborators present a sequence of
systems~\cite{Chambers:1992:OMC,Chambers:1994:TMM,MS02} with statically-typed multimethods
and modular type checking.  In contrast to Typed Clojure, in these
system methods declare the types of arguments that they expect which
corresponds to exclusively using \clj{class} as the dispatch function
in Typed Clojure. However, Typed Clojure does not attempt to rule out
failed dispatches.







\paragraph{Record Types} Row polymorphism~\cite{Wand89typeinference,CM91,HP91}, used
in systems such as the OCaml object system, provides many of the
features of HMap types, but defined using universally-quantified row
variables. HMaps in Typed Clojure are instead designed to be used with
subtyping, but nonetheless provide similar expressiveness, including
the ability to require presence and absence of certain keys. 

Dependent JavaScript~\cite{Chugh:2012:DTJ} can track similar
invariants as HMaps with types for JS objects. They must deal with
mutable objects, they feature refinement types and strong updates to
the heap to track changes to objects.

TeJaS~\cite{TeJaS}, another type system for JavaScript,
also supports similar HMaps, with the ability to
record the presence and absence of entries, but lacks a compositional
flow-checking approach like occurrence typing.

Typed Lua~\cite{Maidl:2014:TLO} has \emph{table types} which track
entries in a mutable Lua table.  Typed Lua changes the dynamic
semantics of Lua to accommodate mutability: Typed Lua raises a runtime
error for lookups on missing keys---HMaps consider lookups on missing
keys normal.

\paragraph{Java Interoperability in Statically Typed Languages}
Scala~\cite{OCD+} has nullable references for compatibility with Java.
Programmers must manually check for
\java{null} as in Java to avoid null-pointer exceptions.

\paragraph{Other optional and gradual type systems}
Several other gradual type
systems have been developed for existing
dynamically-typed languages.  Reticulated Python~\cite{Vitousek14} is
an experimental gradually typed system for Python, implemented as a
source-to-source translation that inserts dynamic checks at language
boundaries and supporting Python's first-class object system. 
Clojure's nominal classes avoids the need to support
first-class object system in Typed Clojure, however HMaps offer an alternative to
the structural objects offered by Reticulated. Similarly,
Gradualtalk~\cite{gradualtalk} offers gradual typing for Smalltalk,
with nominal classes.

Optional types
have been  adopted in industry, including Hack~\cite{hack}, and Flow~\cite{flow} and
TypeScript~\cite{typescript}, two extensions of JavaScript. These
systems  support  limited forms of occurrence typing,
and do not include the other features we
present.


\section{Conclusion}
\label{sec:conclusion}

Optional type systems must be designed with close attention to the
language that they are intended to work for.
We have therefore designed Typed Clojure, an optionally-typed version of
Clojure, with a type system that works with a wide variety of distinctive
Clojure idioms and features. Although based on the foundation of Typed
Racket's occurrence typing approach, Typed Clojure both extends the
fundamental control-flow based reasoning as well as applying it to
handle seemingly unrelated features such as multi-methods. In
addition, Typed Clojure supports crucial features such as
heterogeneous maps and Java interoperability while integrating these
features into the core type system. Not only are each of these
features important in isolation to Clojure and Typed Clojure
programmers, but they must fit together smoothly to ensure that
existing untyped programs are easy to convert to Typed Clojure.

The result is a sound, expressive, and useful type system which, as
implemented in \coretyped with appropriate extensions, is suitable for
typechecking a significant amount of existing Clojure programs.
As a result, Typed Clojure is already successful: it is used in
the Clojure community among both enthusiasts and professional
programmers.

Our empirical analysis of existing Typed Clojure programs bears out
our design choices. Multimethods, Java interoperation, and
heterogeneous maps are indeed common in both Clojure and Typed Clojure,
meaning that our type system must accommodate them. Furthermore, they
are commonly used together, and the features of each are mutually
reinforcing. Additionally, the choice to make Java's \clj{null}
explicit in the type system is validated by the many Typed Clojure
programs that  specify non-nullable types.


%




\newpage

\bibliography{ms}

\clearpage

\end{document}




\special{papersize=8.5in,11in}
\setlength{\pdfpageheight}{\paperheight}
\setlength{\pdfpagewidth}{\paperwidth}

%

\newcommand{\clj}[1]{\lstinline{#1}}
\newcommand{\java}[1]{\lstinline{#1}}
\newcommand{\rkt}[1]{\lstinline{#1}}

\counterwithin{figure}{section}
\counterwithin{assumption}{section}
\counterwithin{theorem}{section}
\counterwithin{lemma}{section}
\counterwithin{definition}{section}

\onecolumn
\appendix

\section{Soundness for Typed Clojure}

{\javaassumptionsall{appendix}}

\begin{lemma} \label{appendix:lemma:envagree}
  If \openv{} and \openvp{} agree on \fv{\prop{}}
  and \satisfies{\openv{}}{\prop{}}
  then \satisfies{\openvp{}}{\prop{}}.
\begin{proof}
  Since the relevant parts of \openv{} and \openvp{} agree, the proof follows trivially.
\end{proof}
\end{lemma}

\begin{lemma} \label{appendix:lemma:substfilter}
  If 
  \begin{itemize}
    \item \prop{1} = {\replacefor {\prop{2}} {\object{}} {\x{}}},
    \item
  {\satisfies{\openv{2}}{\prop{2}}},
    \item
  $\forall v \in \fv{\prop{2}} - \x{}$.
                              {\inopenvnoeq{\openv{1}}{v}} = {\inopenvnoeq {\openv{2}}{v}},
    \item
  and {\inopenvnoeq{\openv{2}}{\x{}}} = {\inopenvnoeq{\openv{1}}{\object{}}}
  \end{itemize}
  then \satisfies{\openv{1}}{\prop{1}}.

  \begin{proof}
    By induction on the derivation of the model judgement.
  \end{proof}
\end{lemma}

\begin{lemma} \label{appendix:lemma:satisfies}
  If \satisfies{\openv{}}{\propenv{}} and \inpropenv{\propenv{}}{\prop{}} then \satisfies{\openv{}}{\prop{}}.

  \begin{proof}
    By structural induction on \inpropenv{\propenv{}}{\prop{}}.
%
%
%
%
%
%
%
%
  \end{proof}
\end{lemma}

\begin{lemma} \label{appendix:lemma:goodobjects+ve}
  If \inpropenv{\propenv{}}{\isprop{\t{}}{\path{\pathelem{}}{\x{}}}},
  \satisfies{\openv{}}{\propenv{}}
  and \inopenv{\openv{}}{\path{\pathelem{}}{\x{}}}{\v{}}
  then
  \judgementselfrewrite{}{\v{}}{\t{}}{\filterset{\thenprop{\propp{}}}{\elseprop{\propp{}}}}{\objectp{}}
  for some {\thenprop{\propp{}}}, {\elseprop{\propp{}}} and {\objectp{}}.
  \begin{proof}
    Corollary of lemma~\ref{appendix:lemma:satisfies}.
  \end{proof}
\end{lemma}

\begin{lemma}[Paths are independent] \label{appendix:lemma:pathindependent}
  If \inopenvnoeq{\openv{}}{\object{}} = \inopenvnoeq{\openv{1}}{\objectp{}}
  then \inopenvnoeq{\openv{}}{\path{\pathelem{}}{\object{}}} =
       \inopenvnoeq{\openv{1}}{\path{\pathelem{}}{\objectp{}}}
 \begin{proof}
   By induction on \pathelem{}.
%
%
%
%
 \end{proof}
\end{lemma}

\begin{lemma}[\classconst]\label{appendix:lemma:classconst}
  If
  {\opsem{\openv{}}{\appexp{\classconst{}}{\openv{}({\path{\pathelem{}}{\x{}}})}}{\class{}}} then
  {\satisfies{\openv{}}{\isprop{\class{}}{\path{\pathelem{}}{\x{}}}}}.

  \begin{proof}
    Induction on the definition of {\classconst{}}.
  \end{proof}
\end{lemma}

{\consistentwithdefinition{appendix}}

{\istruefalsedefinitions{appendix}}

%

\begin{lemma}[isa? has correct propositions] \label{appendix:lemma:isa}
  If
  \begin{itemize}
    \item
  \judgementrewrite {\propenv{}} {\v{1}} {\t{1}}
             {\filterset {\thenprop {\prop{1}}}
                         {\elseprop {\prop{1}}}}
                       {\object{1}}
                       {\v{1}},
    \item
  \judgementrewrite {\propenv{}} {\v{2}} {\t{2}}
             {\filterset {\thenprop {\prop{2}}}
                         {\elseprop {\prop{2}}}}
                       {\object{2}}
                       {\v{2}},
    \item
        \isaopsem{\v{1}}{\v{2}} = {\v{}}, 
    \item
        \satisfies{\openv{}}{\propenv{}},
    \item
  \isacompare{\t{1}}{\object{1}}{\t{2}}{\filterset {\thenprop {\propp{}}} {\elseprop {\propp{}}}},
    \item
        \inpropenv{\thenprop{\propp{}}}{\thenprop{\prop{}}}, and
    \item
        \inpropenv{\elseprop{\propp{}}}{\elseprop{\prop{}}},
    \end{itemize}
  then either
\begin{itemize}
  \item
        if
        \istrueval{\v{}}
        then {\satisfies{\openv{}}{\thenprop{\prop{}}}}, or
  \item
        if
        \isfalseval{\v{}}
        then {\satisfies{\openv{}}{\elseprop{\prop{}}}}.
\end{itemize}
\begin{proof}
        By cases on the definition of \isaopsemliteral
        and subcases on \isaopsemliteral.

        \begin{itemize} 
          \item[]
            \begin{subcase}[\isaopsem{\v{1}}{\v{1}} = {\true{}}, \text{if} \v{1} \notequal\ {\class{}}]
              \ 

              \v{1} = \v{2}, \v{1} \notequal\ {\class{}}, \v{2} \notequal\ {\class{}}, \istrueval{\v{}}
              
              Since \istrueval{\v{}} we prove {\satisfies{\openv{}}{\thenprop{\prop{}}}}
              by cases on the definition of \isacompareliteral{}:
              \begin{itemize} 
                \item[]
                  \begin{subcase}[\isacompare{\s{}}{\path{\classpe{}}{\path{\pathelem{}}{\x{}}}}{\Value{\class{}}}
                                 {\filterset{\isprop{\class{}} {\path{\pathelem{}}{\x{}}}}
                                            {\notprop{\class{}}{\path{\pathelem{}}{\x{}}}}}]
                    \

                    \object{1} = {\path{\classpe{}}{\path{\pathelem{}}{\x{}}}},
                    \t{2} = {\Value{\class{}}},
                    \inpropenv{\isprop{\class{}} {\path{\pathelem{}}{\x{}}}}{\thenprop{\prop{}}}

                    Unreachable by inversion on the typing relation, since \t{2} = {\Value{\class{}}},
                    yet \v{2} \notequal\ {\class{}}.

%
%
%

                  \end{subcase}
                \item[]
                  \begin{subcase}[\isacompare{\s{}}{\object{}}{\Value{\singletonmeta{}}}
                    {\replacefor
                      {\filtersetparen{\isprop{\Value{\singletonmeta{}}} {\x{}}}
                        {\notprop{\Value{\singletonmeta{}}}{\x{}}}}
                      {\object{}}
                      {\x{}}}\ 
                    \text{if}\ {\singletonmeta{}} \notequal \class{}]
                    \ 

                    \t{2} = {\Value{\singletonmeta{}}}, 
                    {\singletonmeta{}} \notequal \class{},
                    \inpropenv{\replacefor{\isprop{\Value{\singletonmeta{}}} {\x{}}}
                                           {\object{1}}
                                           {\x{}}}{\thenprop{\prop{}}}

                    Since \t{2} = {\Value{\singletonmeta{}}} where {\singletonmeta{}} \notequal \class{},
                    by inversion on the typing judgement 
                    {\v{2}} is either \true{}, \false{}, \nil{} or \k{}
                    by T-True, T-False, T-Nil or T-Kw.

                    Since \v{1} = {\v{2}} then \t{1} = \t{2}, and since \t{2} = {\Value{\singletonmeta{}}}
                      then \t{1} = {\Value{\singletonmeta{}}}, so
                    \judgementtwo {} {\v{1}} {\Value{\singletonmeta{}}}

                    If \object{1} = \emptyobject{} then \thenprop{\prop{}} = \topprop{} and
                    we derive
                    {\satisfies{\openv{}}{\topprop{}}} with M-Top.

                    Otherwise \object{1} = \path{\pathelem{}}{\x{}} and 
                    \inpropenv{\isprop{\Value{\singletonmeta{}}}{\path{\pathelem{}}{\x{}}}}{\thenprop{\prop{}}},
                    and since
                    \judgementtwo {} {\v{1}} {\Value{\singletonmeta{}}}
                    then
                    \judgementtwo {} {{\openv{}}(\path{\pathelem{}}{\x{}})} {\Value{\singletonmeta{}}},
                    which we can use M-Type to derive
                    {\satisfies{\openv{}}{\isprop{\Value{\singletonmeta{}}}{\path{\pathelem{}}{\x{}}}}}.
                  \end{subcase}
                \item[]
                  \begin{subcase}[\isacompare{\s{}}{\object{}}{\t{}} {\filterset{\topprop{}} {\topprop{}}}]
                    \ 

                    {\thenprop{\prop{}}} = {\topprop{}}

                    {\satisfies{\openv{}}{\topprop{}}} holds by M-Top.

                  \end{subcase}
              \end{itemize}
            \end{subcase}
          \item[]
            \begin{subcase}[\isaopsem{\class{1}}{\class{2}} = {\true{}}, \text{if}\ \issubtypein{}{\class{1}}{\class{2}}]
              \ 

              \v{1} = \class{1}, \v{2} = \class{2},
              \issubtypein{}{\class{1}}{\class{2}},
              \istrueval{\v{}}
              
              Since \istrueval{\v{}} we prove {\satisfies{\openv{}}{\thenprop{\prop{}}}}
              by cases on the definition of \isacompareliteral{}:
              \begin{itemize} 
                \item[]
                  \begin{subcase}[\isacompare{\s{}}{\path{\classpe{}}{\path{\pathelem{}}{\x{}}}}{\Value{\class{}}}
                                 {\filterset{\isprop{\class{}} {\path{\pathelem{}}{\x{}}}}
                                            {\notprop{\class{}}{\path{\pathelem{}}{\x{}}}}}]
                    \

                    \object{1} = {\path{\classpe{}}{\path{\pathelem{}}{\x{}}}},
                    \t{2} = {\Value{\class{2}}},
                    \inpropenv{\isprop{\class{2}} {\path{\pathelem{}}{\x{}}}}{\thenprop{\prop{}}}

                    By inversion on the typing relation, since \classpe{} is the last path element of \object{1}
                    then \opsem{\openv{}}{\appexp{\classconst{}}{\openv{}({\path{\pathelem{}}{\x{}}})}}{\v{1}}.

                    Since {\opsem{\openv{}}{\appexp{\classconst{}}{\openv{}({\path{\pathelem{}}{\x{}}})}}{\class{1}}},
                    as {\v{1}} = {\class{1}},
                    we can derive from lemma~\ref{appendix:lemma:classconst}
                    {\satisfies{\openv{}}{\isprop{\class{1}} {\path{\pathelem{}}{\x{}}}}}.

                    By the induction hypothesis we can derive 
                    {\inpropenv{\propenv{}}{\isprop{\class{1}} {\path{\pathelem{}}{\x{}}}}},
                    and with the fact {\issubtypein{}{\class{1}}{\class{2}}}
                    we can use L-Sub to conclude 
                    {\inpropenv{\propenv{}}{\isprop{\class{2}} {\path{\pathelem{}}{\x{}}}}},
                    and finally by lemma~\ref{appendix:lemma:satisfies}
                    we derive
                    {\satisfies{\openv{}}{\isprop{\class{2}} {\path{\pathelem{}}{\x{}}}}}.

                  \end{subcase}
                \item[]
                  \begin{subcase}[\isacompare{\s{}}{\object{}}{\Value{\singletonmeta{}}}
                    {\replacefor
                      {\filtersetparen{\isprop{\Value{\singletonmeta{}}} {\x{}}}
                        {\notprop{\Value{\singletonmeta{}}}{\x{}}}}
                      {\object{}}
                      {\x{}}}\ 
                    \text{if}\ {\singletonmeta{}} \notequal \class{}]
                    \ 

                    \t{2} = {\Value{\singletonmeta{}}}, 
                    {\singletonmeta{}} \notequal \class{},
                    \inpropenv{\replacefor{\isprop{\Value{\singletonmeta{}}} {\x{}}}
                                           {\object{1}}
                                           {\x{}}}{\thenprop{\prop{}}}

                    Unreachable case since 
                    \t{2} = {\Value{\singletonmeta{}}} where 
                    {\singletonmeta{}} \notequal \class{},
                    but \v{2} = \class{2}.
                  \end{subcase}
                \item[]
                  \begin{subcase}[\isacompare{\s{}}{\object{}}{\t{}} {\filterset{\topprop{}} {\topprop{}}}]
                    \ 

                    {\thenprop{\prop{}}} = {\topprop{}}

                    {\satisfies{\openv{}}{\topprop{}}} holds by M-Top.

                  \end{subcase}
              \end{itemize}
            \end{subcase}
          \item[]
            \begin{subcase}[\isaopsem{\v{1}}{\v{2}} = {\false{}}, otherwise]
              \ 

              \v{1} \notequal\ \v{2},
              \isfalseval{\v{}}
              
              Since \isfalseval{\v{}} we prove {\satisfies{\openv{}}{\elseprop{\prop{}}}}
              by cases on the definition of \isacompareliteral{}:
              \begin{itemize} 
                \item[]
                  \begin{subcase}[\isacompare{\s{}}{\path{\classpe{}}{\path{\pathelem{}}{\x{}}}}{\Value{\class{}}}
                                 {\filterset{\isprop{\class{}} {\path{\pathelem{}}{\x{}}}}
                                            {\notprop{\class{}}{\path{\pathelem{}}{\x{}}}}}]
                    \

                    \object{1} = {\path{\classpe{}}{\path{\pathelem{}}{\x{}}}},
                    \t{2} = {\Value{\class{}}},
                    \inpropenv{\notprop{\class{}} {\path{\pathelem{}}{\x{}}}}{\elseprop{\prop{}}}

                    By inversion on the typing relation, since \classpe{} is the last path element of \object{1}
                    then \opsem{\openv{}}{\appexp{\classconst{}}{\openv{}({\path{\pathelem{}}{\x{}}})}}{\v{1}}.
                    
                    By the definition of {\classconst{}} either {\v{1}} = {\class{}} or {\v{1}} = \nil{}.

                    If {\v{1}} = \nil{}, then we know from the definition of \isaopsemliteral that 
                    {\openv{}({\path{\pathelem{}}{\x{}}})} = \nil{}.

                    Since \judgementtwo{}{\openv{}({\path{\pathelem{}}{\x{}}})}{\Nil},
                    and there is no \v{1} such that both \judgementtwo{}{\openv{}({\path{\pathelem{}}{\x{}}})}{\class} and
                    \judgementtwo{}{\openv{}({\path{\pathelem{}}{\x{}}})}{\Nil{}},
                    we use M-NotType to derive 
                    \satisfies{\openv{}}{\notprop{\class{}} {\path{\pathelem{}}{\x{}}}}.

                    Similarly if {\v{1}} = \class{1}, by the definition of \isacompareliteral
                    we know {\notsubtypein{}{\class{1}}{\class{}}} and 
                    {\openv{}({\path{\pathelem{}}{\x{}}})} = \class{1}.

                    Since \judgementtwo{}{\openv{}({\path{\pathelem{}}{\x{}}})}{\class{1}},
                    and there is no \v{1} such that both 
                    \judgementtwo{}{\v{1}}{\class{}} and
                    \judgementtwo{}{\v{1}}{\class{1}},
                    we use M-NotType to derive 
                    \satisfies{\openv{}}{\notprop{\class{}} {\path{\pathelem{}}{\x{}}}}.

                  \end{subcase}
                \item[]
                  \begin{subcase}[\isacompare{\s{}}{\object{}}{\Value{\singletonmeta{}}}
                    {\replacefor
                      {\filtersetparen{\isprop{\Value{\singletonmeta{}}} {\x{}}}
                        {\notprop{\Value{\singletonmeta{}}}{\x{}}}}
                      {\object{}}
                      {\x{}}}\ 
                    \text{if}\ {\singletonmeta{}} \notequal \class{}]
                    \ 

                    \t{2} = {\Value{\singletonmeta{}}}, 
                    {\singletonmeta{}} \notequal \class{},
                    \inpropenv{\replacefor{\notprop{\Value{\singletonmeta{}}} {\x{}}}
                                           {\object{1}}
                                           {\x{}}}{\elseprop{\prop{}}}

                    Since \t{2} = {\Value{\singletonmeta{}}} where {\singletonmeta{}} \notequal \class{},
                    by inversion on the typing judgement 
                    {\v{2}} is either \true{}, \false{}, \nil{} or \k{}
                    by T-True, T-False, T-Nil or T-Kw.

                    If \object{1} = \emptyobject{} then \elseprop{\prop{}} = \topprop{} and
                    we derive
                    {\satisfies{\openv{}}{\topprop{}}} with M-Top.

                    Otherwise \object{1} = \path{\pathelem{}}{\x{}} and 
                    \inpropenv{\notprop{\Value{\singletonmeta{}}}{\path{\pathelem{}}{\x{}}}}{\elseprop{\prop{}}}.
                    Noting that \v{1} \notequal\ \v{2},
                    we know \judgementtwo{}{\openv{}({\path{\pathelem{}}{\x{}}})}{\s{}}
                    where \s{} \notequal\ {\Value{\singletonmeta{}}},
                    and there is no \v{1} such that both 
                    \judgementtwo{}{\v{1}}{\Value{\singletonmeta{}}} and
                    \judgementtwo{}{\v{1}}{\s{}}
                    so we can use M-NotType to derive
                    {\satisfies{\openv{}}{\notprop{\Value{\singletonmeta{}}}{\path{\pathelem{}}{\x{}}}}}.
                  \end{subcase}
                \item[]
                  \begin{subcase}[\isacompare{\s{}}{\object{}}{\t{}} {\filterset{\topprop{}} {\topprop{}}}]
                    \ 

                    {\elseprop{\prop{}}} = {\topprop{}}

                    {\satisfies{\openv{}}{\topprop{}}} holds by M-Top.

                  \end{subcase}
              \end{itemize}
            \end{subcase}
        \end{itemize}
      \end{proof}
\end{lemma}

\begin{lemma} \label{appendix:lemma:soundness}
{\soundnesslemmahypothesis}
\begin{proof}
By induction and cases on the derivation of \opsem {\openv{}} {\e{}} {\a{}},
and subcases on the penultimate rule of the derivation of
\judgementrewrite{\propenv{}}{\ep{}}{\t{}}{\filterset{\thenprop{\prop{}}}{\elseprop{\prop{}}}}{\object{}}{\e{}}
followed by T-Subsume as the final rule.


\begin{case}[B-Val]

  \begin{itemize}
    \item[] 
      \begin{subcase}[T-True]
        \v{} = \true{},
  \ep{} = \true{},
  \e{} = \true{}, \issubtypein{}{\True}{\t{}}, \inpropenv{\topprop{}}{\thenprop{\prop{}}}, 
  \inpropenv{\botprop{}}{\elseprop{\prop{}}}, \issubtypein{}{\emptyobject{}}{\object{}}

        Proving part 1 is trivial: \object{} is a superobject of \emptyobject{}, which can only be \emptyobject{}.

        To prove part 2, we note that \v{} = \true{}
        and \inpropenv{\topprop{}}{\thenprop{\prop{}}},
        so \satisfies{\openv{}}{\thenprop{\prop{}}} by M-Top.

        Part 3 holds as \e{} can only be reduced to itself via B-Val.

        Part 4 holds vacuously.
      \end{subcase}
    \item[]
      \begin{subcase}[T-HMap] \v{} = {\curlymapvaloverright{\v{k}}{\v{v}}},
  \ep{} = {\curlymapvaloverright{\v{k}}{\v{v}}},
  \e{} = {\curlymapvaloverright{\v{k}}{\v{v}}},
  \issubtypein{}{\HMapc {\mandatory{}}}{\t{}},
  \inpropenv{\topprop{}}{\thenprop{\prop{}}},
  \inpropenv{\botprop{}}{\elseprop{\prop{}}},
  \issubtypein{}{\emptyobject{}}{\object{}},
  $\overrightarrow{\judgementtwo {} {\v{k}}{\Value \k{}}}$,
  $\overrightarrow{\judgementtwo {} {\v{v}}{\t{v}}}$,
  \mandatory{} = \mandatorysetoverright{\k{}}{\t{v}}

        Similar to T-True.

        Part 4 holds by the induction hypothese on {\overr{\v{k}}} and {\overr{\v{v}}}.
      \end{subcase}
    \item[]
      \begin{subcase}[T-Kw] \v{} = {\k{}},
  \ep{} = {\k{}},
  \e{} = {\k{}},
  \issubtypein{}{\Value{\k{}}}{\t{}},
  \inpropenv{\topprop{}}{\thenprop{\prop{}}},
  \inpropenv{\botprop{}}{\elseprop{\prop{}}},
  \issubtypein{}{\emptyobject{}}{\object{}}

        Similar to T-True.
      \end{subcase}
      \begin{subcase}[T-Str]
        Similar to T-Kw.
      \end{subcase}
  \item[] 
    \begin{subcase}[T-False]
      \v{} = \false{},
\ep{} = \false, 
\e{} = \false, 
\issubtypein{}{\False}{\t{}},
\inpropenv{\botprop{}}{\thenprop{\prop{}}},
\inpropenv{\topprop{}}{\elseprop{\prop{}}},
\issubtypein{}{\emptyobject{}}{\object{}}

Proving part 1 is trivial: \object{} is a superobject of \emptyobject{}, which must be \emptyobject{}. 

To prove part 2, we note that \v{} = \false{}
and \inpropenv{\topprop{}}{\elseprop{\prop{}}}, so \satisfies{\openv{}}{\elseprop{\prop{}}} by M-Top. 

Part 3 holds as \e{} can only be reduced to itself via B-Val.

Part 4 holds vacuously.
\end{subcase}
    \item[]
      \begin{subcase}[T-Class] \v{} = {\class{}},
  \ep{} = {\class{}},
  \e{} = {\class{}},
  \issubtypein{}{\Value{\class{}}}{\t{}},
  \inpropenv{\topprop{}}{\thenprop{\prop{}}},
  \inpropenv{\botprop{}}{\elseprop{\prop{}}},
  \issubtypein{}{\emptyobject{}}{\object{}}

        Similar to T-True.
      \end{subcase}
    \item[]
      \begin{subcase}[T-Instance] \v{} = {\classvalue{\classhint{}} {\overrightarrow {\classfieldpair{\fld{i}} {\v{i}}}}},
  \ep{} = {\classvalue{\classhint{}} {\overrightarrow {\classfieldpair{\fld{}} {\v{}}}}},
  \e{} = {\classvalue{\classhint{}} {\overrightarrow {\classfieldpair{\fld{}} {\v{}}}}},
  \issubtypein{}{\class{}}{\t{}},
  \inpropenv{\topprop{}}{\thenprop{\prop{}}},
  \inpropenv{\botprop{}}{\elseprop{\prop{}}},
  \issubtypein{}{\emptyobject{}}{\object{}}

        Similar to T-True.

        Part 4 holds by the induction hypotheses on ${\overrightarrow{\v{i}}}$.
      \end{subcase}
  \item[] 
    \begin{subcase}[T-Nil] 
      \v{} = \nil{},
\ep{} = \nil, 
\e{} = \nil, 
\issubtypein{}{\Nil}{\t{}},
\inpropenv{\botprop{}}{\thenprop{\prop{}}},
\inpropenv{\topprop{}}{\elseprop{\prop{}}},
\issubtypein{}{\emptyobject{}}{\object{}}

      Similar to T-False.
\end{subcase}
\item[]
\begin{subcase}[T-Multi] 
  \v{} = {\multi {\v{1}} {\curlymapvaloverright{\v{k}}{\v{v}}}}
  \ep{} = {\multi {\v{1}} {\curlymapvaloverright{\v{k}}{\v{v}}}},
  \judgementtworewrite {} {\v{1}} {\t{1}}{\v{1}},
  \overr{\judgementtworewrite{}{\v{k}}{\Top{}}{\v{k}}},
  \overr{\judgementtworewrite{}{\v{v}}{\s{}}{\v{v}}},
  \e{} = {\multi {\v{1}} {\curlymapvaloverright{\v{k}}{\v{v}}}},
  \issubtypein{}{\MultiFntype {\s{}} {\t{1}}}{\t{}},
  \inpropenv{\topprop{}}{\thenprop{\prop{}}},
  \inpropenv{\botprop{}}{\elseprop{\prop{}}},
  \issubtypein{}{\emptyobject{}}{\object{}}

        Similar to T-True.
\end{subcase}
\item[]
\begin{subcase}[T-Const]
  \e{} = {\const{}},
  \issubtypein{}{\constanttype{\const{}}}{\t{}},
  \inpropenv{\topprop{}}{\thenprop{\prop{}}},
  \inpropenv{\botprop{}}{\elseprop{\prop{}}},
  \issubobjin{}{\emptyobject{}}{\object{}}

        Parts 1, 2 and 3 hold for the same reasons as T-True. 
\end{subcase}

  \end{itemize}
\end{case}

\begin{case}[B-Local]
{ \inopenv {\openv{}} {\x{}} {\v{}} },
{ \opsem {\openv{}} {\x{}} {\v{}} }

\begin{itemize}
  \item[]
\begin{subcase}[T-Local]
  \ep{} = \x{}, 
  \e{} = \x{}, 
  \inpropenv{\notprop {\falsy{}} {\x{}}}{\thenprop{\prop{}}},
  \inpropenv{\isprop {\falsy{}} {\x{}}}{\elseprop{\prop{}}},
\issubtypein{}{\x{}}{\object{}},
\inpropenv{\propenv{}}{\isprop{\t{}}{\x{}}}

Part 1 follows from \inopenv{\openv{}}{\object{}} {\v{}}, since either {\object{}} = \x{}
and \inopenv{\openv{}}{\x{}} {\v{}} is a premise of B-Local, or {\object{}} = {\emptyobject{}} which also
satisfies the goal.

Part 2 considers two cases: if \istrueval{\v{}}, then 
\satisfies{\openv{}}{\notprop{\falsy}{\x{}}} holds by M-NotType; if \isfalseval{\v{}}, then 
\satisfies{\openv{}}{\isprop{\falsy}{\x{}}} holds by M-Type.

We prove part 3 by observing
\inpropenv{\propenv{}}{\isprop{\t{}}{\x{}}},
\satisfies{\openv{}}{\propenv{}},
and
\inopenv {\openv{}} {\x{}} {\v{}}
(by B-Local)
which gives us the desired result.

Part 4 holds vacuously.
\end{subcase}
\end{itemize}

\end{case}

\begin{case}[B-Do]
  \opsem {\openv{}} {\e{1}} {\v{1}},
  \opsem {\openv{}} {\e{2}} {\v{}}

\begin{itemize}
  \item[] \begin{subcase}[T-Do]
\ep{} = {\doexp {\ep1} {\ep2}},
  \judgementrewrite {\propenv{}} 
             {\ep1} {\t1}
             {\filterset {\thenprop {\prop{1}}} {\elseprop {\prop1}}} 
             {\object{1}}
             {\e1},
\judgementrewrite {\propenv{}, {\orprop {\thenprop {\prop{1}}} {\elseprop {\prop{1}}}}}
           {\ep{}} {\t{}}
           {\filterset {\thenprop {\prop{}}} {\elseprop {\prop{}}}} 
           {\object{}}
           {\e{}},
\e{} = {\doexp {\e1} {\e2}}

For all parts we note 
    since {\e{1}} can be either a true or false value
    then
    {\satisfies{\openv{}}{{\propenv{}},{\orprop {\thenprop {\prop{1}}} {\elseprop {\prop{1}}}}}}
    by M-Or,
    which together with 
\judgement {\propenv{}, {\orprop {\thenprop {\prop{1}}} {\elseprop {\prop{1}}}}}
           {\e{2}} {\t{}}
           {\filterset {\thenprop {\prop{}}} {\elseprop {\prop{}}}} 
           {\object{}},
    and
  \opsem {\openv{}} {\e{2}} {\v{}}
    allows us to apply the induction hypothesis on \e{2}.

To prove part 1 we use the induction hypothesis on \e{2}
to show either \object{} = \emptyobject{} 
or \inopenv {\openv{}} {\object{}} {\v{}}, since \e{} always
evaluates to the result of \e{2}.

For part 2 we use the induction hypothesis on \e{2}
to show if \istrueval{\v{}} then
        {\satisfies{\openv{}}{\thenprop{\prop{}}}}
        or
  if \isfalseval{\v{}} then
        {\satisfies{\openv{}}{\elseprop{\prop{}}}}.

Parts 3 and 4 follow from the induction hypothesis on \e{2}.
    \end{subcase}
\end{itemize}
\end{case}

\begin{case}[BE-Do1]
  \opsem {\openv{}} {\e{1}} {\errorval{\v{e}}},
  \opsem {\openv{}} {\e{}} {\errorval{\v{}}}

        Trivially reduces to an error.
\end{case}

\begin{case}[BE-Do2]
  \opsem {\openv{}} {\e{1}} {\v{1}},
  \opsem {\openv{}} {\e{2}} {\errorvalv{}},
  \opsem {\openv{}} {\e{}} {\errorvalv{}}

        As above.
\end{case}

\begin{case}[B-New]
  $
  \overrightarrow{
  \opsem {\openv{}}
         {\e{i}}
         {\v{i}}
       }$,
         $\newjava {\classhint{1}}
                  {\overrightarrow{\classhint{i}}}
                  {\overrightarrow{\v{i}}}
                  {\v{}}$

\begin{itemize}
  \item[]
\begin{subcase}[T-New]
  \ep{} = {\newexp {\class{}} {\overrightarrow{\ep{i}}}},
  \inct{\ctctorentry{\overr{\class{i}}}}{\ctlookupctors{\ct{}}{\classhint{}}},
  \overr{\javatotcnil{\classhint{i}}{\t{i}}},
  \overr{
  \judgementtworewrite {\propenv{}}
                    {\ep{i}} {\t{i}}
                    {\e{i}}
                  },
  \e{} = {\newstaticexp {\overrightarrow{\classhint{i}}} {\classhint{}} 
                                                          {\class{}} {\overrightarrow{\e{i}}}},
  \issubtypein{}{\javatotcexp{\classhint{}}}{\t{}},
  \inpropenv{\topprop{}}{\thenprop{\prop{}}},
  \inpropenv{\botprop{}}{\elseprop{\prop{}}},
  \issubobjin{}{\emptyobject{}}{\object{}}

Part 1 follows \object{} = \emptyobject{}.

Part 2 requires some explanation. The two false values in Typed Clojure
cannot be constructed with \newliteral{}, so the only case is \v{} $\not=$ \false\ (or \nil)
where \thenprop{\prop{}} = \topprop{} so \satisfies{\openv{}}{\thenprop{\prop{}}}.
\Void{} also lacks a constructor.

Part 3 holds as B-New reduces to a \emph{non-nilable}
instance of \class{} via \newjavaliteral (by assumption~\ref{appendix:assumption:new}), 
and {\t{}} is a supertype of \javatotcexp{\classhint{}}.

\end{subcase}
\item[]
\begin{subcase}[T-NewStatic]
  {\ep{}} = {\newstaticexp {\overrightarrow{\classhint{i}}} {\classhint{}}
                                                          {\class{}} {\overrightarrow{\e{i}}}}

  Non-reflective constructors cannot be written directly by the user, so we can assume
  the class information attached to the syntax corresponds to an actual constructor by inversion
  from T-New.

  The rest of this case progresses like T-New.
\end{subcase}
\end{itemize}
\end{case}

\begin{case}[BE-New1] $\overrightarrow{
  \opsem {\openv{}}
         {\e{i-1}}
         {\v{i-1}}
       }$,
  \opsem {\openv{}}
         {\e{i}}
         {\errorvalv{}},
  \opsem {\openv{}} {\e{}} {\errorvalv{}}

        Trivially reduces to an error.

\end{case}

\begin{case}[BE-New2] 
  $\overrightarrow{
  \opsem {\openv{}}
         {\e{i}}
         {\v{i}}
       }$,
         \newjava {\classhint{1}}
                  {\overrightarrow{\classhint{i}}}
                  {\overrightarrow{\v{i}}}
                  {\errorvalv{}},
        \opsem {\openv{}} {\e{}} {\errorvalv{}}

        As above.

\end{case}

\begin{case}[B-Field]
  \opsem {\openv{}}
         {\e{1}} 
         {\classvalue{\classhint{1}} {\classfieldpair{\fld{}} {\v{}}}}

\begin{itemize}
  \item[]
\begin{subcase}[T-Field]
  \ep{} = {\fieldexp {\fld{}} {\ep{1}}},
  \judgementtworewrite {\propenv{}} {\ep{}} {\s{}} {\e{}},
  \issubtypein{}{\s{}}{\Object{}},
  \tctojava{\s{}}{\classhint{1}},
  \inct{\ctfldentry{\fld{}}{\classhint{2}}}{\ctlookupfields{\ct{}}{\classhint{1}}},
  \e{} = {\fieldstaticexp {\classhint{1}} {\classhint{2}} {\fld{}} {\e{1}}}
  \issubtypein{}{\javatotcnilexp{\classhint{2}}}{\t{}},
  \inpropenv{\topprop{}}{\thenprop{\prop{}}},
  \inpropenv{\topprop{}}{\elseprop{\prop{}}},
  \issubobjin{}{\emptyobject{}}{\object{}}

Part 1 is trivial as \object{} is always \emptyobject{}.

Part 2 holds trivially; \v{} can be either a true or false value
and both {\thenprop{\prop{}}} and {\elseprop{\prop{}}}
are \topprop{}.

Part 3 relies on the semantics of \getfieldliteral (assumption~\ref{appendix:assumption:field})
in B-Field, which returns a \emph{nilable} instance of \classhint{2},
and \t{} is a supertype of \javatotcnilexp{\classhint{2}}.
Notice \issubtypein{}{\s{}}{\Object{}} is required to guard from dereferencing \nil{},
as {\classhint{1}} erases occurrences of \Nil{} in \s{} via  \tctojava{\s{}}{\classhint{1}}.
\end{subcase}
  \item[]

\begin{subcase}[T-FieldStatic]
  {\ep{}} = {\fieldstaticexp {\classhint{1}} {\classhint{2}} {\fld{}} {\e{1}}}

  Non-reflective field lookups cannot be written directly by the user, so we can assume
  the class information attached to the syntax corresponds to an actual field by inversion
  from T-Field.

  The rest of this case progresses like T-Field.
\end{subcase}

\end{itemize}
\end{case}

\begin{case}[BE-Field]
  \opsem {\openv{}}
         {\e{1}} 
         {\errorvalv{}},
  \opsem {\openv{}}
         {\e{}}
         {\errorvalv{}}

         Trivially reduces to an error.

\end{case}

\begin{case}[B-Method]
  \opsem {\openv{}}
         {\e{m}}
         {\v{m}},
  $\overrightarrow{
  \opsem {\openv{}}
         {\e{a}}
         {\v{a}}}$,
  \invokejavamethod {\classhint{1}} {\v{m}} {mth}
                    {\overrightarrow{\classhint{a}}} {\overrightarrow{\v{a}}}
                    {\classhint{2}}
                    {\v{}}

\begin{itemize}
  \item[]
\begin{subcase}[T-Method]
  \judgementtworewrite {\propenv{}} {\ep{}} {\s{}} {\e{}},
             \issubtypein{}{\s{}}{\Object{}},
  \tctojava{\s{}}{\classhint{1}},
                  \inct{\ctmthentry{\mth{}}{\overrightarrow{\classhint{i}}}{\classhint{2}}}{\ctlookupmethods{\ct{}}{\classhint{1}}},
                  \overr{\javatotcnil{\classhint{i}}{\t{i}}},
             \overr{
  \judgementtworewrite {\propenv{}} {\ep{i}} {\t{i}} {\e{i}}
                  },
  \e{} = {\methodstaticexp {\classhint{1}} 
                          {\overr {\classhint{i}}} 
                          {\classhint{2}}
                          {\mth{}} {\e{m}} {\overr{\e{a}}}},
                        \issubtypein{}{\javatotcnilexp{\classhint{2}}}{\t{}},
  \inpropenv{\topprop{}}{\thenprop{\prop{}}},
  \inpropenv{\topprop{}}{\elseprop{\prop{}}},
  \issubobjin{}{\emptyobject{}}{\object{}}

Part 1 is trivial as \object{} is always \emptyobject{}.

Part 2 holds trivially, \v{} can be either a true or false value
and both {\thenprop{\prop{}}} and {\elseprop{\prop{}}}
are \topprop{}.

Part 3 relies on the semantics of \invokejavamethodliteral (assumption~\ref{appendix:assumption:method})
in B-Method, which returns a \emph{nilable} instance of \classhint{2},
and \t{} is a supertype of \javatotcnil{\classhint{2}}.
Notice \issubtypein{}{\s{}}{\Object{}} is required to guard from dereferencing \nil{},
as {\classhint{1}} erases occurrences of \Nil{} in \s{} via  \tctojava{\s{}}{\classhint{1}}.
\end{subcase}
\item[]
\begin{subcase}[T-MethodStatic]
  \ep{} = 
  {\methodstaticexp {\classhint{1}} 
        {\overrightarrow {\classhint{i}}} 
        {\classhint{2}}
        {\mth{}} {\e{1}} {\overrightarrow{\e{i}}}}

  Non-reflective method invocations cannot be written directly by the user, so we can assume
  the class information attached to the syntax corresponds to an actual method by inversion
  from T-Method.

  The rest of this case progresses like T-Method.
\end{subcase}

\end{itemize}

\end{case}

\begin{case}[BE-Method1]
  \opsem {\openv{}}
         {\e{m}}
         {\errorval{\v{}}},
  \opsem {\openv{}}
         {\e{}}
         {\errorval{\v{}}}

         Trivially reduces to an error.
\end{case}
\begin{case}[BE-Method2]
  \opsem {\openv{}}
         {\e{m}}
         {\v{m}},
 $\overrightarrow{
  \opsem {\openv{}}
         {\e{n-1}}
         {\v{n-1}}
       }$,
  \opsem {\openv{}}
         {\e{n}}
         {\errorval{\v{}}},
  \opsem {\openv{}}
         {\e{}}
         {\errorval{\v{}}}

  As above.
\end{case}
\begin{case}[BE-Method3]
  \opsem {\openv{}}
         {\e{m}}
         {\v{m}},
  $\overrightarrow{
  \opsem {\openv{}}
         {\e{a}}
         {\v{a}}
       }$,
  \invokejavamethod {\classhint{1}} {\v{m}} {mth}
                    {\overrightarrow{\classhint{a}}} {\overrightarrow{\v{a}}}
                    {\classhint{2}}
                    {\errorvalv{}},
  \opsem {\openv{}} {\e{}} {\errorvalv{}}

  As above.

\end{case}

\begin{case}[B-DefMulti]
  \v{} = {\multi {\v{d}} {\emptydisptable}},
  \opsem {\openv{}} {\e{d}} {\v{d}}

\begin{itemize}
  \item[]
\begin{subcase}[T-DefMulti]
  \ep{} = {\createmultiexp {\s{}} {\ep{d}}},
  \s{} = {\ArrowOne {\x{}} {\t{1}} {\t{2}}
                          {\filterset {\thenprop {\prop{1}}}
                                      {\elseprop {\prop{1}}}}
                          {\object{1}}},
  \t{d} = {\ArrowOne {\x{}} {\t{1}} {\t{3}}
                          {\filterset {\thenprop {\prop{2}}}
                                      {\elseprop {\prop{2}}}}
                          {\object{2}}},
\judgementtworewrite {\propenv{}} {\ep{}} {\sp{}} {\e{}},
  \e{} = {\createmultiexp {\s{}} {\e{d}}},
  \issubtypein{}{\MultiFntype {\s{}} {\t{d}}}{\t{}},
  \inpropenv{\topprop{}}{\thenprop{\prop{}}},
  \inpropenv{\botprop{}}{\elseprop{\prop{}}},
  \issubobjin{}{\emptyobject{}}{\object{}}

Part 1 and 2 hold for the same reasons as T-True.
For part 3 we show \judgementtwo{}{\multi {\v{d}} {\emptydisptable}}{\MultiFntype {\s{}} {\t{d}}}
by T-Multi, since \judgementtwo {} {\v{d}} {\t{d}} by the inductive hypothesis on {\e{d}}
and {\emptydisptable} vacuously satisfies the other premises of T-Multi, so we are done.

\end{subcase}
\end{itemize}
\end{case}

\begin{case}[BE-DefMulti] \opsem {\openv{}} {\e{d}} {\errorvalv{}},
        \opsem {\openv{}} {\e{}} {\errorvalv{}}

        Trivially reduces to an error.

\end{case}

\begin{case}[B-DefMethod]

        \ 

        \begin{enumerate}
          \item
       \v{} = {\multi {\v{d}} {\disptablep{}}},
          \item
        \opsem {\openv{}}
               {\e{m}}
               {\multi {\v{d}} {\disptable{}}},
          \item
  \opsem {\openv{}}
         {\e{v}}
         {\v{v}},
          \item
  \opsem {\openv{}}
         {\e{f}}
         {\v{f}},
          \item
         \disptablep{} = {\extenddisptable {\disptable{}} 
                                {\v{v}}
                                {\v{f}}}
        \end{enumerate}

  \begin{itemize}
    \item[]
      \begin{subcase}[T-DefMethod]
        \ 
        
        \begin{enumerate}[resume]

          \item
  \ep{} = {\extendmultiexp {\ep{m}} {\ep{v}} {\ep{f}}},
          \item
  \t{m} = {\ArrowOne {\x{}} {\t{1}} {\s{}}
                     {\filterset {\thenprop {\prop{m}}}
                                 {\elseprop {\prop{m}}}}
                     {\object{m}}},
          \item
  \t{d} = {\ArrowOne {\x{}} {\t{1}} {\sp{}}
                     {\filterset {\thenprop {\prop{d}}}
                                 {\elseprop {\prop{d}}}}
                     {\object{d}}},
          \item
\judgementtworewrite {\propenv{}}
                  {\ep{m}} {\MultiFntype {\t{m}} {\t{d}}}
                  {\e{m}}
          \item
  \isacompare{\sp{}}{\object{d}}{\t{v}}{\filterset {\thenprop {\prop{i}}} {\elseprop {\prop{i}}}},
          \item
\judgementtworewrite {\propenv{}}
           {\e{v}} {\t{v}}
           {\e{v}}
          \item
  \judgementrewrite {\propenv{}, {\isprop{\t{1}} {\x{}}}, {\thenprop {\prop{i}}}}
           {\ep{f}} {\s{}}
           {\filterset {\thenprop {\prop{m}}}
                       {\elseprop {\prop{m}}}}
           {\object{m}}
           {\e{f}}
          \item
  \e{} = {\extendmultiexp {\e{m}} {\e{v}} {\e{f}}},
          \item
  \e{f} = {\abs {\x{}} {\t{1}} {\e{b}}},
\item
  \issubtypein{}{\MultiFntype {\t{m}} {\t{d}}}{\t{}},
\item
  \inpropenv{\topprop{}}{\thenprop{\prop{}}},
\item
  \inpropenv{\botprop{}}{\elseprop{\prop{}}},
\item
  \issubobjin{}{\emptyobject{}}{\object{}}
        \end{enumerate}

                                Part 1 and 2 hold for the same reasons as T-True, noting that the propositions
                                and object agree with T-Multi.

For part 3 we show
\judgementtwo{}{\multi {\v{d}} {\extenddisptable {\disptable{}}{\v{v}}{\v{f}}}}{\MultiFntype {\t{m}} {\t{d}}}
by noting \judgementtwo {} {\v{d}} {\t{d}},
  \judgementtwo{}{\v{v}}{\Top{}}
  and
  \judgementtwo{}{\v{f}}{\t{m}}, and since \disptable{} is in the correct form by the inductive
  hypothesis on {\e{m}} we can satisfy all premises of T-Multi, so we are done.

      \end{subcase}

  \end{itemize}
\end{case}

      \begin{case}[BE-DefMethod1]
        \opsem {\openv{}}
               {\e{m}}
               {\errorval{\v{}}},
        \opsem {\openv{}}
                  {\e{}}
                {\errorval{\v{}}}

                Trivially reduces to an error.

      \end{case}
      \begin{case}[BE-DefMethod2]
        \opsem {\openv{}}
         {\e{m}}
         {\multi {\v{d}} {\disptable{}}},
  \opsem {\openv{}}
         {\e{v}}
         {\errorval{\v{}}},
        \opsem {\openv{}}
                  {\e{}}
                {\errorval{\v{}}}

                Trivially reduces to an error.
      \end{case}
      \begin{case}[BE-DefMethod3]
        \opsem {\openv{}}
         {\e{m}}
         {\multi {\v{d}} {\disptable{}}},
  \opsem {\openv{}}
         {\e{v}}
         {\v{v}},
  \opsem {\openv{}}
         {\e{f}}
         {\errorval{\v{}}},
        \opsem {\openv{}}
                  {\e{}}
                {\errorval{\v{}}}

                Trivially reduces to an error.

      \end{case}

\begin{case}[B-BetaClosure]
  \ 

  \begin{itemize}
    \item
  \opsem{\openv{}}{\e{}}{\v{}},
    \item
  \opsem {\openv{}}
         {\e{1}}
         {\closure {\openv{c}} {\abs {\x{}} {\s{}} {\e{b}}}},
    \item
  \opsem {\openv{}}
         {\e{2}}
         {\v{2}},
    \item
  \opsem {\extendopenv {\openv{c}} {\x{}} {\v{2}}}
         {\e{b}}
         {\v{}}
     \end{itemize}

\begin{itemize}
  \item[]
\begin{subcase}[T-App]
  \ 

  \begin{itemize}
    \item
  \ep{} = {\appexp {\ep{1}} {\ep{2}}},
    \item
  \judgementrewrite {\propenv{}} {\ep{1}} {\ArrowOne {\x{}} {\s{}}
                                                       {\t{f}}
                                                       {\filterset {\thenprop {\prop{f}}}
                                                                   {\elseprop {\prop{f}}}}
                                                       {\object{f}}}
                {\filterset {\thenprop {\prop{1}}}
                            {\elseprop {\prop{1}}}}
                {\object{1}}
                {\e{1}},
    \item
  \judgementrewrite {\propenv{}}
                 {\ep{2}} {\s{}}
                 {\filterset {\thenprop {\prop{2}}}
                             {\elseprop {\prop{2}}}}
                 {\object{2}}
                 {\e{2}},
    \item
  \e{} = {\appexp {\e{1}} {\e{2}}},
    \item
      \issubtypein{}  {\replacefor {\t{f}} {\object{2}} {\x{}}}{\t{}},
    \item
      \inpropenv{\replacefor {\thenprop {\prop{f}}} {\object{2}} {\x{}}} {\thenprop {\prop{}}},
    \item
      \inpropenv{\replacefor {\elseprop {\prop{f}}} {\object{2}} {\x{}}} {\elseprop {\prop{}}},
    \item
      \issubobjin{}{\replacefor {\object{f}} {\object{2}} {\x{}}} {\object{}}
  \end{itemize}

         By inversion on \e{1} from T-Clos
         there is some environment {\propenvc{}} such that
         \begin{itemize}
           \item
              \satisfies{\openv{c}}{\propenvc{}} and
            \item
  \judgement {\propenvc{}} {\abs {\x{}} {\s{}} {\e{b}}} {\ArrowOne {\x{}} {\s{}}
                                                       {\t{f}}
                                                       {\filterset {\thenprop {\prop{f}}}
                                                                   {\elseprop {\prop{f}}}}
                                                       {\object{f}}}
                {\filterset {\thenprop {\prop{1}}}
                            {\elseprop {\prop{1}}}}
                {\object{1}},
         \end{itemize}
         and also by inversion on \e{1} from T-Abs
         \begin{itemize}
           \item
  { \judgementrewrite {\propenvc{}, {\isprop {\s{}} {\x{}}}}
              {\ep{b}} {\t{f}}
               {\filterset {\thenprop {\prop{f}}}
                           {\elseprop {\prop{f}}}}
               {\object{f}}
               {\e{b}}}.
         \end{itemize}

          From 
          \begin{itemize}
            \item
              \satisfies{\openv{c}}{\propenvc{}},
            \item
  \judgementrewrite {\propenv{}}
                 {\ep{2}} {\s{}}
                 {\filterset {\thenprop {\prop{2}}}
                             {\elseprop {\prop{2}}}}
                 {\object{2}}
                 {\e{2}} and 
            \item
  \opsem {\openv{}}
         {\e{2}}
         {\v{2}},
     \end{itemize}
              we know (by substitution)
              \satisfies{\extendopenv {\openv{c}} {\x{}} {\v{2}}}{\propenvc{},{\isprop{\s{}}{\x{}}}}.

              We want to prove
        \judgementrewrite {\propenvc{}}
                          {\replacefor{\ep{b}}{\v{2}}{\x{}}}
                          {\replacefor{\t{f}}{\object{2}}{\x{}}}
               {\replacefor{\filterset {\thenprop {\prop{f}}}
                                       {\elseprop {\prop{f}}}}{\object{2}}{\x{}}}
                          {\replacefor{\object{f}}{\object{2}}{\x{}}}
                          {\replacefor{\e{b}}{\v{2}}{\x{}}}, 
                          which can be justified by noting 
          \begin{itemize}
            \item
  \judgementtworewrite {\propenvc{},{\isprop{\s{}}{\x{}}}}{\ep{b}}{\t{f}}{\e{b}},
            \item
  \judgementrewrite {\propenv{}}
                 {\ep{2}} {\s{}}
                 {\filterset {\thenprop {\prop{2}}}
                             {\elseprop {\prop{2}}}}
                 {\object{2}}
                 {\e{2}} and 
            \item
  \opsem {\openv{}}
         {\e{2}}
         {\v{2}}.
     \end{itemize}

     From the previous fact and \satisfies{\openv{c}}{\propenvc{}},
              we know
  \opsem {\openv{c}}
         {\replacefor{\e{b}}{\v{2}}{\x{}}}
         {\v{}}.

                    Noting that 
      \issubtypein{}  {\replacefor {\t{f}} {\object{2}} {\x{}}}{\t{}},
      \inpropenv{\replacefor {\thenprop {\prop{f}}} {\object{2}} {\x{}}} {\thenprop {\prop{}}},
      \inpropenv{\replacefor {\elseprop {\prop{f}}} {\object{2}} {\x{}}} {\elseprop {\prop{}}}
      and
      \issubobjin{}{\replacefor {\object{f}} {\object{2}} {\x{}}} {\object{}},
                    we can use
         \begin{itemize}
           \item
        \judgementrewrite {\propenvc{}}
                          {\replacefor{\ep{b}}{\v{2}}{\x{}}}
                          {\replacefor{\t{f}}{\object{2}}{\x{}}}
               {\replacefor{\filterset {\thenprop {\prop{f}}}
                                       {\elseprop {\prop{f}}}}{\object{2}}{\x{}}}
                          {\replacefor{\object{f}}{\object{2}}{\x{}}}
                          {\replacefor{\e{b}}{\v{2}}{\x{}}}, 
           \item
              \satisfies{\openv{c}}{\propenvc{}},
           \item
\isconsistent{\openv{c}} (via induction hypothesis on {\ep{1}}), and
           \item 
  \opsem {\openv{c}}
         {\replacefor{\e{b}}{\v{2}}{\x{}}}
         {\v{}}.
         \end{itemize}
         to apply the induction hypothesis on {\replacefor{\ep{b}}{\v{2}}{\x{}}} and satisfy
         all conditions.

\end{subcase}
\end{itemize}
\end{case}

\begin{case}[B-Delta]
  \opsem {\openv{}} {\e{1}} {\const{}},
  \opsem {\openv{}} {\e{2}} {\v{2}},
  \constantopsem{\const{}}{\v{2}} = \v{}

\begin{itemize}
  \item[]
\begin{subcase}[T-App]
  \ 

  \begin{itemize}
    \item
  \ep{} = {\appexp {\ep{1}} {\ep{2}}},
    \item
  \judgementrewrite {\propenv{}} {\ep{1}} {\ArrowOne {\x{}} {\s{}}
                                                       {\t{f}}
                                                       {\filterset {\thenprop {\prop{f}}}
                                                                   {\elseprop {\prop{f}}}}
                                                       {\object{f}}}
                {\filterset {\thenprop {\prop{1}}}
                            {\elseprop {\prop{1}}}}
                {\object{1}}
                {\e{1}},
    \item
  \judgementrewrite {\propenv{}}
                 {\ep{2}} {\s{}}
                 {\filterset {\thenprop {\prop{2}}}
                             {\elseprop {\prop{2}}}}
                 {\object{2}}
                 {\e{2}},
    \item
  \e{} = {\appexp {\e{1}} {\e{2}}},
    \item
      \issubtypein{}  {\replacefor {\t{f}} {\object{2}} {\x{}}}{\t{}},
    \item
      \inpropenv{\replacefor {\thenprop {\prop{f}}} {\object{2}} {\x{}}} {\thenprop {\prop{}}},
    \item
      \inpropenv{\replacefor {\elseprop {\prop{f}}} {\object{2}} {\x{}}} {\elseprop {\prop{}}},
    \item
      \issubobjin{}{\replacefor {\object{f}} {\object{2}} {\x{}}} {\object{}}
  \end{itemize}


  Prove by cases on \const{}.
  \begin{itemize}
    \item[] \begin{subcase}[\const{} = \classconst]
        \issubtypein{}
  {\ArrowOne {\x{}} {\Top{}}
                                      {\Union{\Nil}{\Class}}
                                      {\filterset {\topprop{}}
                                                  {\topprop{}}}
                                      {\path {\classpe{}} {\x{}}}}
    {\ArrowOne {\x{}} {\s{}}
                                                       {\t{f}}
                                                       {\filterset {\thenprop {\prop{f}}}
                                                                   {\elseprop {\prop{f}}}}
                                                       {\object{f}}}

    Prove by cases on \v{2}.

        \begin{itemize}
          \item[] \begin{subcase}[\v{2} = \classvalue{\class{}} {\overrightarrow {\classfieldpair{\fld{i}} {\v{i}}}}]
                    \v{} = \class{}

                    To prove part 1,
                    note
                    \issubobjin{}{\replacefor {\object{f}} {\object{2}} {\x{}}} {\object{}},
                    and \issubobjin{}{\path {\classpe{}} {\x{}}}{\object{f}}.
                    Then either \object{} = \emptyobject{} and we are done,
                    or \object{} = {\path {\classpe{}}{\object{2}}} and
                    by the induction hypothesis on \e{2} we know \inopenv {\openv{}} {\object{2}} {\v{2}}
                    and by the definition of path translation we know
                    {\openv{}}({\path {\classpe{}} {\object{2}}}) = {\appexp {\classconst{}} {{\openv{}}(\object{2})}},
                    which evaluates to \v{}.

                    Part 2 is trivial since both propositions can only be \topprop{}.
                    
                    Part 3 holds because 
                    \v{} = \class{},
                    \issubtypein{}{\Union{\Nil}{\Class}}{\replacefor {\t{f}} {\object{2}} {\x{}}}
                    and
                    \issubtypein{}{\replacefor {\t{f}} {\object{2}} {\x{}}}{\t{}},
                    so
                    {\judgementtwo{}{\v{}}{\t{}}}
                    since
                    {\judgementtwo{}{\class{}}{\Union{\Nil}{\Class}}}.
                  \end{subcase}
          \item[] \begin{subcase}[\v{2} = \class{}] \v{} = \Class{}

              As above.
                  \end{subcase}
          \item[] \begin{subcase}[\v{2} = \true{}] \v{} = \Boolean{}

              As above.
                  \end{subcase}
          \item[] \begin{subcase}[\v{2} = \false{}] \v{} = \Boolean{}

              As above.
                  \end{subcase}
          \item[] \begin{subcase}[\v{2} = {\closure {\openv{}} {\abs {\x{}} {\t{}} {\e{}}}}] \v{} = \IFn{}

              As above.
                  \end{subcase}
          \item[] \begin{subcase}[\v{2} = {\multi {\v{d}} {\disptable{}}}] \v{} = \HMapInstance{}

              As above.
                  \end{subcase}
          \item[] \begin{subcase}[\v{2} = {\curlymapvaloverright{\v{1}}{\v{2}}}] \v{} = \Keyword{}

              As above.
                  \end{subcase}
          \item[] \begin{subcase}[\v{2} = {\nil{}}] \v{} = \nil{}

             Parts 1 and 2 as above.
                    Part 3 holds because \v{} = \nil{}
                    and {\judgementtwo{}{\nil{}}{\Union{\Nil}{\Class}}}.
                  \end{subcase}
        \end{itemize}
      \end{subcase}


  \end{itemize}

\end{subcase}
\end{itemize}
\end{case}

\begin{case}[B-BetaMulti]
  \ 

  \begin{itemize}
    \item
  \opsem {\openv{}}
         {\e{1}}
         {\multi {\v{d}} {\disptable{}}},
    \item
  \opsem {\openv{}}
         {\e{2}}
         {\v{2}},
    \item
  \opsem {\openv{}}
         {\appexp {\v{d}} {\v{2}}}
         {\v{e}},
    \item
  \getmethod {\disptable{}}
             {\v{e}}
             {\v{l}}
             {\v{g}},
    \item
  \opsem {\openv{}}
         {\appexp {\v{g}} {\v{2}}}
         {\v{}},
       \item {\disptable{}} = {\curlymapvaloverright{\v{k}}{\v{v}}}
     \end{itemize}
     \begin{itemize}
       \item[]
\begin{subcase}[T-App]
  \ 

  \begin{itemize}
    \item
  \ep{} = {\appexp {\ep{1}} {\ep{2}}},
    \item
  \judgementrewrite {\propenv{}} {\ep{1}} {\ArrowOne {\x{}} {\s{}}
                                                       {\t{f}}
                                                       {\filterset {\thenprop {\prop{f}}}
                                                                   {\elseprop {\prop{f}}}}
                                                       {\object{f}}}
                {\filterset {\thenprop {\prop{1}}}
                            {\elseprop {\prop{1}}}}
                {\object{1}}
                {\e{1}},
    \item
  \judgementrewrite {\propenv{}}
                 {\ep{2}} {\s{}}
                 {\filterset {\thenprop {\prop{2}}}
                             {\elseprop {\prop{2}}}}
                 {\object{2}}
                 {\e{2}},
    \item
  \e{} = {\appexp {\e{1}} {\e{2}}},
    \item
      \issubtypein{}  {\replacefor {\t{f}} {\object{2}} {\x{}}}{\t{}},
    \item
      \inpropenv{\replacefor {\thenprop {\prop{f}}} {\object{2}} {\x{}}} {\thenprop {\prop{}}},
    \item
      \inpropenv{\replacefor {\elseprop {\prop{f}}} {\object{2}} {\x{}}} {\elseprop {\prop{}}},
    \item
      \issubobjin{}{\replacefor {\object{f}} {\object{2}} {\x{}}} {\object{}},
  \end{itemize}

     By inversion on \e{1} via T-Multi we know 
     \begin{itemize}
       \item
         \judgementrewrite{\propenv{}}{\ep{1}}{\MultiFntype{\s{t}}{\s{d}}}
                {\filterset {\thenprop {\prop{1}}}
                            {\elseprop {\prop{1}}}}
                {\object{1}}{\e{1}},
           
         \item \s{t} = {\ArrowOne {\x{}} {\s{}}
                                                       {\t{f}}
                                                       {\filterset {\thenprop {\prop{f}}}
                                                                   {\elseprop {\prop{f}}}}
                                                       {\object{f}}},
         \item \s{d} = {\ArrowOne {\x{}} {\s{}}
                                                       {\t{d}}
                                                       {\filterset {\thenprop {\prop{d}}}
                                                                   {\elseprop {\prop{d}}}}
                                                       {\object{d}}},
       \item
         \judgementtwo{}{\v{d}}{\s{d}}
              \item
  $\overrightarrow{\judgementtwo{}{\v{k}}{\Top{}}}$, and 
\item
  $\overrightarrow{\judgementtwo{}{\v{v}}{\s{t}}}$.
  \end{itemize}

  By the inductive hypothesis on 
  \opsem {\openv{}}
         {\e{2}}
         {\v{2}}
  we know 
  \judgementrewrite {\propenv{}} {\v{2}} {\s{}}
             {\filterset {\thenprop {\prop{2}}}
                         {\elseprop {\prop{2}}}}
                       {\object{2}}
                       {\v{2}}.

We then consider applying the evaluated argument to the dispatch function:
  \opsem {\openv{}}
         {\appexp {\v{d}} {\v{2}}}
         {\v{e}}.

         Since we can satisfy T-App with
       \begin{itemize}
         \item
         \judgementtwo{}{\v{d}}{\ArrowOne {\x{}} {\s{}}
                                                       {\t{d}}
                                                       {\filterset {\thenprop {\prop{d}}}
                                                                   {\elseprop {\prop{d}}}}
                                                       {\object{d}}}, and
         \item
  \judgementrewrite {\propenv{}} {\v{2}} {\s{}}
             {\filterset {\thenprop {\prop{2}}}
                         {\elseprop {\prop{2}}}}
                       {\object{2}}
                       {\v{2}}.
       \end{itemize}
       we can then apply the inductive hypothesis
       to derive
  \judgementrewrite {\propenv{}} {\v{e}} 
  {\replacefor{\t{d}}
              {\object{2}}
              {\x{}}}
             {\replacefor{\filterset {\thenprop {\prop{d}}}
                                     {\elseprop {\prop{d}}}}
                         {\object{2}}
                         {\x{}}}
                       {\replacefor
                         {\object{d}}
                         {\object{2}}
                         {\x{}}}
                       {\v{e}}.

 Now we consider how we choose which method to dispatch to.

 As 
  \getmethod {\disptable{}}
             {\v{e}}
             {\v{l}}
             {\v{g}}, by inversion on \getmethodliteral
             we know
   there exists exactly one \v{k} such that 
   \entryinmap{\mapvalentry{\v{k}}{\v{g}}}{\disptable{}} and \isaopsem{\v{e}}{\v{k}} = {\true{}}.

   By inversion we know T-DefMethod must have extended \disptable{} 
   with the well-typed dispatch value \v{k},
   thus {\judgementtwo{}{\v{k}}{\t{k}}}, and
   the well-typed method \v{g},
   so {\judgementtwo{}{\v{g}}{\s{t}}}.

  We can also prove that given
        \begin{itemize}
          \item
  \judgementrewrite {\propenv{}} {\v{e}} 
  {\replacefor{\t{d}}
              {\object{2}}
              {\x{}}}
             {\replacefor{\filterset {\thenprop {\prop{d}}}
                                     {\elseprop {\prop{d}}}}
                         {\object{2}}
                         {\x{}}}
                       {\replacefor
                         {\object{d}}
                         {\object{2}}
                         {\x{}}}
                       {\v{e}}.
    \item
  \judgementtwo {\propenv{}} {\v{k}} {\t{k}},
                     \item
        \isaopsem{\v{e}}{\v{k}} = {\true{}}, 
      \item
        \satisfies{\openv{}}{\propenv{}},
    \item
  \isacompare{\t{d}}
                       {\replacefor
                         {\object{d}}
                         {\object{2}}
                         {\x{}}}
  {\t{k}}{\filterset {\thenprop {\propp{}}} {\elseprop {\propp{}}}},
          \item
        \inpropenv{\thenprop{\propp{}}}{\thenprop{\propp{}}}, and
    \item
        \inpropenv{\elseprop{\propp{}}}{\elseprop{\propp{}}}.
        \end{itemize}
  we can apply \lemref{appendix:lemma:isa} to derive
  then {\satisfies{\openv{}}{\thenprop{\propp{}}}}.

   Now we consider applying the evaluated argument to the chosen method:
  \opsem {\openv{}}
         {\appexp {\v{g}} {\v{2}}}
         {\v{}}.

  By inversion via B-DefMethod we can assume {\v{g}} = {\abs{\x{}}{\s{}}{\e{b}}}, 
  ie. that we have chosen a method to dispatch to that is a closure.

  Because 
  \opsem {\openv{}}
         {\appexp {\v{g}} {\v{2}}}
         {\v{}}
         and
    \judgementtwo{\propenv{}}{\v{2}}{\s{}},
  by inversion via B-BetaClosure we know {\v{}} = {\replacefor{\e{b}}{\v{2}}{\x{}}}.

  With the following premises:
\begin{itemize}
  \item
{\judgementrewrite{{\propenv{}},{\thenprop{\propp{}}}}
                  {\replacefor{\ep{b}}{\v{2}}{\x{}}}
                  {\replacefor{\t{f}} {\object{2}}{\x{}}}
                  {\replacefor
                   {\filterset {\thenprop {\prop{f}}}
                               {\elseprop {\prop{f}}}}
                             {\object{2}}
                             {\x{}}}
                  {\replacefor
                          {\object{f}}
                             {\object{2}}
                             {\x{}}}
                  {\replacefor{\e{b}}{\v{2}}{\x{}}}
                        },
    \begin{itemize}
      \item From
{\judgementrewrite{{\propenv{}},{\isprop{\s{}}{\x{}}}}
                  {\e{b}}
                  {\t{f}}
               {\filterset {\thenprop {\prop{f}}}
                                       {\elseprop {\prop{f}}}}
                          {\object{f}}
                          {\e{b}}}
          via the inductive hypothesis on 
  \opsem {\openv{}}
         {\appexp {\abs{\x{}}{\s{}}{\e{b}}} {\v{2}}}
         {\v{}},
      \item then we can derive
{\judgementrewrite{{\propenv{}}}
                  {\replacefor{\ep{b}}{\v{2}}{\x{}}}
                  {\replacefor{\t{f}} {\object{2}}{\x{}}}
                  {\replacefor
                   {\filterset {\thenprop {\prop{f}}}
                               {\elseprop {\prop{f}}}}
                             {\object{2}}
                             {\x{}}}
                  {\replacefor
                          {\object{f}}
                             {\object{2}}
                             {\x{}}}
                  {\replacefor{\e{b}}{\v{2}}{\x{}}}
                        } via substitution and the fact that {\x{}} is fresh 
                        therefore \x{} $\not\in$ \fv{\propenv{}} so we do not need to substitution for \x{} in \propenv{}. 
                        
      \item 
        \satisfies{\openv{}}{\propenv{}, {\thenprop{\propp{}}}}
        because
        \satisfies{\openv{}}{\propenv{}} and {\satisfies{\openv{}}{\thenprop{\propp{}}}} via M-And.
    \end{itemize}
  \item
              \satisfies{\openv{}}{{\propenv{}},{\thenprop{\propp{}}}},
    \begin{itemize}
      \item From \satisfies{\openv{}}{\propenv{}} and \item {\satisfies{\openv{}}{\thenprop{\propp{}}}}  via M-And.
    \end{itemize}
           \item
\isconsistent{\openv{}}, and
           \item 
  \opsem {\openv{}}
         {\replacefor{\e{b}}{\v{2}}{\x{}}}
         {\v{}}.
\end{itemize}

we can apply the inductive hypothesis to satisfy our overall goal for this subcase.
\end{subcase}
     \end{itemize}
\end{case}

\begin{case}[BE-Beta1]
  \ 

  Reduces to an error.
\end{case}
\begin{case}[BE-Beta2]
  \ 

  Reduces to an error.
\end{case}
\begin{case}[BE-BetaClosure]
  \ 

  Reduces to an error.
\end{case}
\begin{case}[BE-BetaMulti1]
  \ 

  Reduces to an error.
\end{case}
\begin{case}[BE-BetaMulti2]
  \ 

  Reduces to an error.
\end{case}
\begin{case}[BE-Delta]
  \ 

  Reduces to an error.
\end{case}

\begin{case}[B-IsA]
        \opsem {\openv{}} {\e{1}} {\v{1}},
        \opsem {\openv{}} {\e{2}} {\v{2}},
        \isaopsem{\v{1}}{\v{2}} = {\v{}}

  \begin{itemize}
    \item[]
      \begin{subcase}[T-IsA]
  \ep{} = {\isaapp {\ep{1}} {\ep{2}}},
  \judgementrewrite {\propenv{}} {\ep{1}} {\t{1}}
             {\filterset {\thenprop {\prop{1}}}
                         {\elseprop {\prop{1}}}}
                       {\object{1}}
                       {\e{1}},
  \judgementrewrite {\propenv{}} {\ep{2}} {\t{2}}
             {\filterset {\thenprop {\prop{2}}}
                         {\elseprop {\prop{2}}}}
                       {\object{2}}
                       {\e{2}},
  \e{} = {\isaapp {\e{1}} {\e{2}}},
  \issubtypein{}{\Boolean}{\t{}},
  \isacompare{\t{1}}{\object{1}}{\t{2}}{\filterset {\thenprop {\propp{}}} {\elseprop {\propp{}}}},
  \inpropenv{\thenprop {\propp{}}}{\thenprop {\prop{}}},
  \inpropenv{\elseprop {\propp{}}}{\elseprop {\prop{}}},
  \issubobjin{}{\emptyobject{}}{\object{}}

        Part 1 holds trivially with \object{} = \emptyobject{}.

        For part 2, by the induction hypothesis on \e{1} and \e{2}
        we know
  \judgementrewrite {\propenv{}} {\v{1}} {\t{1}}
             {\filterset {\thenprop {\prop{1}}}
                         {\elseprop {\prop{1}}}}
                       {\object{1}}
                       {\v{1}} and
  \judgementrewrite {\propenv{}} {\v{2}} {\t{2}}
             {\filterset {\thenprop {\prop{2}}}
                         {\elseprop {\prop{2}}}}
                       {\object{2}}
                       {\v{2}},
                       so we can then apply
        \lemref{appendix:lemma:isa}
        to reach our goal.

        Part 3 holds because by the definition of \isaopsemliteral
        \v{} can only be \true\ or \false, 
        and since \judgementtwo{\propenv{}}{\true}{\t{}}
        and
        \judgementtwo{\propenv{}}{\false}{\t{}}
        we are done.
      \end{subcase}
  \end{itemize}
\end{case}

      \begin{case}[BE-IsA1]
        \opsem {\openv{}} {\e{1}} {\errorvalv{}}

        Trivially reduces to an error.
      \end{case}
      \begin{case}[BE-IsA2]
       \opsem {\openv{}} {\e{1}} {\v{1}},
       \opsem {\openv{}} {\e{2}} {\errorvalv{}}

        Trivially reduces to an error.
      \end{case}

\begin{case}[B-Get]
      $\opsem {\openv{}} {\e{m}}{\v{m}}$,
        $\v{m} = {\curlymap{\overrightarrow{({\v{a}}\ {\v{b}})}}}$,
         \opsem {\openv{}} {\e{k}} {\k{}},
         $\keyinmap{\k{}}{\curlymap{\overrightarrow{({\v{a}}\ {\v{b}})}}}$,
         \getmap{\curlymap{\overrightarrow{({\v{a}}\ {\v{b}})}}} {\k{}} = {\v{}}

  \begin{itemize}
    \item[]
      \begin{subcase}[T-GetHMap]
  \ep{} = {\getexp {\ep{m}} {\ep{k}}},
  \judgementrewrite {\propenv{}} {\ep{m}} {\Unionsplice {\overrightarrow {\HMapgeneric {\mandatory{}} {\absent{}}}}}
           {\filterset {\thenprop {\prop{m}}} {\elseprop {\prop{m}}}}
           {\object{m}}
           {\e{m}},
  \judgementtworewrite {\propenv{}} {\ep{k}} {\Value {k}}{\e{k}},
  \overr{\inmandatory{\k{}}{\t{i}}{\mandatory{}},}
  \e{} = {\getexp {\e{m}} {\e{k}}},
  \issubtypein{}{\Unionsplice {\overrightarrow {\t{i}}}}{\t{}} ,
  \thenprop{\prop{}} = {\topprop{}},
  \elseprop{\prop{}} = {\topprop{}},
  \issubobjin{}{\replacefor {\path {\keype{k}} {\x{}}}
                          {\object{m}}
                          {\x{}}}
                        {\object{}}

         To prove part 1 we consider two cases on the form of \object{m}: 
         \begin{itemize}
           \item
         if {\object{m}} = \emptyobject{}
         then \object{} = \emptyobject{} by substitution, which gives the desired result;
           \item
         if \object{m} = {\path {\pathelem{m}} {\x{m}}}
         then \issubobjin{}{\path {\keype{k}} {\object{m}}}{\object{}} by substitution.
         We note by the definition of path translation
         {\openv{}}({\path {\keype{k}} {\object{m}}}) =
         {\getexp {{\openv{}}(\object{m})}{\k{}}}
         and by the induction hypothesis on \e{m}
         {{\openv{}}(\object{m})} = {\curlymap{\overrightarrow{({\v{a}}\ {\v{b}})}}},
         which together imply 
         \inopenv {\openv{}} {\object{}} {\getexp {\curlymap{\overrightarrow{({\v{a}}\ {\v{b}})}}} {\k{}}}.
         Since this is the same form as B-Get, we can apply the premise
         \getmap{\curlymap{\overrightarrow{({\v{a}}\ {\v{b}})}}} {\k{}} = {\v{}}
         to derive \inopenv {\openv{}} {\object{}} {\v{}}.
         \end{itemize}
         
         Part 2 holds trivially as \thenprop{\prop{}} = {\topprop{}}
         and \elseprop{\prop{}} = {\topprop{}}.

         To prove part 3 we note that (by the induction hypothesis on \e{m})
         $\judgementtwo{}{\v{m}}{\Unionsplice{\overrightarrow {\HMapgeneric {\mandatory{}} {\absent{}}}}}$,
         where $\overrightarrow{\inmandatory{\k{}}{\t{i}}{\mandatory{}}}$, and 
         both
         $\keyinmap{\k{}}{\curlymap{\overrightarrow{({\v{a}}\ {\v{b}})}}}$
         and
         \getmap{\curlymap{\overrightarrow{({\v{a}}\ {\v{b}})}}} {\k{}} = {\v{}}
         imply \judgementtwo{}{\v{}}{\Unionsplice {\overrightarrow {\t{i}}}}.

      \end{subcase}
    \item[]
      \begin{subcase}[T-GetHMapAbsent]
  \ep{} = {\getexp {\ep{m}} {\ep{k}}},
  \judgementtworewrite {\propenv{}} {\ep{k}} {\Value {k}} {\e{k}},
  \judgementrewrite {\propenv{}} {\ep{m}} {\HMapgeneric {\mandatory{}} {\absent}}
           {\filterset {\thenprop {\prop{m}}} {\elseprop {\prop{m}}}}
           {\object{m}}
           {\e{m}},
  {\inabsent{\k{}}{\absent{}}},
  \e{} = {\getexp {\e{m}} {\e{k}}},
  \issubtypein{}{\Nil}{\t{}},
  \thenprop{\prop{}} = {\topprop{}},
  \elseprop{\prop{}} = {\topprop{}},
  \issubobjin{}{\replacefor
               {\path {\keype{k}} {\x{}}}
                          {\object{m}}
                          {\x{}}}
                        {\object{}}

       Unreachable subcase because 
         $\keyinmap{\k{}}{\curlymap{\overrightarrow{({\v{a}}\ {\v{b}})}}}$,
         contradicts
                {\inabsent{\k{}}{\absent{}}}.
      \end{subcase}
    \item[]
      \begin{subcase}[T-GetHMapPartialDefault]
  \ep{} = {\getexp {\ep{m}} {\ep{k}}},
  \judgementtworewrite {\propenv{}} {\ep{k}} {\Value {k}}{\e{k}},
 \judgementrewrite {\propenv{}} {\ep{m}} {\HMapp {\mandatory{}} {\absent}}
           {\filterset {\thenprop {\prop{m}}} {\elseprop {\prop{m}}}}
           {\object{m}}
           {\e{m}},
             {\notinmandatory{\k{}}{\t{}}{\mandatory{}}},
             {\notinabsent{\k{}}{\absent{}}},
  \e{} = {\getexp {\e{m}} {\e{k}}},
  \t{} = \Top,
  \thenprop{\prop{}} = {\topprop{}},
  \elseprop{\prop{}} = {\topprop{}},
  \issubobjin{}{\replacefor
               {\path {\keype{k}} {\x{}}}
                          {\object{m}}
                          {\x{}}}{\object{}}

         Parts 1 and 2 are the same as the B-Get subcase.
         Part 3 is trivial as \t{} = \Top.
      \end{subcase}
  \end{itemize}
\end{case}

\begin{case}[B-GetMissing]
        \v{} = \nil,
        $\opsem {\openv{}}
        {\e{m}} {\curlymap{\overrightarrow{({\v{a}}\ {\v{b}})}}}$,
       \opsem {\openv{}} {\e{k}} {\k{}},
       \keynotinmap{\k{}}{\curlymap{\overrightarrow{({\v{a}}\ {\v{b}})}}}

  \begin{itemize}
    \item[]
      \begin{subcase}[T-GetHMap]
  \ep{} = {\getexp {\ep{m}} {\ep{k}}},
  \judgementrewrite {\propenv{}} {\ep{m}} {\Unionsplice {\overrightarrow {\HMapgeneric {\mandatory{}} {\absent{}}}}}
           {\filterset {\thenprop {\prop{m}}} {\elseprop {\prop{m}}}}
           {\object{m}}
           {\e{m}},
  \judgementtworewrite {\propenv{}} {\ep{k}} {\Value {k}}{\e{k}},
  \overr{\inmandatory{\k{}}{\t{i}}{\mandatory{}},}
  \e{} = {\getexp {\e{m}} {\e{k}}},
  \issubtypein{}{\Unionsplice {\overrightarrow {\t{i}}}}{\t{}},
  \thenprop{\prop{}} = {\topprop{}},
  \elseprop{\prop{}} = {\topprop{}},
  \issubobjin{}{\replacefor {\path {\keype{k}} {\x{}}}
                          {\object{m}}
                          {\x{}}}{\object{}}

       Unreachable subcase because 
       \keynotinmap{\k{}}{\curlymap{\overrightarrow{({\v{a}}\ {\v{b}})}}}
       contradicts ${\inmandatory{\k{}}{\t{}}{\mandatory{}}}$.
      \end{subcase}
    \item[]
      \begin{subcase}[T-GetHMapAbsent]
  \ep{} = {\getexp {\ep{m}} {\ep{k}}},
  \judgementtworewrite {\propenv{}} {\ep{k}} {\Value {k}} {\e{k}},
  \judgementrewrite {\propenv{}} {\ep{m}} {\HMapgeneric {\mandatory{}} {\absent}}
           {\filterset {\thenprop {\prop{m}}} {\elseprop {\prop{m}}}}
           {\object{m}}
           {\e{m}},
  {\inabsent{\k{}}{\absent{}}},
  \e{} = {\getexp {\e{m}} {\e{k}}},
  \issubtypein{}{\Nil}{\t{}},
  \thenprop{\prop{}} = {\topprop{}},
  \elseprop{\prop{}} = {\topprop{}},
  \issubobjin{}{\replacefor
               {\path {\keype{k}} {\x{}}}
                          {\object{m}}
                          {\x{}}}{\object{}}

         To prove part 1 we consider two cases on the form of \object{m}: 
         \begin{itemize}
           \item
         if {\object{m}} = \emptyobject{}
         then \object{} = \emptyobject{} by substitution, which gives the desired result;
           \item
         if \object{m} = {\path {\pathelem{m}} {\x{m}}}
         then \issubobjin{}{\path {\keype{k}} {\object{m}}}{\object{}} by substitution.
         We note by the definition of path translation
         {\openv{}}({\path {\keype{k}} {\object{m}}}) =
         {\getexp {{\openv{}}(\object{m})}{\k{}}}
         and by the induction hypothesis on \e{m}
         {{\openv{}}(\object{m})} = {\curlymap{\overrightarrow{({\v{a}}\ {\v{b}})}}},
         which together imply 
         \inopenv {\openv{}} {\object{}} {\getexp {\curlymap{\overrightarrow{({\v{a}}\ {\v{b}})}}} {\k{}}}.
         Since this is the same form as B-GetMissing, we can apply the premise
        \v{} = \nil\ 
         to derive \inopenv {\openv{}} {\object{}} {\v{}}.
         \end{itemize}
         
         Part 2 holds trivially as \thenprop{\prop{}} = {\topprop{}}
         and \elseprop{\prop{}} = {\topprop{}}.

         To prove part 3 we note that \e{m} has type {\HMapgeneric {\mandatory{}} {\absent{}}}
         where {\inabsent{\k{}}{\absent{}}}, and
         the premises of B-GetMissing
         \keynotinmap{\k{}}{\curlymap{\overrightarrow{({\v{a}}\ {\v{b}})}}}
         and
          \v{} = \nil\ 
         tell us {\v{}} must be of type {\t{}}.
      \end{subcase}
    \item[]
      \begin{subcase}[T-GetHMapPartialDefault]
  \ep{} = {\getexp {\ep{m}} {\ep{k}}},
  \judgementtworewrite {\propenv{}} {\ep{k}} {\Value {k}}{\e{k}},
 \judgementrewrite {\propenv{}} {\ep{m}} {\HMapp {\mandatory{}} {\absent}}
           {\filterset {\thenprop {\prop{m}}} {\elseprop {\prop{m}}}}
           {\object{m}}
           {\e{m}},
             {\notinmandatory{\k{}}{\t{}}{\mandatory{}}},
             {\notinabsent{\k{}}{\absent{}}},
  \e{} = {\getexp {\e{m}} {\e{k}}},
  \t{} = \Top,
  \thenprop{\prop{}} = {\topprop{}},
  \elseprop{\prop{}} = {\topprop{}},
  \issubobjin{}{\replacefor
               {\path {\keype{k}} {\x{}}}
                          {\object{m}}
                          {\x{}}}{\object{}}

         Parts 1 and 2 are the same as the B-GetMissing subcase of T-GetHMapAbsent.
         Part 3 is trivial as \t{} = \Top.
      \end{subcase}
  \end{itemize}
\end{case}

\begin{case}[BE-Get1]
  \ 

  Reduces to an error.
\end{case}

\begin{case}[BE-Get2]
  \ 

  Reduces to an error.
\end{case}

\begin{case}[B-Assoc]
        \v{} = 
        {\extendmap{\curlymap{\overrightarrow{({\v{a}}\ {\v{b}})}}}
                {\k{}}{\v{v}}},
        \opsem {\openv{}}
        {\e{m}} {\curlymap{\overrightarrow{({\v{a}}\ {\v{b}})}}},
        \opsem {\openv{}} {\e{k}} {\k{}},
        \opsem {\openv{}} {\e{v}} {\v{v}}

  \begin{itemize}
    \item[]
      \begin{subcase}[T-AssocHMap]
  \judgementtworewrite {\propenv{}} {\ep{m}} {\HMapgeneric {\mandatory{}} {\absent}} {\e{m}},
  \judgementtworewrite {\propenv{}} {\ep{k}} {\Value{\k{}}}{\e{k}},
  \judgementtworewrite {\propenv{}} {\ep{v}} {\t{}}{\e{v}},
  {\k{}} $\not\in$ {\absent{}},
  \ep{} = {\assocexp {\ep{m}} {\ep{k}} {\ep{v}}},
  \e{} = {\assocexp {\e{m}} {\e{k}} {\e{v}}},
  \issubtypein{}{\HMapgeneric {\extendmandatoryset {\mandatory{}}{\k{}}{\t{}}} {\absent}}{\t{}},
  \thenprop{\prop{}} = {\topprop{}},
  \elseprop{\prop{}} = {\botprop{}},
  \object{} = \emptyobject{}

        Parts 1 and 2 hold for the same reasons as T-True.
      \end{subcase}
  \end{itemize}
\end{case}

\begin{case}[BE-Assoc1]
  \ 

  Reduces to an error.
\end{case}

\begin{case}[BE-Assoc2]
  \ 

  Reduces to an error.
\end{case}

\begin{case}[BE-Assoc3]
  \ 

  Reduces to an error.
\end{case}

\begin{case}[B-IfFalse]
        \opsem {\openv{}} {\e{1}} {\false}
        \ \ \text{or}\ \ 
        \opsem {\openv{}} {\e{1}} {\nil},
        \opsem {\openv{}} {\e{3}} {\v{}}

  \begin{itemize}
    \item[]
      \begin{subcase}[T-If]
  \ep{} = {\ifexp {\ep1} {\ep2} {\ep3}},
  \judgementrewrite {\propenv{}} {\ep1} {\t{1}} {\filterset {\thenprop {\prop{1}}} {\elseprop {\prop{1}}}}
                 {\object{1}}
                 {\e1},
  \judgementrewrite {\propenv{}, {\thenprop {\prop{1}}}}
                 {\ep2} {\t{}} {\filterset {\thenprop {\prop{2}}} {\elseprop {\prop{2}}}}
                 {\object{}}
                 {\e2},
  \judgementrewrite {\propenv{}, {\elseprop {\prop{1}}}}
                 {\ep3} {\t{}} {\filterset {\thenprop {\prop{3}}} {\elseprop {\prop{3}}}}
                 {\object{}}
                 {\e3},
  \e{} = {\ifexp {\e1} {\e2} {\e3}},
  \inpropenv{\orprop {\thenprop {\prop{2}}} {\thenprop {\prop{3}}}}{\thenprop{\prop{}}},
  \inpropenv{\orprop {\elseprop {\prop{2}}} {\elseprop {\prop{3}}}}{\elseprop{\prop{}}}

              For part 1, either \object{} = \emptyobject{}, or \e{} evaluates to the
              result of \e{3}.

              To prove part 2, we consider two cases:
              \begin{itemize}
                \item if \isfalseval{\v{}}
                  then \e{3} evaluates to a false value so {\satisfies{\openv{}}{\elseprop {\prop{3}}}}, and thus
                  {\satisfies{\openv{}}{\orprop {\elseprop {\prop{2}}} {\elseprop {\prop{3}}}}} by M-Or, 
                \item otherwise
                  \istrueval{\v{}},
                  so \e{3} evaluates to a true value so {\satisfies{\openv{}}{\thenprop {\prop{3}}}}, and thus
                  {\satisfies{\openv{}}{\orprop {\thenprop {\prop{2}}} {\thenprop {\prop{3}}}}} by M-Or.
              \end{itemize}

              Part 3 is trivial as
              \opsem {\openv{}} {\e{3}} {\v{}}
              and {\judgementtwo{}{\v{}}{\t{}}} by the induction hypothesis on {\e{3}}.
      \end{subcase}
  \end{itemize}
\end{case}

\begin{case}[B-IfTrue]
        \opsem {\openv{}} {\e{1}} {\v{1}},
              ${\v{1}} \not= {\false}$,
              ${\v{1}} \not= {\nil}$,
              \opsem {\openv{}} {\e{2}} {\v{}}

  \begin{itemize}
    \item[]
      \begin{subcase}[T-If]
  \ep{} = {\ifexp {\ep1} {\ep2} {\ep3}},
  \judgementrewrite {\propenv{}} {\ep1} {\t{1}} {\filterset {\thenprop {\prop{1}}} {\elseprop {\prop{1}}}}
                 {\object{1}}
                 {\e1},
  \judgementrewrite {\propenv{}, {\thenprop {\prop{1}}}}
                 {\ep2} {\t{}} {\filterset {\thenprop {\prop{2}}} {\elseprop {\prop{2}}}}
                 {\object{}}
                 {\e2},
  \judgementrewrite {\propenv{}, {\elseprop {\prop{1}}}}
                 {\ep3} {\t{}} {\filterset {\thenprop {\prop{3}}} {\elseprop {\prop{3}}}}
                 {\object{}}
                 {\e3},
  \e{} = {\ifexp {\e1} {\e2} {\e3}},
  \inpropenv{\orprop {\thenprop {\prop{2}}} {\thenprop {\prop{3}}}}{\thenprop{\prop{}}},
  \inpropenv{\orprop {\elseprop {\prop{2}}} {\elseprop {\prop{3}}}}{\elseprop{\prop{}}}

              For part 1, either \object{} = \emptyobject{}, or \e{} evaluates to the
              result of \e{2}.

              To prove part 2, we consider two cases:
              \begin{itemize}
                \item if \isfalseval{\v{}}
                  then \e{2} evaluates to a false value so {\satisfies{\openv{}}{\elseprop {\prop{2}}}}, and thus
                  {\satisfies{\openv{}}{\orprop {\elseprop {\prop{2}}} {\elseprop {\prop{3}}}}} by M-Or, 
                \item otherwise
                  \istrueval{\v{}},
                  so \e{2} evaluates to a true value so {\satisfies{\openv{}}{\thenprop {\prop{2}}}}, and thus
                  {\satisfies{\openv{}}{\orprop {\thenprop {\prop{2}}} {\thenprop {\prop{3}}}}} by M-Or.
              \end{itemize}

              Part 3 is trivial as
              \opsem {\openv{}} {\e{2}} {\v{}}
              and {\judgementtwo{}{\v{}}{\t{}}} by the induction hypothesis on {\e{2}}.

      \end{subcase}
  \end{itemize}
\end{case}

\begin{case}[BE-If]
  \ 

  Reduces to an error.
\end{case}

\begin{case}[BE-IfFalse]
  \ 

  Reduces to an error.
\end{case}

\begin{case}[BE-IfTrue]
  \ 

  Reduces to an error.
\end{case}

\begin{case}[B-Let]
  \e{} = {\letexp {\x{}} {\e{1}} {\e{2}}},
        \opsem {\openv{}} {\e{1}} {\v{1}},
        \opsem {\extendopenv{\openv{}}{\x{}}{\v{1}}} {\e{2}} {\v{}}

  \begin{itemize}
    \item[]
      \begin{subcase}[T-Let]
  \ep{} = {\letexp {\x{}} {\ep{1}} {\ep{2}}},
  \judgementrewrite {\propenv{}} {\ep{1}} {\s{}} {\filterset {\thenprop {\prop{1}}} {\elseprop {\prop{1}}}}
             {\object{1}}
             {\ep{1}},
             \propp{} = {\impprop {\notprop {\falsy{}} {\x{}}} {\thenprop {\prop{1}}}},
             \proppp{} = {\impprop {\isprop {\falsy{}} {\x{}}} {\elseprop {\prop{1}}}},
  \judgementrewrite
       {\propenv{}, {\isprop {\s{}} {\x{}}},
         {\propp{}},
         {\proppp{}}}
             {\ep{2}} {\t{}} {\filterset {\thenprop {\prop{}}} {\elseprop {\prop{}}}}
             {\object{}} 
             {\e{2}}

        For all the following cases (with a reminder that \x{} is fresh)
        we apply the induction hypothesis on \e{2}. We justify this by noting
        that occurrences of \x{} inside \e{2} have the same type as \e{1} and 
        simulate the propositions of \e{1}
        because 
        \begin{itemize}
          \item
  \judgementrewrite
       {\propenv{}, {\isprop {\s{}} {\x{}}},
         {\propp{}},
         {\proppp{}}}
             {\ep{2}} {\t{}} {\filterset {\thenprop {\prop{}}} {\elseprop {\prop{}}}}
             {\object{}} 
             {\e{2}},
           \item
        \satisfies{\extendopenv{\openv{}}{\x{}}{\v{1}}}{\propenv{}, {\isprop {\s{}} {\x{}}}, \propp{}, \proppp{}},
           \item
        {\isconsistent{\extendopenv{\openv{}}{\x{}}{\v{1}}}},
        and
           \item
        \opsem {\extendopenv{\openv{}}{\x{}}{\v{1}}} {\e{2}} {\v{}}.
    \end{itemize}

        We prove parts 1, 2 and 3 by directly using the induction hypothesis on \e{2}.
      \end{subcase}
  \end{itemize}
\end{case}

\begin{case}[BE-Let]
  \ 
  
  Reduces to an error.
\end{case}

\begin{case}[B-Abs] 
        \v{} = {\closure {\openv{}} {\abs {\x{}} {\s{}} {\e{1}}}}

  \begin{itemize}
    \item[]
      \begin{subcase}[T-Clos]
  \ep{} = {\closure {\openv{}} {\abs {\x{}} {\s{}} {\e{1}}}},
  $\exists {\propenvp{}}. \satisfies{\openv{}}{\propenvp{}}$
  \ \text{and}\ 
\judgementrewrite {\propenvp{}} {\abs {\x{}} {\s{}} {\e{1}}} {\t{}}
                 {\filterset {\thenprop {\prop{f}}}
                             {\elseprop {\prop{f}}}}
                 {\object{f}}
                 {\abs {\x{}} {\s{}} {\e{1}}},
  \e{} = {\closure {\openv{}} {\abs {\x{}} {\s{}} {\e{1}}}},
                 {\thenprop{\prop{}}} = \topprop{},
                 {\elseprop{\prop{}}} = \botprop{},
                 {\object{}} = \emptyobject{}

        We assume some \propenvp{}, such that
        \begin{itemize}
          \item \satisfies{\openv{}}{\propenvp{}}
          \item \judgement {\propenvp{}} {\abs {\x{}} {\s{}} {\e{1}}} {\t{}}
                           {\filterset {\thenprop {\prop{}}}
                                       {\elseprop {\prop{}}}}
                           {\object{}}.
       \end{itemize}
       Note the last rule in the derivation of
          \judgement {\propenvp{}} {\abs {\x{}} {\s{}} {\e{1}}} {\t{}}
                           {\filterset {\thenprop {\prop{}}}
                                       {\elseprop {\prop{}}}}
                           {\object{}}
                           must be T-Abs, so 
                           {\thenprop {\prop{}}} = {\topprop{}},
                           {\elseprop {\prop{}}} = {\botprop{}}
                           and {\object{}} = {\emptyobject{}}.
         Thus parts 1 and 2 hold for the same reasons as T-True.
         Part 3 holds as \v{} has the same type as {\abs {\x{}} {\s{}} {\e{1}}}
         under \propenvp{}.

      \end{subcase} 
  \end{itemize}
\end{case}

\begin{case}[B-Abs]
        \v{} = ${\closure {\openv{}} {\abs {\x{}} {\s{}} {\e{1}}}}$,
          { \opsem {\openv{}}
                   {\abs {\x{}} {\t{}} {\e{1}}}
                   {\closure {\openv{}} {\abs {\x{}} {\s{}} {\e{1}}}}}

  \begin{itemize}
    \item[]
      \begin{subcase}[T-Abs]
  \ep{} = {\abs {\x{}} {\s{}} {\ep{1}}},
{ \judgementrewrite {\propenv{}, {\isprop {\s{}} {\x{}}}}
            {\ep{1}} {\t{}}
             {\filterset {\thenprop {\prop{1}}}
                         {\elseprop {\prop{1}}}}
             {\object{1}}
             {\e{1}}},
           \issubtypein{}
           {\ArrowOne {\x{}} {\s{}}
                      {\t{1}}
                      {\filterset {\thenprop {\prop{1}}}
                                  {\elseprop {\prop{1}}}}
                      {\object{1}}}
          {\t{}},
          \inpropenv{\topprop{}}{\thenprop{\prop{}}},
          \inpropenv{\botprop{}}{\elseprop{\prop{}}},
          {\object{}} = {\emptyobject{}}

        Parts 1 and 2 hold for the same reasons as T-True.
        Part 3 holds directly via T-Clos, since \v{} must be a closure.
      \end{subcase}
  \end{itemize}
\end{case}

\begin{case}[BE-Error]
        \opsem {\openv{}} {\e{}} {\errorval{\v{1}}}

  \begin{itemize}
    \item[]
      \begin{subcase}[T-Error] 
  \ep{} = \errorval{\v{1}},
  \e{} = \errorval{\v{1}},
  \t{} = \Bot,
  \thenprop{\prop{}} = \botprop{}, \elseprop{\prop{}} = \botprop{}, \object{} = \emptyobject{}

        Trivially reduces to an error.
      \end{subcase}
  \end{itemize}
\end{case}

\end{proof}

\end{lemma}

{\wrongtheorem{appendix}}

{\soundnesstheorem{appendix}}

\begin{figure*}
$$
\begin{altgrammar}
  \expd{}, \e{} &::=& \x{}
                      \alt \v{} 
                      \alt {\comb {\e{}} {\e{}}} 
                      \alt {\abs {\x{}} {\t{}} {\e{}}}
                      \alt {\ifexp {\e{}} {\e{}} {\e{}}}
                      \alt {\doexp {\e{}} {\e{}}}
                      \alt {\letexp {\x{}} {\e{}} {\e{}}}
                      \alt {\wrongorerror{}}
                      \alt {\ReflectiveExp{}}
                      \alt {\NonReflectiveExp{}}
                      \alt {\MultimethodExp{}}
                      \alt {\HashMapExp{}}
                &\mbox{Expressions} \\
  \v{} &::=&          \singletonmeta{}
                      \alt \classvaluemeta{}
                      \alt {\emptymap{}}
                      \alt {\const{}}
                      \alt {\num{}}
                      \alt {\str{}}
                      \alt \mapval{}
                      \alt {\closure {\openv{}} {\abs {\x{}} {\t{}} {\e{}}}}
                      \alt {\multi {\v{}} {\disptable{}}}
                &\mbox{Values} \\
  \mapval{} &::=&  {\curlymapvaloverright{\v{}}{\v{}}}
                &\mbox{Map Values} \\
                \constantssyntax{}\\
  \HashMapExp{}                &::=& \hmapexpressionsyntax{}
                &\mbox{Hash Maps} \\
  \NonReflectiveExp{}     &::=& \nonreflectiveexpsyntax{}
                &\mbox{Non-Reflective Java Interop} \\
  \ReflectiveExp{}     &::=& \reflectiveexpsyntax{}
                &\mbox{Reflective Java Interop} \\
  \MultimethodExp{}     &::=& \multimethodexpsyntax{}
                &\mbox{Immutable First-Class Multimethods}
                      \\\\
  \s{}, \t{}    &::=& \Top 
                      \alt \class{}
                      \alt {\Value \singletonmeta{}} 
                      \alt {\Unionsplice {\overrightarrow{\t{}}}}
                      \alt
                      {\ArrowOne {\x{}} {\t{}}
                                   {\t{}}
                                   {\filterset {\prop{}} {\prop{}}}
                                   {\object{}}}
                      \alt {\HMapgeneric {\mandatory{}} {\absent{}}}
                      \alt {\MultiFntype{\t{}}{\t{}}}
                      
                &\mbox{Types} \\
                \auxhmapsyntax{}\\
  \singletonallsyntax{}
                \\ \\
                \openvsyntax{}\\\\


  \occurrencetypingsyntax{}
  \propenvsyntax{}
  \\\\

 \disptablesyntax{} \\
\classtableallsyntax{} \\
               \classliteralallsyntax{}\\

               \classvaluesyntaxentry{}\\
                      \\
  \wrongorerror{} &::=& \wrong{} \alt \errorvalv{}
                &\mbox{Wrong or error}
                      \\
  \definedreduction{} &::=& \v{} \alt \wrongorerror{}
                 &\mbox{Defined reductions}
                 \\
  \polaritymeta{} &::=& \pluspolarityliteral \alt \minuspolarityliteral
                 &\mbox{Substitution Polarity}
\end{altgrammar}
$$
\caption{Syntax of Terms, Types, Propositions, and Objects}
\end{figure*}

\begin{figure*}
$$
\begin{array}{lllr}
  \Nil &\equiv& {\ValueNil}\\
  \True &\equiv& {\ValueTrue}\\
  \False &\equiv& {\ValueFalse}\\
\end{array}
$$
\caption{Type abbreviations}
\end{figure*}

\begin{figure*}
$$
\begin{array}{lllr}
  \judgementtwo{\propenv{}}{\e{}}{\t{}} &\equiv& 
  \judgement{\propenv{}}{\e{}}{\t{}}{\filterset{\thenprop{\prop{}}}{\elseprop{\prop{}}}}{\object{}}
  & \text{for some}\ {\thenprop{\prop{}}}, {\elseprop{\prop{}}} \text{and}\ {\object{}}

  \\
  {\replacefor{\t{}}{\object{}}{\x{}}} &\equiv& {\pluspolarity{\replacefor{\t{}}{\object{}}{\x{}}}}
  \\
  {\replacefor{\prop{}}{\object{}}{\x{}}} &\equiv&  {\pluspolarity{\replacefor{\prop{}}{\object{}}{\x{}}}}
  \\
  {\replacefor{\filterset{\prop{}}{\prop{}}}{\object{}}{\x{}}} &\equiv&  {\pluspolarity{\replacefor{\filterset{\prop{}}{\prop{}}}{\object{}}{\x{}}}}
  \\
  {\replacefor{\object{}}{\object{}}{\x{}}} &\equiv& {\pluspolarity{\replacefor{\object{}}{\object{}}{\x{}}}}

\end{array}
$$
\caption{Judgment abbreviations}
\end{figure*}

\begin{figure*}
\begin{mathpar}

  {\TLocal}

{\TConst}

{\TTrue}

{\TFalse}

{\TNil}

{\TNum}

{\TDo}

{\TIf}

{\TLet}
                 
{\TApp}

{\TAbs}

\infer [T-Clos]
{ \exists {\propenv{}}. \satisfies{\openv{}}{\propenv{}}
  \ \text{and}\ 
\judgementrewrite {\propenv{}} {\abs {\x{}} {\t{}} {\e{}}} {\s{}}
                 {\filterset {\thenprop {\prop{}}}
                             {\elseprop {\prop{}}}}
                 {\object{}}
                 {\abs {\x{}} {\t{}} {\ep{}}}
              }
{ \judgementrewrite {}
            {\closure {\openv{}} {\abs {\x{}} {\t{}} {\e{}}}} 
                      {\s{}}
             {\filterset {\thenprop {\prop{}}}
                         {\elseprop {\prop{}}}}
             {\object{}}
            {\closure {\openv{}} {\abs {\x{}} {\t{}} {\ep{}}}}
          }

           {\TError}

         {\TSubsume}
\end{mathpar}
\caption{Standard Typing Rules}
\end{figure*}

\begin{figure*}
\begin{mathpar}

{\TNew}

{\TNewStatic}

{\TField}

{\TFieldStatic}

{\TMethod}

{\TMethodStatic}

{\TClass}

{\TInstance}
\end{mathpar}
\caption{Java Interop Typing Rules}
\end{figure*}

\begin{figure*}
\begin{mathpar}

  \TDefMulti{}

  \TDefMethod{}

\TIsA{}

\infer [T-Multi]
{ \judgementtworewrite {} {\v{}} {\t{}} {\vp{}}
  \\
  \overrightarrow{\judgementtworewrite{}{\v{k}}{\Top}{\vp{k}}}
  \\
  \overrightarrow{\judgementtworewrite{}{\v{v}}{\s{}}{\vp{v}}}
}
{ \judgementrewrite {}
  {\multi {\v{}} {\curlymapvaloverright{\v{k}}{\v{v}}}}
                      {\MultiFntype {\s{}} {\t{}}}
             {\filterset {\topprop{}} {\botprop{}}}
           {\emptyobject{}}
  {\multi {\vp{}} {\curlymapvaloverright{\vp{k}}{\vp{v}}}}
}

\end{mathpar}
\caption{Multimethod Typing Rules}
\end{figure*}

\begin{figure*}
\begin{mathpar}

%

\infer [T-HMap]
{ \overrightarrow{\judgementtworewrite {} {\v{k}}{\Value \k{}}{\vp{k}}}\\
  \overrightarrow{\judgementtworewrite {} {\v{v}}{\t{v}}{\vp{v}}}\\
  \mandatory{} = \mandatorysetoverright{\k{}}{\t{v}}
}
{ \judgementrewrite {}
             {\curlymapvaloverright{\v{k}}{\v{v}}}
                       {\HMapc {\mandatory{}}}
             {\filterset {\topprop{}} {\botprop{}}}
             {\emptyobject{}}
             {\curlymapvaloverright{\vp{k}}{\vp{v}}}
           }

    {\TKw}

    {\TGetHMap}

    {\TGetAbsent}

    {\TGetHMapPartialDefault}

    {\TAssoc}

\end{mathpar}
\caption{Map Typing Rules}
\end{figure*}

\begin{figure*}
\begin{mathpar}
\objectsub{}

\standardsubtyping{}
\SPMultiFn{}
\Multisubtyping{}

\HMapsubtyping{}
\end{mathpar}
\caption{Subtyping rules}
\end{figure*}



{\convertjavatypefigure{figure*}{}}

\begin{figure}
\begin{mathpar}
\constanttypefigure{}
\end{mathpar}
\caption{Constant Typing}
\end{figure}

\constantsemfigure{appendix}

\begin{figure*}
\isapropsfigure{}

\isaopsemfigure{}
\caption{Definition of isa?}
\end{figure*}


\begin{figure*}
  \getmethodfigure{}
\caption{Definition of get-method}
\end{figure*}

\clearpage

\begin{figure*}
\begin{mathpar}

\BLocal{}

\BDo{}

\BLet{}

\BVal{}

\BIfTrue{}

\BIfFalse{}

\BAbs{}

\BBetaClosure{}

\BDelta{}

\BBetaMulti{}

\BField{}

\BMethod{}

\BNew{}

       \BDefMulti{}

       \BDefMethod{}

       \BIsA{}

       {\BAssoc}

       {\BGet}

       {\BGetMissing}
\end{mathpar}
\caption{Operational Semantics}
\label{appendix:figure:opsem}
\end{figure*}

\begin{figure*}
\begin{mathpar}

\infer [BS-MethodRefl]
{}
{\opsem {\openv{}} {\methodexp {mth} {\e{}} {\overrightarrow{\e{}}}}
        {\wrong{}}}

\infer [BS-FieldRefl]
{}
{\opsem {\openv{}} {\fieldexp {\fld{}} {\e{}}}
        {\wrong{}}}

\infer [BS-NewRefl]
{}
{\opsem {\openv{}} {\fieldexp {\fld{}} {\e{}}}
        {\wrong{}}}

\infer [BS-Beta]
{ \opsem {\openv{}}
         {\e{f}}
         {\v{}}
         \\\\
  {\v{}} \not= {\const{}}
  \\
  {\v{}} \not= {\multi {\v{d}} {\disptable{}}}
  \\\\
  {\v{}} \not= {\closure {\openv{c}} {\abs {\x{}} {\t{}} {\e{b}}}}
       }
{ \opsem {\openv{}}
         {\appexp {\e{f}} {\e{a}}}
         {\wrong{}}
       }

\infer [BS-BetaMulti]
{ \opsem {\openv{}}
         {\e{f}}
         {\multi {\v{}} {\disptable{}}}
         \\\\
  {\v{}} \not= {\const{}}
  \\
  {\v{}} \not= {\multi {\v{d}} {\disptable{}}}
  \\\\
  {\v{}} \not= {\closure {\openv{c}} {\abs {\x{}} {\t{}} {\e{b}}}}
       }
{ \opsem {\openv{}}
         {\appexp {\e{f}} {\e{a}}}
         {\wrong{}}
       }

\infer [BS-FieldTarget]
{ \opsem {\openv{}}
         {\e{}} 
       {\v{1}}
         \\\\
         {\v{}} \not= {\classvalue{\classhint{1}} {\overrightarrow {\classfieldpair{\fld{i}} {\v{i}}}}}
       }
{ \opsem {\openv{}}
         {\fieldstaticexp {\classhint{1}} {\classhint{2}} {\fld{}} {\e{}}}
         {\wrong{}}
   }

\infer [BS-FieldMissing]
{ \opsem {\openv{}}
         {\e{}} 
       {\classvalue{\classhint{1}} {\overrightarrow {\classfieldpair{\fld{i}} {\v{i}}}}}
       \\
       \fld{} \not\in \{\overrightarrow{\fld{i}}\}
       }
{ \opsem {\openv{}}
         {\fieldstaticexp {\classhint{1}} {\classhint{2}} {\fld{}} {\e{}}}
         {\wrong{}}
   }

\infer [BS-MethodTarget]
{ \opsem {\openv{}}
         {\e{m}}
         {\v{}}
  \\
         {\v{}} \not= {\classvalue{\classhint{1}} {\overrightarrow {\classfieldpair{\fld{i}} {\v{i}}}}}
}
{\opsem {\openv{}}
        {\methodstaticexp {\classhint{1}} {\overrightarrow{\classhint{a}}} {\classhint{2}} {mth} {\e{m}} {\overrightarrow{\e{a}}}}
        {\wrong{}}
      }

\infer [BS-MethodArity]
{ i \not= a
}
{\opsem {\openv{}}
        {\methodstaticexp {\classhint{1}} {\overrightarrow{\classhint{i}}} {\classhint{2}} {mth} {\e{m}} {\overrightarrow{\e{a}}}}
        {\wrong{}}
      }

\infer [BS-MethodArg]
{ \opsem {\openv{}}
         {\e{m}}
         {\v{m}}
  \\
  \overrightarrow{
  \opsem {\openv{}}
         {\e{a}}
         {\v{a}}
       }
       \\\\
  \exists a.\ 
    \v{a} \not=\ {\classvalue{\classhint{a}} {\overrightarrow {\classfieldpair{\fld{i}} {\v{i}}}}}\ or\ \v{a} \not= \nil{}
}
{\opsem {\openv{}}
        {\methodstaticexp {\classhint{1}} {\overrightarrow{\classhint{a}}} {\classhint{2}} {mth} {\e{m}} {\overrightarrow{\e{a}}}}
        {\wrong{}}
      }

\infer [BS-NewArg]
{ \overrightarrow{
  \opsem {\openv{}}
         {\e{i}}
         {\v{i}}
     }
       \\\\
  \exists i.\ 
    \v{i} \not=\ {\classvalue{\classhint{i}} {\overrightarrow {\classfieldpair{\fld{i}} {\v{i}}}}}\ or\ \v{i} \not= \nil{}
}
{\opsem {\openv{}}
        {\newstaticexp {\overrightarrow{\classhint{i}}} {\classhint{1}} 
                       {\class{}} {\overrightarrow{\e{i}}}}
        {\wrong{}}
      }

\infer [BS-NewArity]
{ i \not= a
}
{\opsem {\openv{}}
        {\newstaticexp {\overrightarrow{\classhint{i}}} {\classhint{1}} 
                       {\class{}} {\overrightarrow{\e{a}}}}
        {\wrong{}}
      }

\infer [BS-AssocMap]
{\opsem {\openv{}}
        {\e{m}} {\v{}}
        \\
        \v{} \not= {\curlymap{\overrightarrow{({\v{a}}\ {\v{b}})}}}
}
{
 \opsem {\openv{}}
        {\assocexp {\e{m}} {\e{k}} {\e{v}}} 
        {\wrong{}}
                }

\infer [BS-AssocKey]
{\opsem {\openv{}}
        {\e{m}} {\curlymap{\overrightarrow{({\v{a}}\ {\v{b}})}}}
        \\
 \opsem {\openv{}} {\e{k}} {\v{k}}
 \\\\
 {\v{k}} \not= \k{}
}
{
 \opsem {\openv{}}
        {\assocexp {\e{m}} {\e{k}} {\e{v}}} 
        {\wrong{}}
                }

\infer [BS-GetMap]
{ \opsem {\openv{}}
         {\e{m}} {\v{}}
        \\
        \v{} \not= {\curlymap{\overrightarrow{({\v{a}}\ {\v{b}})}}}
}
{\opsem {\openv{}}
        {\getexp {\e{m}} {\e{k}}}
        {\wrong{}}
}

\infer [BS-GetKey]
{ \opsem {\openv{}}
         {\e{m}} {\v{}}
        \\
 \opsem {\openv{}}
        {\e{k}} {\v{k}}
        \\\\
      \v{} \not= {\k{}}
}
{\opsem {\openv{}}
        {\getexp {\e{m}} {\e{k}}}
        {\wrong{}}
}

\infer [BS-Local]
{ \notinopenv {\openv{}} {\x{}}}
{ \opsem {\openv{}} {\x{}} {\wrong{}} }

\infer [BS-DefMethod]
{ \opsem {\openv{}}
         {\e{m}}
         {\v{m}}
         \\
         \v{m} \not= {\multi {\v{d}} {\disptable{}}}
}
{\opsem {\openv{}}
        {\extendmultiexp {\e{m}} {\e{v}} {\e{f}}}
        {\wrong{}}
      }

\end{mathpar}
\caption{Stuck programs}
\end{figure*}

\begin{figure*}
\begin{mathpar}
\infer [BE-ErrorWrong]
{}
{ \opsem {\openv{}} 
         {\wrongorerror{}}
         {\wrongorerror{}}}

\infer [BE-Let]
{ \opsem {\openv{}} {\e{a}} {\wrongorerror{}}
 }
{ \opsem {\openv{}} 
         {\letexp {\x{}} {\e{a}} {\e{}}}
       {\wrongorerror{}}}

\infer [BE-Do1]
{ \opsem {\openv{}} {\e{1}} {\wrongorerror{}} }
{ \opsem {\openv{}} {\doexp{\e{1}}{\e{}}} {\wrongorerror{}}}

\infer [BE-Do2]
{ \opsem {\openv{}} {\e{1}} {\v{1}} 
  \\\\
  \opsem {\openv{}} {\e{}}  {\wrongorerror{}}
}
{ \opsem {\openv{}} {\doexp{\e{1}}{\e{}}} {\wrongorerror{}} }

\infer [BE-If]
{  \opsem {\openv{}} {\e{1}} {\wrongorerror{}}
}
{ \opsem {\openv{}}
         {\ifexp {\e1} {\e2} {\e3}}
         {\wrongorerror{}}
       }

\infer [BE-IfTrue]
{ \opsem {\openv{}} {\e{1}} {\v{1}}
  \\\\
  {\v{1}} \not= {\false{}}
  \\
  {\v{1}} \not= {\nil{}}
  \\\\
  \opsem {\openv{}} {\e{2}} {\wrongorerror{}}
}
{ \opsem {\openv{}}
         {\ifexp {\e1} {\e2} {\e3}}
         {\wrongorerror{}}
       }

\infer [BE-IfFalse]
{  \opsem {\openv{}} {\e{1}} {\false{}}
  \ \ \text{or}\ \ 
  \opsem {\openv{}} {\e{1}} {\nil{}}
  \\\\
  \opsem {\openv{}} {\e{3}} {\wrongorerror{}}
}
{ \opsem {\openv{}}
         {\ifexp {\e1} {\e2} {\e3}}
         {\wrongorerror{}}
       }

\infer [BE-Beta1]
{ \opsem {\openv{}}
         {\e{f}}
         {\wrongorerror{}}
       }
{ \opsem {\openv{}}
         {\appexp {\e{f}} {\e{a}}}
         {\wrongorerror{}}
       }

\infer [BE-Beta2]
{ \opsem {\openv{}}
         {\e{f}}
         {\v{f}}
         \\\\
  \opsem {\openv{}}
         {\e{a}}
         {\wrongorerror{}}
       }
{ \opsem {\openv{}}
         {\appexp {\e{f}} {\e{a}}}
         {\wrongorerror{}}
       }

\infer [BE-BetaClosure]
{ \opsem {\openv{}}
         {\e{f}}
         {\closure {\openv{c}} {\abs {\x{}} {\t{}} {\e{b}}}}
         \\\\
  \opsem {\openv{}}
         {\e{a}}
         {\v{a}}
         \\\\
  \opsem {\extendopenv {\openv{c}} {\x{}} {\v{a}}}
         {\e{b}}
         {\wrongorerror{}}
       }
{ \opsem {\openv{}}
         {\appexp {\e{f}} {\e{a}}}
         {\wrongorerror{}}
       }

\infer [BE-BetaMulti1]
{ \opsem {\openv{}}
         {\e{f}}
         {\multi {\v{d}} {m}}
         \\\\
  \opsem {\openv{}}
         {\e{a}}
         {\v{a}}
         \\\\
  \opsem {\openv{}}
         {\appexp {\v{d}} {\v{a}}}
         {\wrongorerror{}}
       }
{ \opsem {\openv{}}
         {\appexp {\e{f}} {\e{a}}}
         {\wrongorerror{}}
       }

\infer [BE-BetaMulti2]
{ \opsem {\openv{}}
         {\e{f}}
         {\multi {\v{d}} {m}}
         \\\\
  \opsem {\openv{}}
         {\e{a}}
         {\v{a}}
         \\\\
  \opsem {\openv{}}
         {\appexp {\v{d}} {\v{a}}}
         {\v{e}}
         \\\\
  \getmethoderr {\disptable{}}
             {\v{e}}
             {\errorvalv{}}
       }
{ \opsem {\openv{}}
         {\appexp {\e{f}} {\e{a}}}
         {\errorvalv{}}
       }

\infer [BE-Delta]
{ \opsem {\openv{}} {\e{}} {\const{}}
  \\\\
  \opsem {\openv{}} {\ep{}} {\v{}}
  \\\\
  \constantopsem{\const{}}{\v{}} = \wrongorerror{}
}
{ \opsem {\openv{}}
         {\appexp {\e{}} {\ep{}}}
         {\wrongorerror{}}
       }

\infer [BE-Field]
{ \opsem {\openv{}}
         {\e{}} 
         {\wrongorerror{}}
       }
{ \opsem {\openv{}}
         {\fieldstaticexp {\classhint{1}} {\classhint{2}} {\fld{}} {\e{}}}
         {\wrongorerror{}}
   }

\infer [BE-Method1]
{ \opsem {\openv{}}
         {\e{m}}
         {\wrongorerror{}}
}
{\opsem {\openv{}}
        {\methodstaticexp {\classhint{1}} {\overrightarrow{\classhint{a}}} {\classhint{2}} {mth} {\e{m}} {\overrightarrow{\e{}}}}
        {\wrongorerror{}}
      }

\infer [BE-Method2]
{ \opsem {\openv{}}
         {\e{m}}
         {\v{m}}
  \\\\
  \overrightarrow{
  \opsem {\openv{}}
         {\e{n-1}}
         {\v{n-1}}
       }
         \\\\
  \opsem {\openv{}}
         {\e{n}}
         {\wrongorerror{}}
}
{\opsem {\openv{}}
        {\methodstaticexp {\classhint{1}} {\overrightarrow{\classhint{a}}} {\classhint{2}} {mth} {\e{m}} {\overrightarrow{\e{}}}}
        {\wrongorerror{}}
      }

\infer [BE-Method3]
{ \opsem {\openv{}}
         {\e{m}}
         {\v{m}}
  \\
  \overrightarrow{
  \opsem {\openv{}}
         {\e{a}}
         {\v{a}}
       }
  \\\\
  \invokejavamethod {\classhint{1}} {\v{m}} {mth}
                    {\overrightarrow{\classhint{a}}} {\overrightarrow{\v{a}}}
                    {\classhint{2}}
                    {\errorvalv{}}
}
{\opsem {\openv{}}
        {\methodstaticexp {\classhint{1}} {\overrightarrow{\classhint{a}}} {\classhint{2}} {mth} {\e{m}} {\overrightarrow{\e{a}}}}
        {\errorvalv{}}
      }

\infer [BE-New1]
{ \overrightarrow{
  \opsem {\openv{}}
         {\e{n-1}}
         {\v{n-1}}
       }
       \\\\
  \opsem {\openv{}}
         {\e{n}}
         {\wrongorerror{}}
       }
{ \opsem {\openv{}}
         {\newstaticexp {\overrightarrow{\classhint{i}}} {\classhint{1}} 
                        {\class{}} {\overrightarrow{\e{}}}}
         {\wrongorerror{}}
       }

\infer [BE-New2]
{ 
  \overrightarrow{
  \opsem {\openv{}}
         {\e{i}}
         {\v{i}}
       }
         \\\\
         \newjava {\classhint{1}}
                  {\overrightarrow{\classhint{i}}}
                  {\overrightarrow{\v{i}}}
                  {\errorvalv{}}
       }
{ \opsem {\openv{}}
         {\newstaticexp {\overrightarrow{\classhint{i}}} {\classhint{1}} 
                        {\class{}} {\overrightarrow{\e{i}}}}
         {\errorvalv{}}}

\infer [BE-DefMulti]
{ \opsem {\openv{}} {\e{d}} {\wrongorerror{}}
}
{\opsem {\openv{}}
        {\createmultiexp {\t{}}
                         {\e{d}}}
        {\wrongorerror{}}
}

\infer [BE-DefMethod1]
{ \opsem {\openv{}}
         {\e{m}}
         {\wrongorerror{}}
}
{\opsem {\openv{}}
        {\extendmultiexp {\e{m}} {\e{v}} {\e{f}}}
        {\wrongorerror{}}
      }

\infer [BE-DefMethod2]
{ \opsem {\openv{}}
         {\e{m}}
         {\multi {\v{d}} {\disptable{}}}
         \\\\
  \opsem {\openv{}}
         {\e{v}}
         {\wrongorerror{}}
}
{\opsem {\openv{}}
        {\extendmultiexp {\e{m}} {\e{v}} {\e{f}}}
        {\wrongorerror{}}
      }

\infer [BE-DefMethod3]
{ \opsem {\openv{}}
         {\e{m}}
         {\multi {\v{d}} {\disptable{}}}
         \\\\
  \opsem {\openv{}}
         {\e{v}}
         {\v{v}}
         \\\\
  \opsem {\openv{}}
         {\e{f}}
         {\wrongorerror{}}
}
{\opsem {\openv{}}
        {\extendmultiexp {\e{m}} {\e{v}} {\e{f}}}
         {\wrongorerror{}}
      }

\infer [BE-IsA1]
{ \opsem {\openv{}} {\e{1}} {\wrongorerror{}}
}
{\opsem {\openv{}} {\isaapp {\e{1}} {\e{2}}} {\wrongorerror{}}}

\infer [BE-IsA2]
{ \opsem {\openv{}} {\e{1}} {\v{1}}
  \\\\
  \opsem {\openv{}} {\e{2}} {\wrongorerror{}}
}
{\opsem {\openv{}} {\isaapp {\e{1}} {\e{2}}} {\wrongorerror{}}}

\infer [BE-Assoc1]
{\opsem {\openv{}}
        {\e{m}}{\wrongorerror{}} 
}
{
 \opsem {\openv{}}
        {\assocexp {\e{m}} {\e{k}} {\e{v}}} 
        {\wrongorerror{}}
                }

\infer [BE-Assoc2]
{\opsem {\openv{}}
        {\e{m}} {\curlymap{\overrightarrow{({\v{a}}\ {\v{b}})}}}
        \\
 \opsem {\openv{}}
        {\e{k}}{\wrongorerror{}}
}
{
 \opsem {\openv{}}
        {\assocexp {\e{m}} {\e{k}} {\e{v}}} 
        {\wrongorerror{}}
                }

\infer [BE-Assoc3]
{\opsem {\openv{}}
        {\e{m}} {\curlymap{\overrightarrow{({\v{a}}\ {\v{b}})}}}
        \\
 \opsem {\openv{}}
        {\e{k}} {\v{k}}
        \\
 \opsem {\openv{}}
        {\e{v}} {\wrongorerror{}}
}
{
 \opsem {\openv{}}
        {\assocexp {\e{m}} {\e{k}} {\e{v}}} 
        {\wrongorerror{}}
                }

\infer [BE-Get1]
{\opsem {\openv{}}
        {\e{m}} {\wrongorerror{}}
}
{
 \opsem {\openv{}}
        {\getexp {\e{m}} {\e{k}}}
        {\wrongorerror{}}
}

\infer [BE-Get2]
{\opsem {\openv{}}
        {\e{m}} {\curlymap{\overrightarrow{({\v{a}}\ {\v{b}})}}}
        \\
 \opsem {\openv{}}
        {\e{k}} {\wrongorerror{}}
}
{
 \opsem {\openv{}}
        {\getexp {\e{m}} {\e{k}}}
        {\wrongorerror{}}
}
\end{mathpar}
\caption{Error and stuck propagation}
\label{appendix:figure:errorstuck}
\end{figure*}

\begin{figure*}
\begin{mathpar}

\begin{array}{lllll}
  \inopenvalign{\openv{}}{\x{}}{\v{} & {\roundpair{\x{}}{\v{}}} \in \openv{}}\\
  \inopenvalign{\openv{}}{\path {\keype{k}} {\object{}}}{\getexp {{\openv{}}(\object{})}{\k{}}}\\
  \inopenvalign{\openv{}}{\path {\classpe{}} {\object{}}}{\appexp {\classconst{}} {{\openv{}}(\object{})}}

\end{array}

\end{mathpar}
\caption{Path translation}
\end{figure*}

\begin{figure*}
\begin{mathpar}

\begin{array}{lllll}
\updatefigure{}
\\\\
\restrictremovefigure{}
\end{array}

\end{mathpar}
\caption{Type Update}
\label{appendix:updaterestrictremove}
\end{figure*}

\begin{figure*}
\begin{mathpar}
\infer [M-Or]
{ \satisfies{\openv{}}{\prop{1}}\ \text{or}\  \satisfies{\openv{}}{\prop{2}}}
{ \satisfies{\openv{}}{\orprop{\prop{1}}{\prop{2}}}
                   }

\infer [M-Imp]
{ \satisfies{\openv{}}{\prop{1}}\ \text{implies}\ \satisfies{\openv{}}{\prop{2}}}
{ \satisfies{\openv{}}{\impprop{\prop{1}}{\prop{2}}}
                   }

\infer [M-And]
{ \satisfies{\openv{}}{\prop{1}}
\\ \satisfies{\openv{}}{\prop{2}}}
{ \satisfies{\openv{}}{\andprop{\prop{1}}{\prop{2}}}
                   }

\infer [M-Top]
{}
{ \satisfies{\openv{}}{\topprop{}}
                   }

                   \\

\infer [M-Type]
{ \judgement {} {\openv{}({\path{\pathelem{}}{\x{}}})} {\t{}}{\filterset{\thenprop{\prop{}}}{\elseprop{\prop{}}}}{\object{}}}
{ \satisfies{\openv{}}{\isprop{\t{}}{\path{\pathelem{}}{\x{}}}}
                   }

\infer [M-NotType]
{ \judgement {} {\openv{}({\path{\pathelem{}}{\x{}}})} {\s{}}{\filterset{\thenprop{\prop{}}}{\elseprop{\prop{}}}}{\object{}}
\\\\
\text{there is no}\ \v{}\ \text{such that}\ \judgement{}{\v{}}{\t{}}{\filterset{\thenprop{\prop{1}}}{\elseprop{\prop{1}}}}{\object{1}}
\ \text{and}\ \judgement{}{\v{}}{\s{}}{\filterset{\thenprop{\prop{2}}}{\elseprop{\prop{2}}}}{\object{2}}
}
{ \satisfies{\openv{}}{\notprop{\t{}}{\path{\pathelem{}}{\x{}}}}
                   }
\end{mathpar}
\caption{Satisfaction Relation}
\end{figure*}

\begin{figure*}
\begin{mathpar}
\infer [L-Atom]
{ {\prop{}} \in {\propenv{}}}
{ \inpropenv {\propenv{}} {\prop{}}
}

\infer [L-True]
{}
{ \inpropenv {\propenv{}} {\topprop{}}}

\infer [L-False]
{ \inpropenv {\propenv{}} {\botprop{}}}
{ \inpropenv {\propenv{}} {\prop{}}}

\infer [L-AndI]
{ \inpropenv {\propenv{}} {\prop{1}}
  \\\\
  \inpropenv {\propenv{}} {\prop{2}}}
{ \inpropenv {\propenv{}} {\andprop {\prop{1}}{\prop{2}}}}

\infer [L-AndE]
{ \inpropenv {\propenv{}, {\prop{1}}, {\prop{2}}} {\prop{}} }
{ \inpropenv {\propenv{}, {\andprop {\prop{1}}{\prop{2}}}} {\prop{}}}

\infer [L-ImplI]
{ \inpropenv {\propenv{}, {\prop{1}}} {\prop{2}}}
{ \inpropenv {\propenv{}} {\impprop {\prop{1}}{\prop{2}}}}

\infer [L-ImplE]
{ \inpropenv {\propenv{}} {\prop{1}}
  \\\\
  \inpropenv {\propenv{}} {\impprop {\prop{1}}{\prop{2}}}}
{ \inpropenv {\propenv{}} {\prop{2}}}

\infer [L-OrI]
{ \inpropenv {\propenv{}} {\prop{1}}
  \ \text{or}\ 
  \inpropenv {\propenv{}} {\prop{2}}}
{ \inpropenv {\propenv{}} {\orprop {\prop{1}}{\prop{2}}}}

\infer [L-OrE]
{ \inpropenv {\propenv{}, {\prop{1}}}{\prop{}}
  \\\\
  \inpropenv {\propenv{}, {\prop{2}}}{\prop{}}}
{ \inpropenv {\propenv{}, {\orprop {\prop{1}}{\prop{2}}}}{\prop{}}}

\infer [L-Sub]
{ \inpropenv {\propenv{}} {\isprop {\t{}}{\path {\pathelem{}} {\x{}}}}
  \\
  \issubtypein {} {\t{}}{\s{}}
}
{ \inpropenv {\propenv{}} {\isprop {\s{}}{\path {\pathelem{}} {\x{}}}}}

\infer [L-SubNot]
{ \inpropenv {\propenv{}} {\notprop {\s{}}{\path {\pathelem{}} {\x{}}}}
  \\
  \issubtypein {} {\t{}}{\s{}}}
{ \inpropenv {\propenv{}} {\notprop {\t{}}{\path {\pathelem{}} {\x{}}}}}

\infer [L-Bot]
{ \inpropenv {\propenv{}} {\isprop {\Bot} {\path {\pathelem{}} {\x{}}}}}
{ \inpropenv {\propenv{}} {\prop{}}}

{\LUpdate}

\\

\text{(The metavariable \propisnotmeta{} ranges over \t{} and \nottype{\t{}} (without variables).)}

\end{mathpar}
\caption{Proof System}
\label{appendix:figure:proofsystem}
\end{figure*}

\begin{figure*}
$$
\begin{array}{lclr}

{\withpolarity
  {\replacefor
    {\filterset {\thenprop {\prop{}}}{\elseprop {\prop{}}}}
    {\object{}}
    {\x{}}}
  {\polaritymeta{}}}
  &=&
{\filterset 
  {\withpolarity
    {\replacefor
      {\thenprop {\prop{}}}
      {\object{}}
      {\x{}}}
    {\polaritymeta{}}}
  {\withpolarity
    {\replacefor
      {\elseprop {\prop{}}}
      {\object{}}
      {\x{}}}
    {\polaritymeta{}}}}
\\\\
{\withpolarity
  {\replacefor
    {\isprop {\propisnotmeta{}} {\path {\pathelem{}} {\x{}}}}
    {\path {\pathelemp{}} {\y{}}}
    {\x{}}}
  {\polaritymeta{}}}
&=&
  {\isprop {({\withpolarity
              {\replacefor
               {\propisnotmeta{}}
               {\path {\pathelemp{}} {\y{}}}
               {\x{}}}
              {\polaritymeta{}}})}
           {{\pathelem{}}({\path {\pathelemp{}} {\y{}}})}}
           \\

{\pluspolarity
{\replacefor
  {\isprop {\propisnotmeta{}} {\path {\pathelem{}} {\x{}}}}
  {\emptyobject{}}
  {\x{}}}
}
&=&
{\topprop{}}
\\
{\minuspolarity
{\replacefor
  {\isprop {\propisnotmeta{}} {\path {\pathelem{}} {\x{}}}}
  {\emptyobject{}}
  {\x{}}}
}
&=&
{\botprop{}}

\\
{\withpolarity
{\replacefor
  {\isprop {\propisnotmeta{}} {\path {\pathelem{}} {\x{}}}}
  {\object{}}
  {\z{}}}
{\polaritymeta{}}}
&=&
  {\isprop {\propisnotmeta{}} {\path {\pathelem{}} {\x{}}}}
  & \x{} \not= \z{}\ \text{and}\ \z{} \not\in {\fv {\propisnotmeta{}}}

\\
{\pluspolarity
{\replacefor
  {\isprop {\propisnotmeta{}} {\path {\pathelem{}} {\x{}}}}
  {\object{}}
  {\z{}}}
}
&=&
{\topprop{}}
  & \x{} \not= \z{}\ \text{and}\ \z{} \in {\fv {\propisnotmeta{}}}
\\
{\minuspolarity
{\replacefor
  {\isprop {\propisnotmeta{}} {\path {\pathelem{}} {\x{}}}}
  {\object{}}
  {\z{}}}
}
&=&
{\botprop{}}
  & \x{} \not= \z{}\ \text{and}\ \z{} \in {\fv {\propisnotmeta{}}}

\\
{\withpolarity
{\replacefor
  {\topprop{}}
  {\object{}}
  {\x{}}}
{\polaritymeta{}}}
&=&
  {\topprop{}}

\\
{\withpolarity
{\replacefor
  {\botprop{}}
  {\object{}}
  {\x{}}}
{\polaritymeta{}}}
&=&
  {\botprop{}}

\\
{\pluspolarity
{\replacefor
  {({\impprop {\prop{1}} {\prop{2}}})}
  {\object{}}
  {\x{}}}
}
&=&
{\impprop 
  {\minuspolarity {\replacefor {\prop{1}} {\object{}} {\x{}}}}
  {\pluspolarity {\replacefor {\prop{2}} {\object{}} {\x{}}}}}
\\
{\minuspolarity
{\replacefor
  {({\impprop {\prop{1}} {\prop{2}}})}
  {\object{}}
  {\x{}}}
}
&=&
{\impprop 
  {\pluspolarity {\replacefor {\prop{1}} {\object{}} {\x{}}}}
  {\minuspolarity {\replacefor {\prop{2}} {\object{}} {\x{}}}}}
\\
{\withpolarity
{\replacefor
  {({\orprop {\prop{1}} {\prop{2}}})}
  {\object{}}
  {\x{}}}
{\polaritymeta{}}}
&=&
{\orprop 
  {\withpolarity
    {\replacefor {\prop{1}} {\object{}} {\x{}}}
    {\polaritymeta{}}}
  {\withpolarity
    {\replacefor {\prop{2}} {\object{}} {\x{}}}
    {\polaritymeta{}}}}
\\
{\withpolarity
{\replacefor
  {({\andprop {\prop{1}} {\prop{2}}})}
  {\object{}}
  {\x{}}}
{\polaritymeta{}}}
&=&
{\andprop 
{\withpolarity
  {\replacefor {\prop{1}} {\object{}} {\x{}}}
  {\polaritymeta{}}}
{\withpolarity
  {\replacefor {\prop{2}} {\object{}} {\x{}}}
  {\polaritymeta{}}}}

    \\\\

{\withpolarity
{\replacefor
  {\path {\pathelem{}} {\x{}}}
  {\path {\pathelemp{}} {\y{}}}
  {\x{}}}
{\polaritymeta{}}}
           &=&
{\path{\pathelem{}}{\path {\pathelemp{}} {\y{}}}}

    \\

{\withpolarity
{\replacefor
  {\path {\pathelem{}} {\x{}}}
  {\emptyobject{}}
  {\x{}}}
{\polaritymeta{}}}
           &=&
{\emptyobject{}}

    \\

{\withpolarity
{\replacefor
  {\path {\pathelem{}} {\x{}}}
  {\object{}}
  {\z{}}}
{\polaritymeta{}}}
           &=&
{\path {\pathelem{}} {\x{}}}

& \x{} \not= \z{}
    \\

{\withpolarity
{\replacefor
  {\emptyobject{}}
  {\object{}}
  {\x{}}}
{\polaritymeta{}}}
           &=&
{\emptyobject{}}

\end{array}
$$
\center{\text{Substitution on types is capture-avoiding structural recursion.}}
\caption{Substitution}
\end{figure*}